\documentclass[twocolumn,twocolappendix,trackchanges]{aastex631}
\mathchardef\mhyphen="2D
\newcommand\tkband{\mathit{T}_{\mathit{K\mhyphen} \mathrm{band}}}
\newcommand{\Nstars}{\textcolor{black}{52}}
\newcommand{\Nstarsweobserved}{\textcolor{black}{42}}

\newcommand{\Nstarsphot}{\textcolor{black}{44}}
\newcommand{\Nstarsmodeled}{\textcolor{black}{32}}

\submitjournal{ApJ}

\accepted{May 19, 2024}

\shorttitle{Magnetic field strength of Class I and FS sources}
\shortauthors{Flores et al.}
\graphicspath{{./}{figures/}}

\begin{document}

\title{iSHELL $K$-band Survey of Class I and Flat Spectrum Sources: \\Magnetic field measurements in the protostellar phase
}

\email{caflores@asiaa.sinica.edu.tw}
\correspondingauthor{Christian Flores}
\author[0000-0002-8591-472X]{C. Flores}
\affiliation{Institute for Astronomy, University of Hawaii at Manoa, 640 N. Aohoku Place, Hilo, HI 96720, USA \\}
\affiliation{Institute of Astronomy and Astrophysics, Academia Sinica, 11F of AS/NTU Astronomy-Mathematics Building,\\ No.1, Sec. 4, Roosevelt Rd, Taipei 10617, Taiwan, R.O.C. \\}

\author[0000-0002-8293-1428]{M. S. Connelley}
\author[0000-0001-8174-1932]{B. Reipurth}
\affiliation{Institute for Astronomy, University of Hawaii at Manoa, 640 N. Aohoku Place, Hilo, HI 96720, USA \\}

\author[0000-0001-9344-0096]{A. Boogert}
\affil{Institute for Astronomy, University of Hawaii at Manoa, 2680 Woodlawn Drive, Honolulu, HI 96822, USA}

\author{G. Doppmann}
\affil{W. M. Keck Observatory, 65-1120 Mamalahoa Hwy, Kamuela, HI, USA}



\begin{abstract}
We perform the first magnetic field strength survey of Class I and Flat Spectrum (FS) sources using $K$-band observations with iSHELL. We obtained new observations of $\Nstarsweobserved$ Class I and FS sources and additionally included 10 sources from the archive. We detect photospheric lines in $\Nstarsphot$ of the sources, in addition to Br$\gamma$, H$_2$, and CO emission in several objects. We model the photospheric absorption lines of $\Nstarsmodeled$ Class I and FS sources and measure their magnetic field strengths, $K$-band temperatures, gravities, projected rotational velocities, and infrared veiling values. We put the physical properties of Class I and FS sources in context by comparing them to the values derived for a sample of Class II sources. We find that a) the average magnetic field strength of Class I and FS sources $\langle B \rangle=2.0\pm0.15$~kG is consistent with the average magnetic field strength of Class II sources $\langle B \rangle=1.8\pm0.15$~kG, and b) the average gravity of Class I and FS objects $\log{g}=3.43\pm0.07$ is lower than the average gravity of Class II sources $\log{g}=3.75\pm0.04$, although there is significant overlap between both gravity distributions. Furthermore, using stellar evolutionary models, we deduce that Class I and FS sources have a median age of $\sim$0.6~Myr, and are, as a group, younger than the Class II stars with a median age of $\sim$3~Myr. Our results confirm that Class I and FS sources host strong magnetic fields on their photospheres. Thus, it is likely that these sources accrete disk material through a magnetosphere similar to the more evolved T Tauri stars. 
\end{abstract}



\section{Introduction} \label{sec:intro}

Directly observing the star formation process is difficult because it starts in the densest parts of molecular cloud cores, where optical light can hardly escape. However, infrared and (sub-)mm photons can travel through heavily obscured material, allowing us to gather information about the (proto)stars and their surrounding environment \citep{Kenyon1990,Connelley2010,Tobin2015}. For this reason, the discovery and classification of young stellar sources have been significantly more successful at infrared and (sub)-mm wavelengths \citep{Lada1984,Evans2009}, and over the years, a generally accepted classifications scheme of young stellar sources emerged. 

\cite{Lada1987} introduced the infrared spectral index classification, where the slope of the Spectral Energy Distribution (SED) between near and mid-infrared wavelengths is used to divide young objects as Class I, II, and III sources based on their decreasing infrared slope values. This categorization was subsequently expanded to include the Class 0 \citep{Andre1993}, and the Flat Spectrum (FS) sources \citep{Greene1994}.


According to this division, the earliest phases of the stellar evolution are the Class 0 and I sources which are still in a strong accretion phase, thus are considered protostars \citep{White2007}. The Class~II sources are almost fully assembled stars which accrete the last few percent of their material through a disk. Finally, the Class~III sources are young and still contracting pre-main sequence stars that have accreted or dissipated most, if not all, of its gaseous circumstellar material. These infrared observational criteria, however, are based on the SED of the sources, where the circumstellar material is not resolved. Therefore, geometrical effects, unseen companions, and accretion-related processes could alter the inferred spectral class.


In order to test the existence of a real evolutionary sequence in this observational classification, two spectroscopic studies have compared the properties of Class I and Class II sources. \cite{White2004} observed 15 Class~I and 26 Class~II sources from Taurus-Auriga using high-resolution $R\sim34,000$ optical spectra from Keck/HIRES \citep{Vogt1994}. They found that the distribution of stellar masses, optical veiling, rotation rates, and mass accretion rates of Class I sources are similar to those observed in Class II sources. Furthermore, in their study, \cite{White2004} found that the stellar luminosities of Class I sources translate into ages of $\sim10^{6}$ yr, consistent with the typical ages of Class~II in Taurus.

\cite{Doppmann2005} performed a large high-resolution (R~$\sim18,000$) infrared spectroscopic study of 52 Class~I and FS sources with the Keck/NIRSPEC instrument \citep{McLean1998}. They found Class~I and FS sources span similar ranges in temperatures, gravities, and stellar luminosities as the Class~II sources, but the IR-veiling and mass accretion rates were higher in Class~I and FS sources. These new results suggested that Class I and FS sources are self-embedded protostars undergoing significant mass accretion, in disagreement with the conclusions of \cite{White2004}. Thus, from the spectroscopic point of view the question remains: are Class I sources truly different from Class II sources? 



An important aspect to the question of the evolution of low-mass young stars is whether Class~I sources have magnetic fields comparable to the Class~II stars. Over 100 measurements of magnetic fields of Class~II and Class~III stars have been performed \citep[e.g.,][]{Johns-Krull2007,Yang2005,Sokal2020, Donati2019, Lopez-Valdivia2021, Flores2022, Lavail2017}. Yet only a few Class~I and FS sources have measured unpolarized magnetic field strengths, e.g., WL17 \citep{Johns-Krull2009} and V347 Aur \citep{Flores2019}. Therefore, it has not been possible to directly compare the two classes in a meaningful way or find whether there is a change in the magnetic field strengths with evolutionary classes.

In this paper, we conduct the first survey of Class I and FS sources in which magnetic field strengths are measured. Additionally, we determine stellar gravities, temperatures, projected rotational velocity, and infrared veiling, and compare these properties with those derived for Class II sources. The spectroscopic observations obtained with iSHELL are presented in Section \ref{sec:observations}. We derive stellar parameters for the Class~I and FS sources using the radiative transfer code MoogStokes \citep{Deen2013} in Section \ref{sec:stellar_params}. In Section \ref{sec:results}, we present the stellar parameters for the Class~I and FS sources, and use the magnetic stellar evolutionary models of \cite{Feiden2016} to derive stellar masses and ages. Additionally, in sections \ref{sec:accretion_and_lines} and \ref{sec:results_co_narrow_lines}, we analyze emission lines and non-stellar narrow CO absorption which provide information on the circumstellar environment of the stars. We discuss our results in Section \ref{sec:discussion} and summarize our findings in Section \ref{sec:conclusion}.


\begin{deluxetable*}{lccccccccc}
\tabletypesize{\scriptsize}
\tablewidth{0pt}
\tablecolumns{9}
\tablecaption{Observation Summary. \label{table:observation_summary}}
\tablehead{
\colhead{Source name}  & \colhead{Alt. Name } & {Spectral Index } &  {Region} & \colhead{$\alpha$ J(2000)} & \colhead{$\delta$ J(2000)} & \colhead{$K_{mag}$} & \colhead{Obs. date} & \colhead{i.  time (min)} & \colhead{$S/N$}}  
\startdata
IRAS03220+3035(N) &  & 0.51$\pm$0.23 & Perseus & 03$^h$ 25$^m$ 09$^s$.4 & +30$^{\circ}$ 46$\arcmin$ 21$\arcsec$ & 10.8 & 170825 & 50 & 75 \\ 
IRAS03220+3035(S) &  & 0.51$\pm$0.23 & Perseus & 03$^h$ 25$^m$ 09$^s$.4 & +30$^{\circ}$ 46$\arcmin$ 21$\arcsec$ & 9.8 & 191116 & 100 & 80 \\ 
IRAS03260+3111(B) &  & 1.43$\pm$0.33 &  Perseus & 03$^h$ 29$^m$ 10$^s$.0 & +31$^{\circ}$ 21$\arcmin$ 59$\arcsec$ & 8.7 & 161023 & 75 & 40 \\ 
IRAS03301+3111 &   & 0.38$\pm$0.17 &  Perseus & 03$^h$ 33$^m$ 12$^s$.8 & +31$^{\circ}$ 21$\arcmin$ 24$\arcsec$  & 9.0 & 181206 & 35 & 110  \\ \hline
DG Tau &  & 0.17$\pm$0.24 & Taurus & 04$^h$ 27$^m$ 04$^s$.7 & +26$^{\circ}$ 06$\arcmin$ 16$\arcsec$ & 7.0 &  200125 & 25 & 230 \\ 
GV Tau S  &   & 1.11$\pm$0.25 & Taurus & 04$^h$ 29$^m$ 23$^s$.7 & +24$^{\circ}$ 32$\arcmin$ 58$\arcsec$ & 8.1 & 161017 & 20 & 133 \\
Haro 6-13 & & $-$0.02$\pm$0.30 & Taurus & 04$^h$ 32$^m$ 15$^s$.4 & +24$^{\circ}$ 28$\arcmin$ 59$\arcsec$ & 8.1 & Flores & 25 & 87  \\
Haro 6-28 A & V1026 Tau A & 0.09$\pm$0.26 & Taurus & 04$^h$ 35$^m$ 56$^s$.8 & +22$^{\circ}$ 54$\arcmin$ 35$\arcsec$ & 9.5 & 161016 & 25 &  75 \\ 
IRAS04108+2803(E) &   & 1.16$\pm$0.25 & Taurus & 04$^h$ 13$^m$ 54$^s$.7 & +28$^{\circ}$ 11$\arcmin$ 31$\arcsec$ & 11.1 & 200216 & 150 &  62  \\ 
IRAS04108+2803(W) & & 0.34$\pm$0.29 & Taurus & 04$^h$ 13$^m$ 53$^s$.4 & +28$^{\circ}$ 11$\arcmin$ 23$\arcsec$ & 10.4 & 191126 & 25 & 42 \\ 
IRAS04113+2758(S) & [BHS98] MHO 2  & $-$0.02$\pm$0.24 & Taurus & 04$^h$ 14$^m$ 26$^s$.4 & +28$^{\circ}$ 05$\arcmin$ 59$\arcsec$ & 7.8 & 191017 & 50 & 237 \\ 
IRAS04158+2805 & & 0.41$\pm$0.23 & Taurus & 04$^h$ 18$^m$ 58$^s$.1 & +28$^{\circ}$ 12$\arcmin$ 23$\arcsec$ & 11.2 & 161017 & 150 &  70 \\ 
IRAS04181+2654(M) & & 0.38$\pm$0.26  & Taurus & 04$^h$ 21$^m$ 10$^s$.4 & +27$^{\circ}$ 01$\arcmin$ 37$\arcsec$ & 11.1 & 190113 & 120 &  73 \\ 
IRAS04181+2654(S) & & 0.64$\pm$0.30  & Taurus & 04$^h$ 21$^m$ 11$^s$.5 & +27$^{\circ}$ 01$\arcmin$ 09$\arcsec$ & 10.3 & 190101 & 110 & 66  \\ 
IRAS04295+2251 & & 0.66$\pm$0.18  & Taurus & 04$^h$ 32$^m$ 32$^s$.0 & +22$^{\circ}$ 57$\arcmin$ 26$\arcsec$ & 10.1 & 190101 & 100 & 83 \\ 
IRAS04369+2539* & & $-$0.38$\pm$0.21 & Taurus & 04$^h$ 39$^m$ 55$^s$.8 & +25$^{\circ}$ 45$\arcmin$ 02$\arcsec$ & 6.28 & 181005 & 10 & 110  \\ 
IRAS04489+3042 &  & 0.28$\pm$0.23   & Taurus & 04$^h$ 52$^m$ 06$^s$.7 & +30$^{\circ}$ 47$\arcmin$ 17$\arcsec$ & 10.4 & 200215 & 75 &  98 \\ 
V347 Aur & & $-$0.10$\pm$0.27  & Taurus-Aur & 04$^h$ 56$^m$ 57$^s$.0 & +51$^{\circ}$ 30$\arcmin$ 51$\arcsec$ & 7.8 & 180831 & 10 &  70 \\ \hline
IRAS04591-0856 &  & 0.51$\pm$0.29 & Orion & 05$^h$ 01$^m$ 29$^s$.6 & -08$^{\circ}$ 52$\arcmin$ 17$\arcsec$ & 10.6 & 181231 & 100  &  31 \\ 
IRAS05256+3049 & & 1.09$\pm$0.37 & Orion &  05$^h$ 28$^m$ 49$^s$.9 & +30$^{\circ}$ 51$\arcmin$ 29$\arcsec$ & 10.17 & 170317 & 65 & 21 \\ 
IRAS05311-0631 &  & 0.71$\pm$0.26 & Orion & 05$^h$ 33$^m$ 32$^s$.5 & -06$^{\circ}$ 29$\arcmin$ 44$\arcsec$ & 10.1  & 191220 & 75 & 80   \\ 
IRAS05379-0758(2) & [MB91] 54 & 0.74$\pm$0.47  & Orion  &  05$^h$ 40$^m$ 20$^s$.3 & -07$^{\circ}$ 56$\arcmin$ 25$\arcsec$ & 10.9 & 200308 & 120 &  57  \\ 
IRAS05513-1024 & & $-$0.07$\pm$0.18 & Orion & 05$^h$ 53$^m$ 42$^s$.5 & -10$^{\circ}$ 24$\arcmin$ 00$\arcsec$  & 5.96  & 181005 & 6 & 130  \\ 
IRAS05555-1405(4) & [LC2009] 17  & 0.45$\pm$0.33  & Orion  & 05$^h$ 57$^m$ 49$^s$.5 & -14$^{\circ}$ 05$\arcmin$ 33$\arcsec$ & 10.2  & 190101 & 100 & 74  \\  \hline
2MASSJ16263682 &  & 0.17$\pm$0.11 & Ophiuchus & 16$^h$ 26$^m$ 36$^s$.8 & -24$^{\circ}$ 15$\arcmin$ 51$\arcsec$ & 10.3  &  180724 & 60 &  54 \\ 
Elias 2-32 & $\left[ \rm GY92 \right]$ 273 & 0.14$\pm$0.25 & Ophiuchus & 16$^h$ 27$^m$ 28$^s$.4 & -24$^{\circ}$ 27$\arcmin$ 21$\arcsec$ & 10.1 &  200820-21 & 82 &  87 \\ 
$\left[ \rm GY92 \right]$ 21 & ISO-Oph 37 & 0.31$\pm$0.35 & Ophiuchus & 16$^h$ 26$^m$ 23$^s$.6 & -24$^{\circ}$ 24$\arcmin$ 38$\arcsec$ & 10.2 & 170319 & 25 & 50  \\ 
$\left[ \rm GY92 \right]$ 224 & WLY 1-43 & 0.15$\pm$0.26 & Ophiuchus & 16$^h$ 27$^m$ 11$^s$.2 & -24$^{\circ}$ 40$\arcmin$ 47$\arcsec$ & 10.2 & 170516 & 50 & 38\\ 
$\left[ \rm GY92 \right]$ 235 & WLY 2-32b & 0.02$\pm$0.34 & Ophiuchus & 16$^h$ 27$^m$ 13$^s$.8 & -24$^{\circ}$ 43$\arcmin$ 31$\arcsec$ & 9.9 & Sullivan & & 94\\ 
$\left[ \rm GY92 \right]$ 284 & ISO-Oph 151  & $-$0.02$\pm$0.27 & Ophiuchus & 16$^h$ 27$^m$ 30$^s$.8 & -24$^{\circ}$ 24$\arcmin$ 56$\arcsec$ & 10.1 & Sullivan & & 50 \\ 
$\left[ \rm GY92 \right]$ 33 & ISO-Oph 43 & $-$0.22$\pm$0.43 & Ophiuchus & 16$^h$ 26$^m$ 27$^s$.5 & -24$^{\circ}$ 41$\arcmin$ 53$\arcsec$ & 10.0 & Sullivan & & 69 \\ 
IRAS16235-2416 & ISO-Oph 48 & 1.34* &  Ophiuchus & 16$^h$ 26$^m$ 34$^s$.2 & -24$^{\circ}$ 23$\arcmin$ 28$\arcsec$ & 6.26 & 190604 & 23 & 75 \\
IRAS16285-2355 & Oph IRS63 & 0.33$\pm$0.34 & Ophiuchus & 16$^h$ 31$^m$ 35$^s$.6 & -24$^{\circ}$ 01$\arcmin$ 29$\arcsec$ & 9.2 & 200823-24 & 100 & 87 \\
IRAS16285-2358 & & 0.12$\pm$0.18 & Ophiuchus & 16$^h$ 31$^m$ 33$^s$.8 & -24$^{\circ}$ 04$\arcmin$ 47$\arcsec$ & 10 & 200821 & 30 & 61 \\
IRAS16288-2450(W2) & & $-$0.15$\pm$0.18  & Ophiuchus & 16$^h$ 31$^m$ 52$^s$.1 & -24$^{\circ}$ 56$\arcmin$ 16$\arcsec$ & 8.7 & 180723 & 23 &  65 \\ 
IRAS16316-1540 & HBC 650 & 0.15$\pm$0.36  & Ophiuchus & 16$^h$ 34$^m$ 29$^s$.3 & -15$^{\circ}$ 47$\arcmin$ 02$\arcsec$ &  8.3 & 180628 & 81 & 113 \\ 
EM* SR 24A & WSB 41 & $-$0.30$\pm$0.30 & Ophiuchus & 16$^h$ 26$^m$ 58$^s$.4 & -24$^{\circ}$ 45$\arcmin$ 32$\arcsec$ & 7.5 & Sullivan & & 108 \\ 
EM* SR 24B  & WSB 42 & $-$0.30$\pm$0.30 & Ophiuchus & 16$^h$ 26$^m$ 58$^s$.5 & -24$^{\circ}$ 45$\arcmin$ 37$\arcsec$ & 7.0 & Sullivan & & 104 \\ 
VSSG 17 & Elias 2-33 & 0.06$\pm$0.41 & Ophiuchus & 16$^h$ 27$^m$ 30$^s$.2 & -24$^{\circ}$ 27$\arcmin$ 44$\arcsec$ & 9.0 & 200706 & 25 & 103 \\ 
WL20(E) & & 0.85$\pm$0.38 & Ophiuchus & 16$^h$ 27$^m$ 15$^s$.7 & -24$^{\circ}$ 38$\arcmin$ 43$\arcsec$ & 9.5 & Sullivan & & 75 \\ 
WL20(W) & & 0.85$\pm$0.38 & Ophiuchus & 16$^h$ 27$^m$ 15$^s$.7 & -24$^{\circ}$ 38$\arcmin$ 43$\arcsec$ & 9.5  & Sullivan & & 60 \\ 
WL4  & VSSG 26 & 0.01$\pm$0.27 & Ophiuchus & 16$^h$ 27$^m$ 18$^s$.5 & -24$^{\circ}$ 29$\arcmin$ 06$\arcsec$ & 9.7 & Sullivan & & 79  \\ 
WLY2-42  & & 0.10$\pm$0.28 & Ophiuchus & 16$^h$ 27$^m$ 21$^s$.5 & -24$^{\circ}$ 41$\arcmin$ 43$\arcsec$ & 8.5 & Sullivan & & 100 \\ 
WLY2-54  & YLW 52 & 0.14$\pm$0.39 & Ophiuchus & 16$^h$ 27$^m$ 51$^s$.8 & -24$^{\circ}$ 31$\arcmin$ 45$\arcsec$  & 8.7 & Sullivan & & 72 \\ 
YLW 45 & ISO-Oph 167 & 0.17$\pm$0.31 & Ophiuchus & 16$^h$ 27$^m$ 39$^s$.8 & -24$^{\circ}$ 43$\arcmin$ 15$\arcsec$ & 9.0 & 200829 & 50 & 150 \\ \hline
$\left[ \rm EC92 \right]$ 92 &  & 0.92$\pm$0.29  & Serpens & 18$^h$ 29$^m$ 57$^s$.7 & +01$^{\circ}$ 12$\arcmin$ 52$\arcsec$ & 10.5 & 200706 & 80 & 87 \\ 
$\left[ \rm EC92 \right]$ 95 & & 0.92$\pm$0.29  & Serpens & 18$^h$ 29$^m$ 57$^s$.9 & +01$^{\circ}$ 12$\arcmin$ 46$\arcsec$ & 9.7   & 190707 & 50 &  85 \\ 
$\left[ \rm EC92 \right]$ 129 &  & 0.37$\pm$0.28 & Serpens & 18$^h$ 30$^m$ 02$^s$.7 & +01$^{\circ}$ 12$\arcmin$ 28$\arcsec$ & 9.9& 190706  &50 & 113 \\ 
IRAS18270-0153 &  & 0.60$\pm$0.16 & Serpens & 18$^h$ 29$^m$ 36$^s$.9 & -01$^{\circ}$ 51$\arcmin$ 02$\arcsec$ & 10.14 & 200820-21 & 60 & 32 \\ 
\hline
IRAS19247+2238(1) &  & $-$0.13$\pm$0.39 & Vulpecula? & 19$^h$ 26$^m$ 51$^s$.3 & +22$^{\circ}$ 45$\arcmin$ 13$\arcsec$ & 8.4 & 191116 & 65 &  79  \\ 
IRAS19247+2238(2) & & $-$0.13$\pm$0.39 & Vulpecula? & 19$^h$ 26$^m$ 51$^s$.3 & +22$^{\circ}$ 45$\arcmin$ 13$\arcsec$ & 9.1 & 190706 & 50 &  87 \\ 
$\left[ \rm TS84 \right]$ IRS 5 (NE) & 	[B87] 7 & 1.11$\pm$0.35 & Corona Australis & 19$^h$ 01$^m$ 48$^s$.1 & -36$^{\circ}$ 57$\arcmin$ 22$\arcsec$ & 10.1 & 190707 & 50 & 50 \\ 
\enddata
\tablecomments{The S/N reported in the last column corresponds to a S/N per pixel as reported by  Spextool. The S/N per resolution element is $\sqrt{5}$ times higher for the $0\farcs75$ slit width configuration. For Sullivan in Column 8, please refer to \cite{Sullivan2019} for more details about the observations. The Flat Spectrum source Haro 6-13 was already presented in \cite{Flores2022}. *IRAS04369+2539 has an spectral index lower than -0.3, however, we included it has it is an eruptive source resembling an EXor. IRAS16235-2416 has no 20~\micron \, data but it has a measured $\alpha=1.34$ from \cite{Connelley2010} using 10 and 100~\micron \, observations from IRAS. [TS84] IRS5 is a close binary with a separation of $\sim 1\arcsec$, we observed the NE component. Haro 6-28 is a $0.66\arcsec$ binary \citep{Leinert1993}, that we resolved in a $\sim0.4\arcsec$ seeing night. Here, we report the brighter eastern component Haro 6-28 A. For sources with dates in the Obs. date column, we combined the data into a single spectrum.} 
\end{deluxetable*}

\section{Observations} \label{sec:observations}

\subsection{Sample Selection and observations}
We have selected Class~I and FS sources observed by \cite{Connelley2010}, \cite{Doppmann2005} and from the ODISEA sample \citep{Cieza2019}. All of the observed sources have infrared $K$-band magnitudes of $K \lesssim 11$. The sources come from different star-forming regions, including Taurus-Auriga, Serpens, Corona Australis, Perseus, and Ophiuchus. The sources from Ophiuchus were selected based on their spectral index between 2 and 24~$\micron$ from 2MASS \citep{Skrutskie2006} and Spitzer/MIPS \citep{Rieke2004}. The rest of the sources have a spectral index calculated between 2 and 22~$\micron$ from the 2MASS and WISE surveys (see Table \ref{table:observation_summary}). 

We observed $\Nstarsweobserved$ Class~I and FS sources between 2016, October 16, and 2020 August 24, using the high-resolution near-infrared echelle spectrograph iSHELL on IRTF \citep{Rayner2016, Rayner2022}. The observations were performed in the $K2$ mode, thus from 2.09 to 2.38 $\mu$m, using the 0\farcs75 slit to achieve a spectral resolution of $R\sim50,000$. Due to a known fringing issue in the iSHELL spectra, we decided to adopt the following observing strategy. First, we guided on a young star and integrated on it long enough to obtain a spectrum with a signal-to-noise ratio (S/N) $>50$ (per pixel) or for a maximum time of 3 hours (when seeing and sky conditions were unfavorable). We immediately acquired a set of calibration spectra consisting of $K$-band flats and arc-lamp spectra. Then, we observed a nearby telluric A0 standard star typically within 0.1 airmasses of the young source and integrated on it long enough to achieve a S/N$>100$ spectrum. Afterward, we obtained a second set of $K$-band calibration files to correct the telluric standard star. The same process was repeated for each star in the sample. 

In addition, 10 of the sources in our sample came from the IRTF/IRSA archive and were already presented in \cite{Sullivan2019}. In Table \ref{table:observation_summary}, we present the summary of the observations, where we include the host star-forming regions, the coordinates, observing dates, and S/N, among others. For only five sources, we did not achieve the goal S/N of 50. The median S/N of the observations is 79, and it ranges from 21 and 237. The S/N values correspond to the ones reported by \texttt{Spextool v5.0.2} \citep{Cushing2004}

\subsection{Data reduction}
The data were reduced using the IDL-based package \texttt{Spextool v5.0.2}\footnote{\url{http://irtfweb.ifa.hawaii.edu/research/dr_resources/}} \citep{Cushing2004}. We first created normalized flat field images and wavelength calibration with the \texttt{xspextool} task. Then, we extracted the spectra of each star (Class~I + FS sources and telluric standards) using the point source extraction configuration. To increase the S/N of the final spectra, we used \texttt{xcombspec} to combine the individual multi-order stellar spectra. Afterward, we used the \texttt{xtellcor} task \citep{vacca2003} to remove atmospheric absorption lines from the science spectra, and \texttt{xmergeorders} to merge multi-order spectra into a single continuous spectrum. We used \texttt{xcleanspec} to eliminate any deviant or negative pixels caused by imperfections in the infrared array. Finally, we continuum-normalized the $K2$  reduced spectra using a first-order broken power-law function with the  \texttt{specnorm}\footnote{\url{ftp://ftp.ster.kuleuven.be/dist/pierre/Mike/IvSPythonDoc/plotting/specnorm.html}} code. 

\section{Magnetic radiative transfer modeling}
\label{sec:stellar_params}

From our sample of $\Nstars$ sources, $\Nstarsphot$ show photospheric lines in their spectra, and only 35 displayed photospheric lines deep enough to perform a robust analysis. Two of the 35 sources are confirmed unresolved binaries or higher-order systems \citep[WL4 and SR24N see ][]{Sullivan2019}, and one source (IRAS 04108+2803W) resulted in a stellar temperature too cold to verify the derived parameters from our modeling approach. That left us with a total of $\Nstarsmodeled$ sources.

To analyze these data, we used a synthetic stellar spectrum technique that combines MARCS stellar atmospheric models \citep{Gustafsson2008}, the magnetic radiative transfer code MoogStokes \citep{Deen2013}, along with atomic and molecular lines from the VALD3 \citep{Ryabchikova2015,Pakhomov2019} and HITEMP databases \citep{Rothman2010}. We used the same method previously described and presented in \cite{Flores2019,Flores2020, Flores2022}. In short, we took a two-step approach to fit the observations. First, we varied seven stellar parameters: temperature, gravity, magnetic field strength, projected rotational velocity, $K$-band veiling, micro-turbulence, and CO abundance. We performed the fitting using eight wavelength regions, one of which contains the CO overtone starting at 2.294~$\micron$. In the second step, we excluded the CO region to avoid potential CO contamination from the circumstellar material (as seen, for example, in \citealt{Doppmann2005} and \citealt{Flores2020}). We anchored the projected rotational velocity to the best value found in the previous step and re-calculated the temperature, gravity, magnetic field strength, $K$-band veiling, and micro-turbulence. In both steps, we fitted the spectra with a Markov Chain Monte Carlo (MCMC) method using the \texttt{emcee} package \citep{Foreman-Mackey2013}. We ran a total of 100 individual chains advancing a total of 1000 steps. 

We found that the spectra of stars cooler than $\tkband \sim$3500~K, with low gravities are more strongly impacted by a forest of water lines. These water lines appear as a sequence of narrow absorption lines and mainly affect the spectrum red-ward of $\sim$2.2~$\micron$. These water lines makes it difficult for us to reliably define a pseudo-continuum in the stellar spectra. Thus, for [GY92] 284, IRAS04113+2758(S), IRAS16285-2358, and VSSG 17, we had to exclude the aluminum regions at $2.110~\micron$ and $2.117~\micron$ during the model fitting process. To quantify the effects of excluding these wavelength regions on the derived stellar parameters of the stars, we tested our method on a sample of 14 main and post-main sequence stars with precise literature values. In general, we find that excluding the aluminum lines only produce temperature and gravity differences of at most $3\%$ but are often much smaller (see Appendix \ref{appendix:aluminum-no-aluminum} for more details).

Finally, we have quantified the goodness of fit for each model using the reduced $\chi^2$ metric, and we provide their values in the last column of Table \ref{table:Stellar_params}. Details of the computations of $\chi^2$ and illustrative examples are presented in Appendix \ref{appendix:chi-squared}.


\section{Results} \label{sec:results}

\begin{figure*}[ht!]
\epsscale{1.0}
\plotone{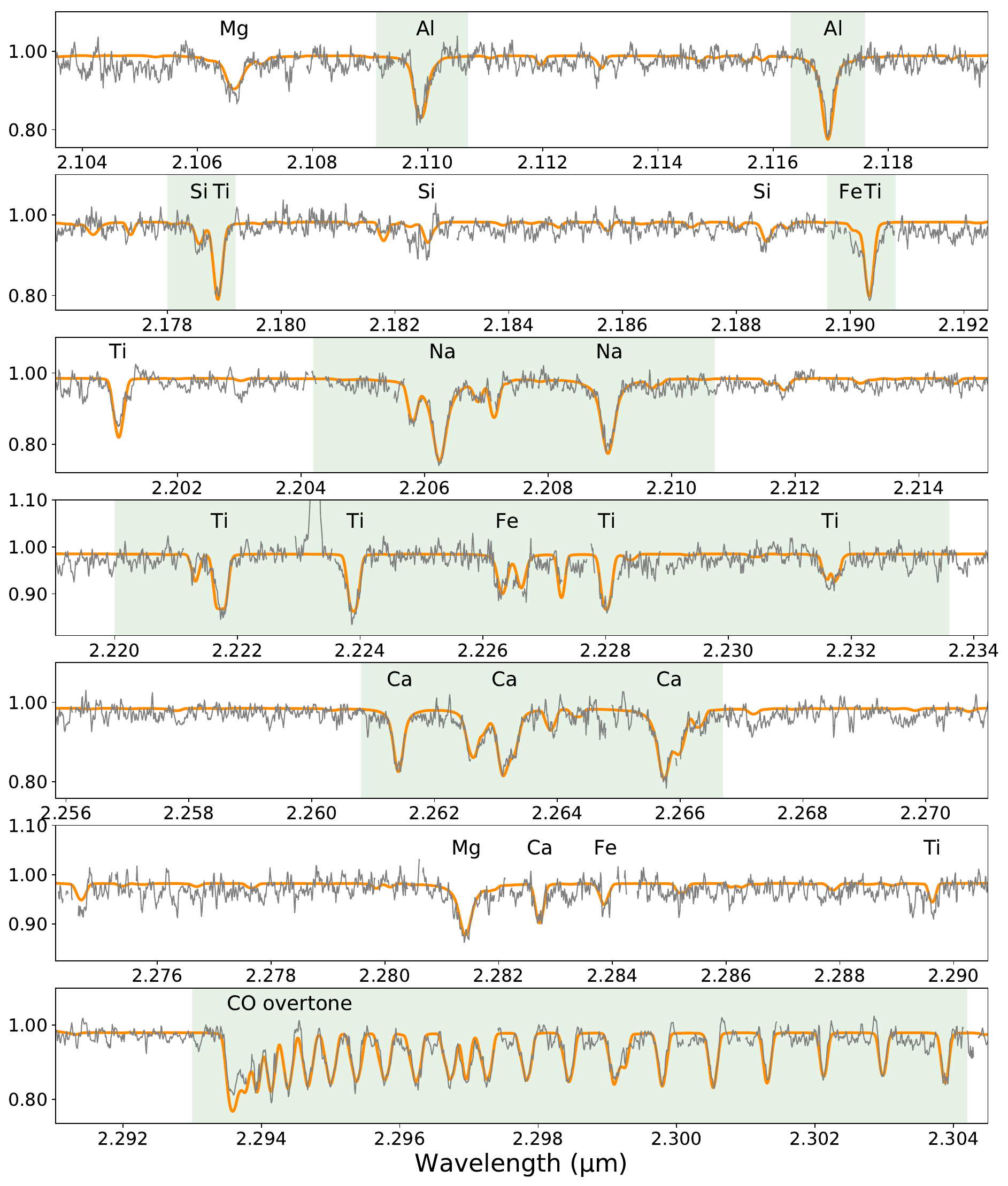}
\caption{Comparison between the spectrum of the Class~I source IRAS03301+3111 (in gray) and its best fit model (in orange). The seven panels show regions with the strongest photospheric absorption in lines. Green faded boxes denote the regions used in the determination of stellar parameters. In each panel, some of the absorption lines are labeled. The best fit model matches very well the spectrum of the Class~I source. This happens in the regions where the fit was performed and also in regions not considered in the fitting process (Mg at 2.1065~$\micron$ and 2.2815~$\micron$, for example). In the fourth panel, a tiny portion of the $\rm H_2$ $v=1-0 \, s(0)$ emission line is shown at 2.2235~$\micron$. \label{fig:spectrum-example}}
\end{figure*}

We used the technique described in the previous section to measure the stellar parameters of $\Nstarsmodeled$ Class~I and FS sources \footnote{Upon deriving stellar parameters, we found that three extra sources: IRAS04292+2422 West, IRAS06393+0913 and IRAS18341-0113(N) appeared to be background giant stars with only visual association to the star-forming regions. These sources are slow rotators ($v\sin{i}< 10 \, \rm km \, s^{-1}$), have no infrared veiling, a null magnetic field strength, and no $\rm Br \gamma$ line in their spectra, i.e., do not display any sign of youth other than a positive infrared spectral index. For this reason, we excluded them from our analysis.}. We present portions of the $K$-band spectrum of IRAS03301+3111 with its best fit model overlaid in Figure \ref{fig:spectrum-example}. The figures for the remaining 31 sources can be accessed in the online journal as figure sets. The plots display seven panels that contain the strongest absorption lines in the spectra of the stars, each covering wavelength regions of about 150~\AA. The most prominent absorption lines are labeled in the figures, and the faded green boxes delimit the regions that were used to derive the stellar parameters of the stars (see Section \ref{sec:stellar_params}). For visual clarity, the spectrum of the stars was smoothed with a 5-pixel wide Savitzky-Golay function of order 2. This smoothing was performed only for figure display and not during the data fitting process.

\figsetstart
\figsetnum{1}
\figsettitle{Class I and FS source spectra}

\figsetend

\begin{deluxetable*}{lcccccc | c  c | c}
\tabletypesize{\scriptsize}
\tablecaption{Stellar parameters of Class I and FS sources derived from \textit{K}-band spectra. \label{table:Stellar_params}}
\tablehead{
\colhead{Source} & \colhead{$\tkband$} & \colhead{log(g) } & \colhead{$\rm r_K$} &\colhead{$v_{micro}$}  & \colhead{$\langle \rm B \rangle $}   & \colhead{$v\sin(i)$} &
\colhead{Masses} & \colhead{Ages} & \colhead{Reduced} \\
\colhead{name} & \colhead{(K)} & \colhead{} & \colhead{} &  \colhead{(km s$^{-1}$)} & \colhead{(kG)}  & \colhead{(km s$^{-1}$)} & \colhead{($\rm M_\odot$)} & \colhead{(Myr)} & \colhead{$\chi^2$} }
\startdata
DG Tau & $4035^{+58}_{-70}$ & $3.28^{+0.13}_{-0.20}$ & $ 3.60^{+0.30}_{-0.38}$ & $3.57^{+0.43}_{-1.30}$ & $1.52^{+0.33}_{-0.34}$ & $27.31^{+1.20}_{-1.34}$ & & & $1.01$ \\
$\rm [EC92]$ 92      & $3747^{+72}_{-78}$ & $2.89^{+0.12}_{-0.11}$ & $0.91^{+0.1}_{-0.1}$ & $0.29^{+0.42}_{-0.19}$ & $2.6^{+0.34}_{-0.35}$ & $46.71^{+1.04}_{-1.25}$ & & & $1.57$ \\ 
$\rm [EC92]$ 95      & $4207^{+116}_{-72}$ & $3.85^{+0.11}_{-0.19}$ & $0.95^{+0.11}_{-0.37}$ & $0.23^{+3.77}_{-0.13}$ & $4.13^{+0.56}_{-4.13}$ & $63.15^{+1.96}_{-2.33}$ & $1.3_{-0.09}^{+0.07} $ & $2.4_{-1.06}^{+2.01} $ & $1.32$ \\
Elias 2-32 & $3387^{+52}_{-59}$ & $2.79^{+0.1}_{-0.12}$ & $0.48^{+0.06}_{-0.05}$ & $1.94^{+0.47}_{-0.46}$ & $1.22^{+0.14}_{-0.15}$ & $25.38^{+0.41}_{-0.42}$ & & & $2.65$ \\ 
GV Tau S    & $3812^{+66}_{-74}$ & $3.26^{+0.12}_{-0.17}$ & $2.24^{+0.15}_{-0.18}$ & $3.86^{+0.14}_{-0.89}$ & $1.15^{+0.32}_{-0.35}$ & $26.38^{+0.77}_{-0.99}$ & & & $1.30$ \\ 
$\rm [GY92]$ 235 & $3374^{+88}_{-162}$ & $3.63^{+0.17}_{-0.34}$ & $2.21^{+0.18}_{-0.17}$ & $2.63^{+1.09}_{-1.02}$ & $<0.94$ & $12.05^{+0.80}_{-0.94}$ & $0.45_{-0.13}^{+0.18} $ & $1.9_{-0.83}^{+2.28} $ & $1.89$ \\ 
$\rm [GY92]$ 284 & $3390^{+64}_{-65}$ & $3.81^{+0.09}_{-0.11}$ & $0.46^{+0.06}_{-0.04}$ & $1.87^{+0.66}_{-0.61}$ & $1.17^{+0.1}_{-0.11}$ & $5.94^{+0.58}_{-0.36}$ & $0.44_{-0.07}^{+0.06} $ & $3.3_{-0.93}^{+1.26} $ & $2.41$ \\
$\rm [GY92]$ 33  & $3755^{+50}_{-62}$ & $3.91^{+0.07}_{-0.09}$ & $0.29^{+0.02}_{-0.02}$ & $0.20^{+0.43}_{-0.10}$ & $2.16^{+0.12}_{-0.11}$ & $13.07^{+0.94}_{-0.37}$ & $0.83_{-0.08}^{+0.08} $ & $3.7_{-1.13}^{+1.53} $ & $2.31$ \\ 
Haro 6-13* & $3694^{+102}_{-103}$ & $3.52^{+0.23}_{-0.21}$ & $1.55^{+0.12}_{-0.13}$ & $1.93^{+1.10}_{-1.07}$ & $1.16^{+0.30}_{-0.30}$ & $22.20^{+0.83}_{-0.80}$ & $0.88_{-0.15}^{+0.13} $ & $0.95_{-0.35}^{+1.17} $ & $1.76$\\ 
Haro 6-28 A & $3287^{+31}_{-39}$ & $3.86^{+0.05}_{-0.06}$ & $ 0.3^{+0.03}_{-0.03}$ & $0.76^{+0.49}_{-0.65}$ & $1.18^{+0.15}_{-0.16}$ & $12.98^{+0.33}_{-0.58}$ & $0.34_{-0.04}^{+0.04} $ & $4.2_{-1.1}^{+1.45} $ & $2.20$ \\ 
IRAS03260+3111(B)* & $3200^{+90}_{-88}$ & $3.16^{+0.16}_{-0.16}$ & $0.42^{+0.05}_{-0.06}$ & $2.32^{+0.85}_{-0.75}$ & $<1.02$ & $23.31^{+1.36}_{-0.78}$ & $0.39_{-0.11}^{+0.07} $ & $0.6_{-0.0}^{+0.58} $ & $2.72$ \\ 
IRAS03301+3111* & $3457^{+73}_{-85}$ & $3.32^{+0.13}_{-0.14}$ & $1.27^{+0.08}_{-0.06}$ & $1.45^{+0.85}_{-0.67}$ & $1.66^{+0.11}_{-0.11}$ & $10.22^{+0.81}_{-0.67}$ & $0.59_{-0.1}^{+0.13} $ & $0.60_{-0.0}^{+0.35} $ & $1.57$ \\ 
IRAS04108+2803(E) & $3779^{+65}_{-84}$ & $3.71^{+0.13}_{-0.16}$ & $0.89^{+0.05}_{-0.06}$ & $1.41^{+0.71}_{-0.6}$ & $2.14^{+0.17}_{-0.17}$ & $17.35^{+0.9}_{-0.68}$ & $0.91_{-0.11}^{+0.1} $ & $1.8_{-0.78}^{+1.29} $ & $2.61$ \\ 
IRAS04113+2758(S)* & $3204^{+41}_{-16}$ & $3.14^{+0.06}_{-0.05}$ & $1.02^{+0.02}_{-0.03}$ & $1.67^{+0.24}_{-0.22}$ & $1.86^{+0.1}_{-0.11}$ & $26.31^{+0.2}_{-0.58}$ & $0.32_{-0.01}^{+0.14} $ & $0.60_{-0.0}^{+0.20} $ & $1.54$\\ 
IRAS04181+2654(M) & $3520^{+68}_{-45}$ & $3.71^{+0.11}_{-0.12}$ & $0.21^{+0.03}_{-0.03}$ & $3.65^{+0.35}_{-0.84}$ & $<0.92$ & $36.94^{+0.72}_{-0.9}$ & $0.6_{-0.07}^{+0.08} $ & $2.21_{-0.76}^{+1.18} $ & $2.06$\\ 
IRAS04181+2654(S) & $3376^{+84}_{-128}$ & $3.75^{+0.13}_{-0.36}$ & $1.98^{+0.18}_{-0.19}$ & $ 3.7^{+0.3}_{-1.25}$ & $1.90^{+0.33}_{-0.38}$ & $15.2^{+1.31}_{-1.36}$ & $0.43_{-0.10}^{+0.13} $ & $2.75_{-1.47}^{+3.39} $ & $1.96$\\ 
IRAS04295+2251* & $3429^{+136}_{-134}$ & $3.41^{+0.33}_{-0.22}$ & $1.98^{+0.17}_{-0.18}$ & $3.69^{+0.31}_{-1.2}$ & $3.56^{+0.79}_{-0.86}$ & $50.6^{+3.28}_{-2.93}$ & $0.60_{-0.20}^{+0.15} $ & $0.87_{-0.27}^{+2.0} $ & $1.68$\\ 
IRAS04489+3042 & $3324^{+73}_{-210}$ & $ 3.7^{+0.12}_{-0.47}$ & $1.58^{+0.13}_{-0.1}$ & $2.66^{+0.84}_{-1.27}$ & $1.26^{+0.47}_{-0.34}$ & $22.32^{+0.83}_{-0.74}$ & $0.39_{-0.12}^{+0.17} $ & $2.57_{-1.49}^{+3.65} $ & $2.81$\\ 
IRAS04591-0856* & $3339^{+129}_{-122}$ & $3.26^{+0.18}_{-0.22}$ & $0.71^{+0.1}_{-0.09}$ & $3.62^{+0.38}_{-0.92}$ & $<1.21$ & $24.6^{+1.66}_{-1.18}$ & $0.5_{-0.20}^{+0.13} $ & $0.65_{-0.05}^{+0.8} $ & $1.78$ \\ 
IRAS05379-0758(2) & $3459^{+97}_{-94}$ & $3.13^{+0.17}_{-0.12}$ & $0.58^{+0.06}_{-0.06}$ & $1.45^{+0.55}_{-0.41}$ & $2.36^{+0.23}_{-0.27}$ & $30.34^{+0.91}_{-0.95}$ & & & $2.57$ \\ 
IRAS05555-1405(4) & $3966^{+155}_{-164}$ & $3.27^{+0.53}_{-0.27}$ & $0.87^{+0.43}_{-0.27}$ & $ 1.2^{+2.79}_{-1.1}$ & $1.17^{+0.79}_{-1.13}$ & $63.4^{+1.19}_{-1.06}$ & & & $1.56$\\ 
IRAS16285-2358 & $3318^{+93}_{-76}$ & $3.77^{+0.14}_{-0.16}$ & $1.69^{+0.13}_{-0.12}$ & $1.89^{+0.95}_{-0.83}$ & $1.71^{+0.26}_{-0.28}$ & $17.25^{+0.73}_{-1.14}$ & $0.38_{-0.07}^{+0.08} $ & $3.11_{-1.25}^{+1.78} $ & $1.90$\\ 
IRAS16288-2450(W2) & $3473^{+41}_{-70}$ & $3.78^{+0.07}_{-0.1}$ & $0.41^{+0.03}_{-0.04}$ & $1.08^{+0.49}_{-0.62}$ & $3.33^{+0.28}_{-0.28}$ & $30.53^{+1.64}_{-1.59}$ & $0.53_{-0.06}^{+0.07} $ & $3.01_{-0.98}^{+0.85} $ & $1.09$\\ 
IRAS19247+2238(1)* & $3635^{+83}_{-71}$ & $3.43^{+0.19}_{-0.11}$ & $ 2.0^{+0.08}_{-0.08}$ & $ 0.2^{+0.45}_{-0.10}$ & $2.08^{+0.18}_{-0.19}$ & $15.85^{+0.73}_{-0.86}$ &  $0.82_{-0.14}^{+0.1} $ & $0.60_{-0.0}^{+1.45} $ & $2.62$\\ 
IRAS19247+2238(2) & $3580^{+97}_{-95}$ & $3.52^{+0.17}_{-0.11}$ & $1.41^{+0.06}_{-0.06}$ & $0.25^{+0.51}_{-0.14}$ & $2.39^{+0.28}_{-0.29}$ & $25.87^{+0.85}_{-1.16}$ & $0.75_{-0.14}^{+0.13} $ & $0.95_{-0.34}^{+1.02} $ & $2.33$\\ 
$\rm [TS84]$ IRS5 (NE) & $3295^{+89}_{-81}$ & $3.08^{+0.16}_{-0.08}$ & $0.38^{+0.05}_{-0.04}$ & $1.16^{+0.61}_{-0.54}$ & $2.01^{+0.36}_{-0.34}$ & $40.55^{+0.81}_{-1.16}$ & & & $2.30$ \\ 
EM* SR24A*  & $4010^{+122}_{-141}$ & $3.57^{+0.32}_{-0.46}$ & $3.03^{+0.53}_{-0.65}$ & $ 1.8^{+1.56}_{-1.63}$ & $<0.78$ & $35.14^{+1.61}_{-1.46}$ & $1.2_{-0.2}^{+0.12} $ & $0.95_{-0.35}^{+2.08} $ & $1.44$ \\ 
V347 Aur & $3190^{+105}_{-93}$ & $2.94^{+0.25}_{-0.15}$ & $1.17^{+0.09}_{-0.08}$ & $1.56^{+0.78}_{-0.65}$ & $1.27^{+0.16}_{-0.17}$ & $12.01^{+0.61}_{-0.3}$ & & & $2.30$ \\ 
VSSG 17 & $3290^{+87}_{-90}$ & $2.94^{+0.21}_{-0.12}$ & $ 1.5^{+0.13}_{-0.13}$ & $2.61^{+0.96}_{-0.67}$ & $2.13^{+0.44}_{-0.53}$ & $45.45^{+0.63}_{-0.9}$ & & & $1.67$ \\ 
WL20(E)     & $3621^{+78}_{-48}$ & $3.65^{+0.12}_{-0.11}$ & $0.32^{+0.03}_{-0.03}$ & $0.18^{+0.36}_{-0.08}$ & $2.65^{+0.29}_{-0.29}$ & $42.82^{+1.59}_{-1.25}$ & $0.73_{-0.09}^{+0.09} $ & $1.66_{-0.62}^{+0.9} $ & $1.66$\\ 
WL20(W)  & $3390^{+53}_{-52}$ & $ 3.8^{+0.1}_{-0.11}$ & $0.14^{+0.03}_{-0.02}$ & $1.08^{+1.2}_{-0.9}$ & $1.47^{+0.42}_{-0.59}$ & $34.14^{+0.53}_{-1.06}$ & $0.44_{-0.05}^{+0.06} $ & $3.19_{-0.86}^{+1.19} $ & $2.06$\\ 
WLY2-42   & $3308^{+67}_{-46}$ & $3.45^{+0.17}_{-0.07}$ & $ 0.1^{+0.02}_{-0.03}$ & $0.43^{+0.5}_{-0.33}$ & $2.27^{+0.12}_{-0.11}$ & $22.04^{+1.01}_{-0.51}$ & $0.42_{-0.07}^{+0.09} $ & $1.23_{-0.13}^{+1.14}$ & $2.58$
\enddata
\tablecomments{
The reported uncertainties correspond to 3$\sigma$ deviations from the median value obtained from the MCMC distributions. Masses and ages in columns 8 and 9, respectively, were derived using the magnetic stellar evolutionary models from \cite{Feiden2016}. Sources with an $*$ symbol at the end of their names have ages truncated due to the lower limit of 0.6 Myrs in the  \cite{Feiden2016} models.}
\end{deluxetable*}

\subsection{Measured Class I stellar parameters}

In columns 2 through 7 of Table \ref{table:Stellar_params}, we report the stellar parameters of the Class~I and FS sources measured using our magnetic radiative transfer model. Although \cite{Doppmann2005} measured the temperatures, gravities, and projected rotational velocities of a similar number of Class~I and FS objects, this is the first time the magnetic field strengths have been measured for a significant number of Class~I and FS sources. Prior to this work, only a couple of Class~I objects and FS sources had their magnetic field strengths measured, e.g.,  WL17 \citep{Johns-Krull2007} and V347 Aur \citep{Flores2019}.

\begin{figure*}
\gridline{\fig{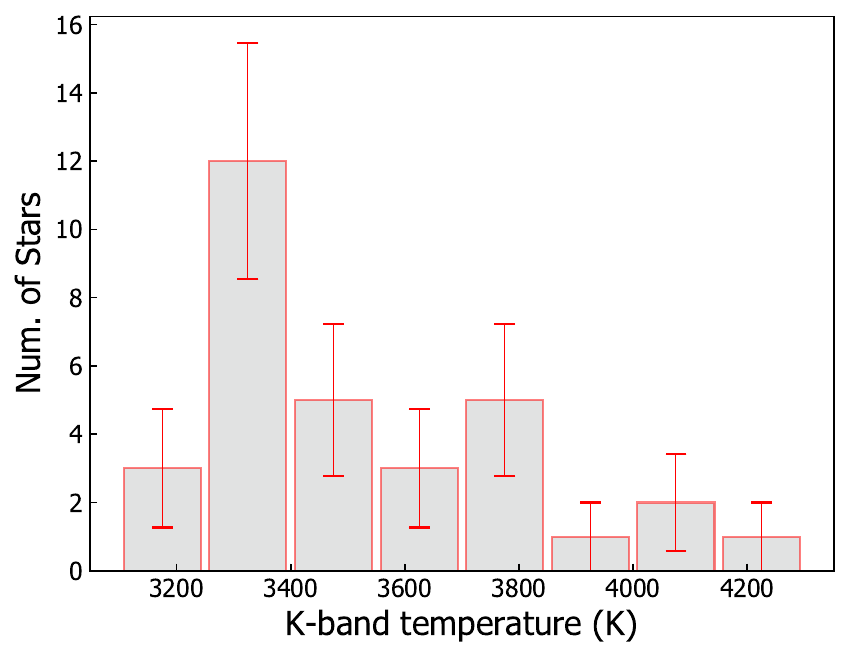}{0.45\textwidth}{(a)}
          \fig{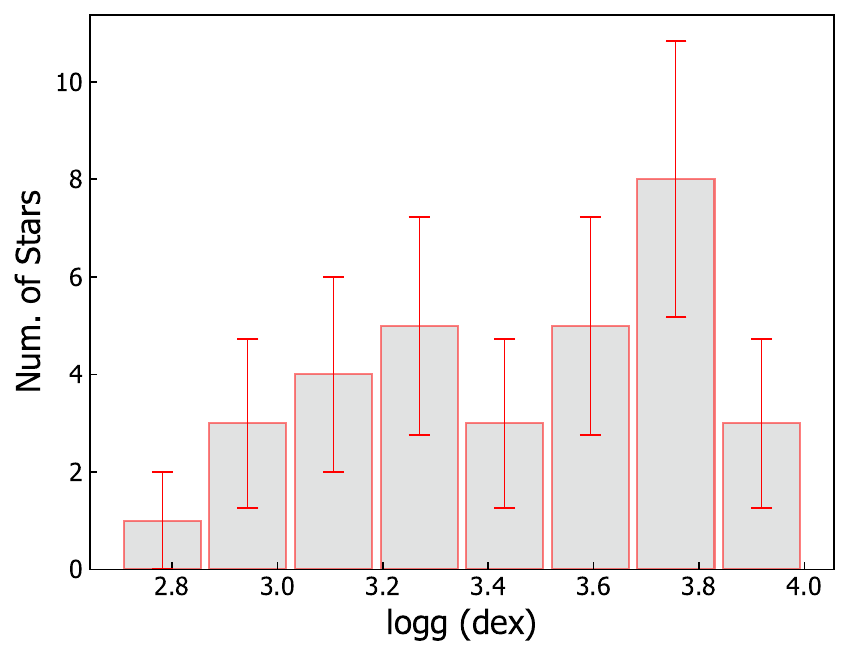}{0.45\textwidth}{(b)}
          }
\gridline{\fig{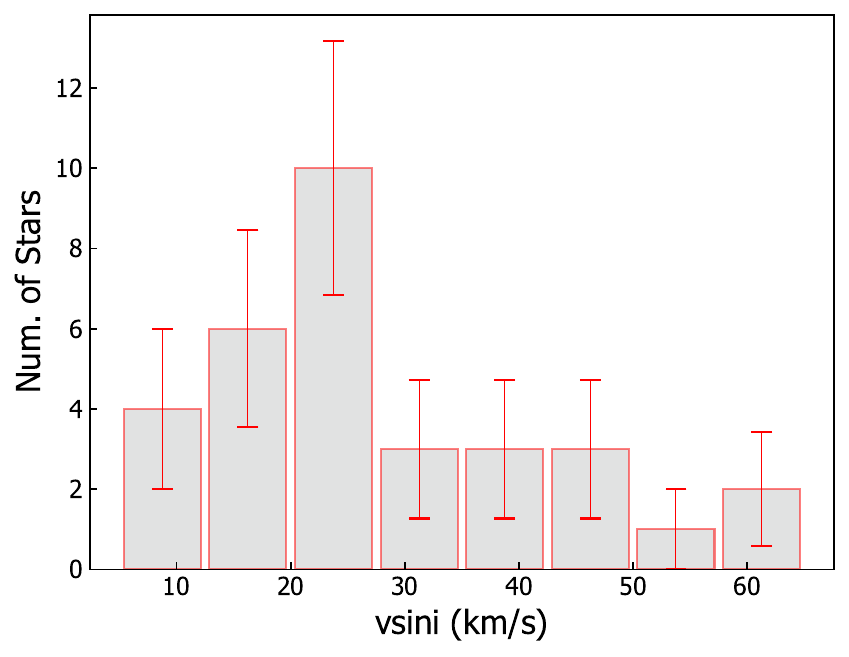}{0.45\textwidth}{(c)}
          \fig{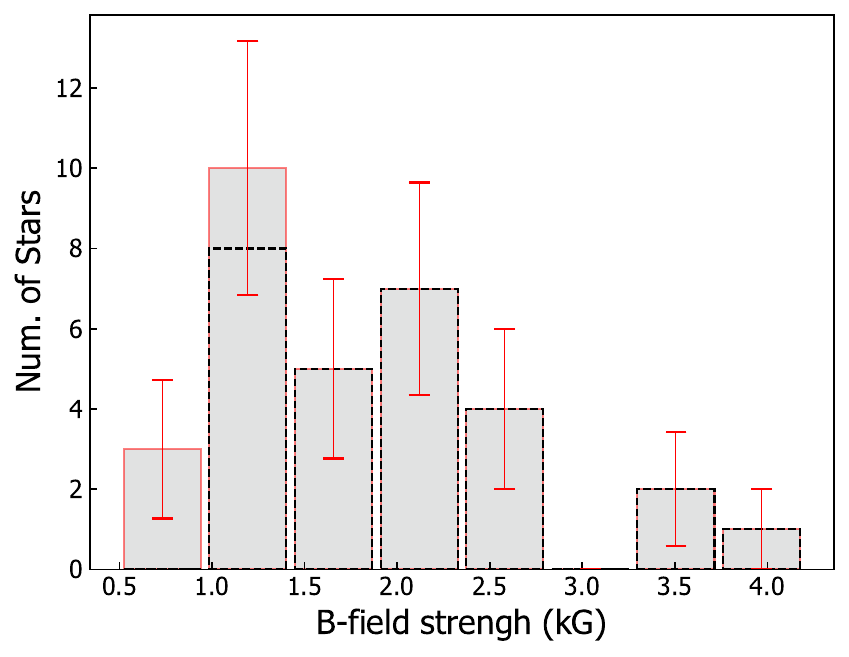}{0.45\textwidth}{(d)}
          }
\caption{We present the results of our modeling technique in form of histograms for (a) $K$-band temperature, (b) surface gravity in log of $\rm cm \, s^{-2}$, (c) projected rotational velocity, and (d) the average magnetic field strength of the stars. Unlike the results presented in \cite{Flores2022} for Class~II sources, the gravity of the Class~I sources have a secondary peak at $\sim3.3$~$\rm cm \, s^{-2}$ with a tail extending to gravities as low as $\log{g} =2.8$ in log of $\rm cm \, s^{-2}$. In panel (d), we also show the histogram excluding the five sources for which we only know their magnetic field upper limits. We show this using black dashed lines.
\label{fig:histogram_Class_I_parameters}}
\end{figure*}

In Figure \ref{fig:histogram_Class_I_parameters}, we present histograms of four of the measured stellar parameters: $K$-band temperature, gravity, $v\sin{i}$, and magnetic field strength. The $K$-band temperatures of Class I and FS sources range from 3190~K to 4207~K, have a median value of 3443~K and a standard deviation of 264~K. The surface gravities of the sources have a median value of $\log{g}=3.48$ in log of $\rm cm \, s^{-2}$ and a standard deviation of 0.3 dex. The main peak of the gravity values is close to $\log{g}=3.8$, but there is non-significant secondary peak at around $\log{g} = 3.3$ and a tail extending down to $\log{g}=2.8$. The projected rotational velocities have a median value of 25~$\rm km \, s^{-1}$, a standard deviation of 14~$\rm km \, s^{-1}$, but they reach values as high as 63~$\rm km \, s^{-1}$.  

\begin{figure*}[ht!]
\epsscale{1.0}
\plotone{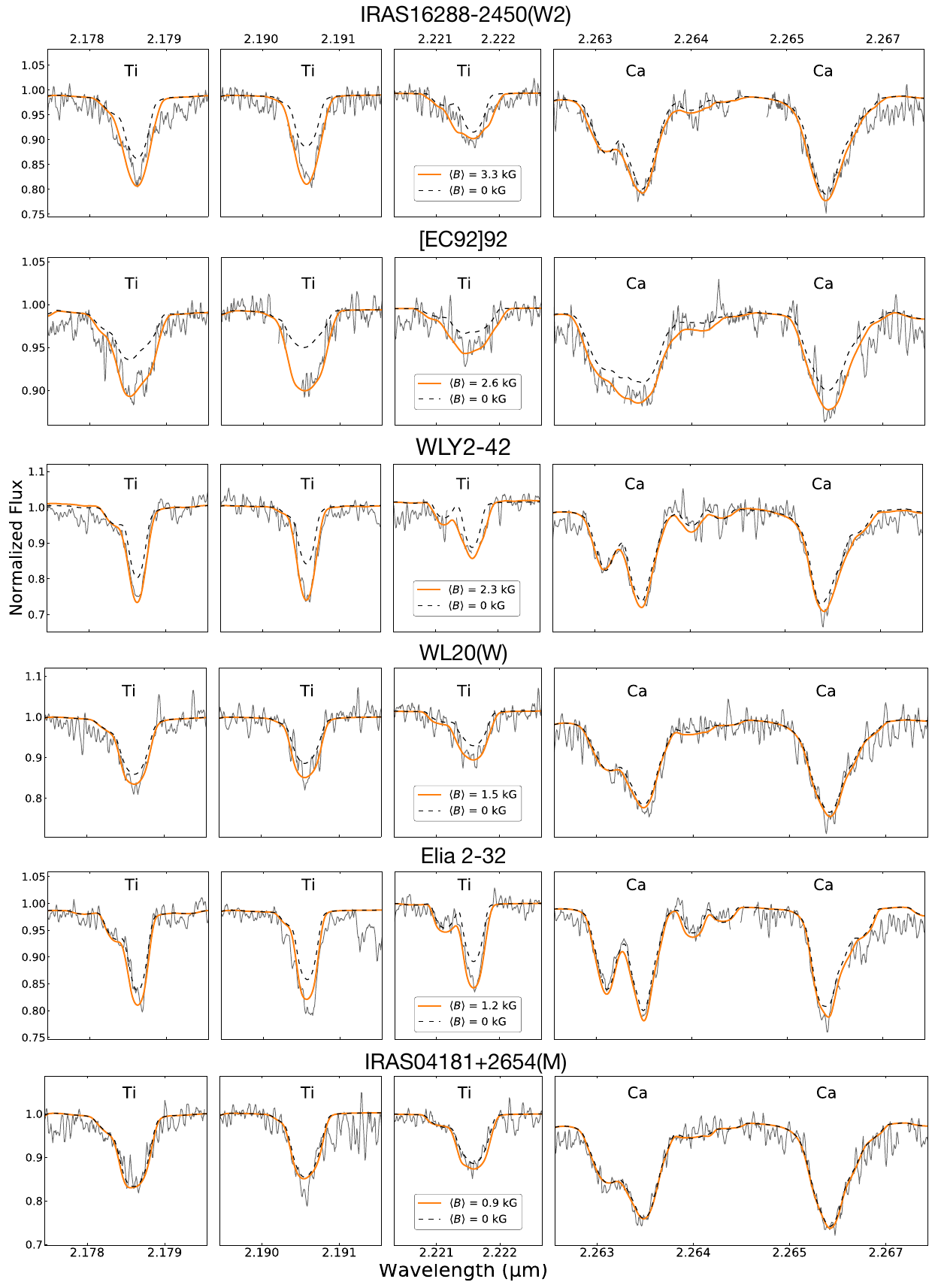}
\caption{Demonstration of the effects of magnetic fields on the spectra for six Class I and FS sources. The first three panels include highly magnetically sensitive Ti lines, while the fourth panel contains the less magnetically sensitive Ca lines. The spectra of the stars are displayed in gray, the best-fit synthetic models are shown in orange, and comparative synthetic models with null field strengths ($\langle B \rangle$ = 0 kG) are displayed as dashed black lines. The top five rows show sources where the differences between the magnetic and null-field models can be visually assessed. \label{fig:B-fields_illustration}}
\end{figure*}

\subsection{Magnetic field of Class I and FS sources}
We find a median magnetic field strength of $1.9$~kG with a standard deviation of 0.8~kG for our sample of Class I and FS sources. In Figure \ref{fig:B-fields_illustration}, we show how magnetic fields affect the spectra of six sources by plotting photospheric lines with different magnetic sensitivities. In most sources, we can visually identify the effect of the magnetic field from the plotted Ti lines due to their higher magnetic sensitivity. The Ca lines with their lower magnetic sensitivity (rightmost panel), allow us to confirm that the spectral changes in the Ti lines are indeed induced by magnetic effects, and not other physical process such as rotation or veiling.

In almost all cases, models including magnetic fields provide a clear better match to the observations. As shown by previous works \citep[e.g.,][]{Basri1992,Flores2022}, the effect of magnetic fields is not only to split spectral lines but also to create deeper lines due to a desaturation of energy levels. This magnetic intensification effect is most conspicuous for the Ti lines at 2.1785 $\mu$m and 2.1908 $\mu$m, displayed in the two leftmost panels of Figure \ref{fig:B-fields_illustration}. Although the figure illustrates a few lines, the magnetic field strengths (along with all the other parameters) are calculated by simultaneously fitting over 20 individual metallic lines with different magnetic sensitivities.

The source at the bottom of the figure is IRAS04181+2654(M). The best fit for this source yields a magnetic field strength of $B\sim0.9$ kG, yet the visual improvement of the fit is barely noticeable.  It is likely that systematic errors in the selection of continuum normalization (which worsen for sources with limited signal-to-noise ratio) made the code to prefer magnetic models over null-field models. Such continuum-normalization systematics mainly affect sources with weaker magnetic fields, as line shapes get increasingly distorted with increasing magnetic field strengths (as shown in Figure \ref{fig:B-fields_illustration}).

Similar to the case of IRAS04181+2654(M), there are other four sources where the visual improvement of magnetic models is so marginal that we instead decided to report their values as upper limits. In all these cases, the predicted magnetic field from our MCMC analysis is of $\sim$1 kG or lower. Higher signal-to-noise ratio observations would allow future studies to put more stringent constraints on the magnetic fields of these sources, although variable IR veiling can make this process difficult \citep[e.g.,][]{Connelley2014}.

\subsection{Derived parameters from stellar evolutionary models}
\label{sec:stellar-params-from-feiden-models}

\begin{figure*}
\gridline{\fig{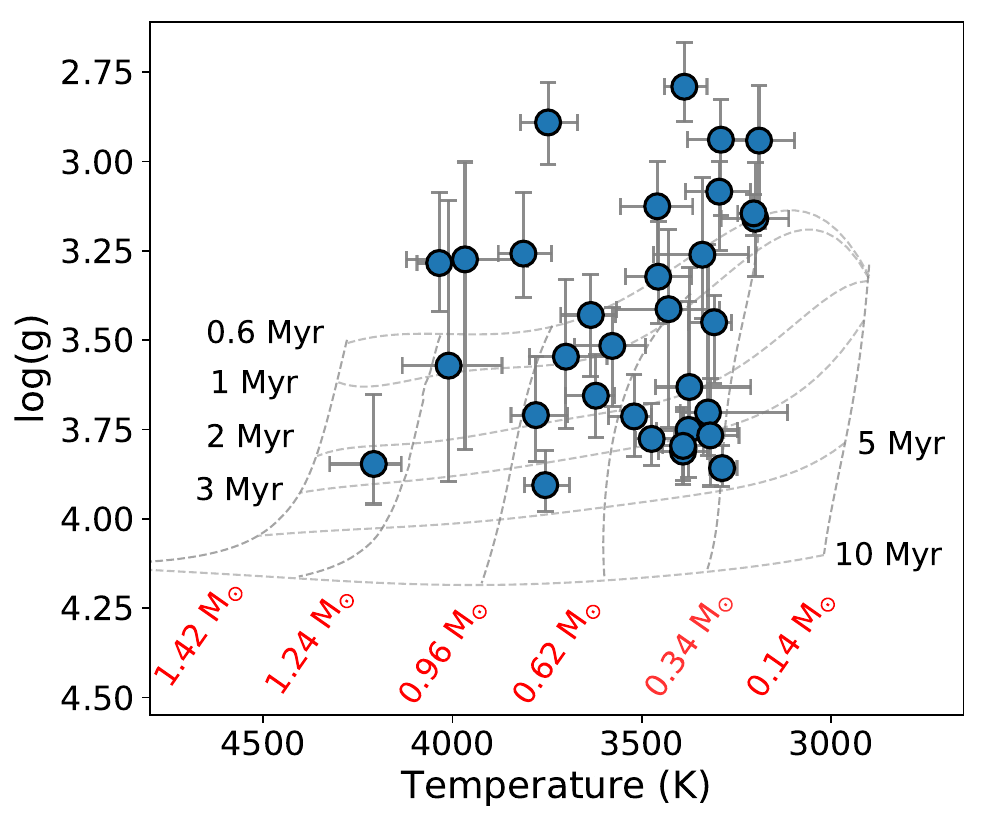}{0.5\textwidth}{(a)}
          \fig{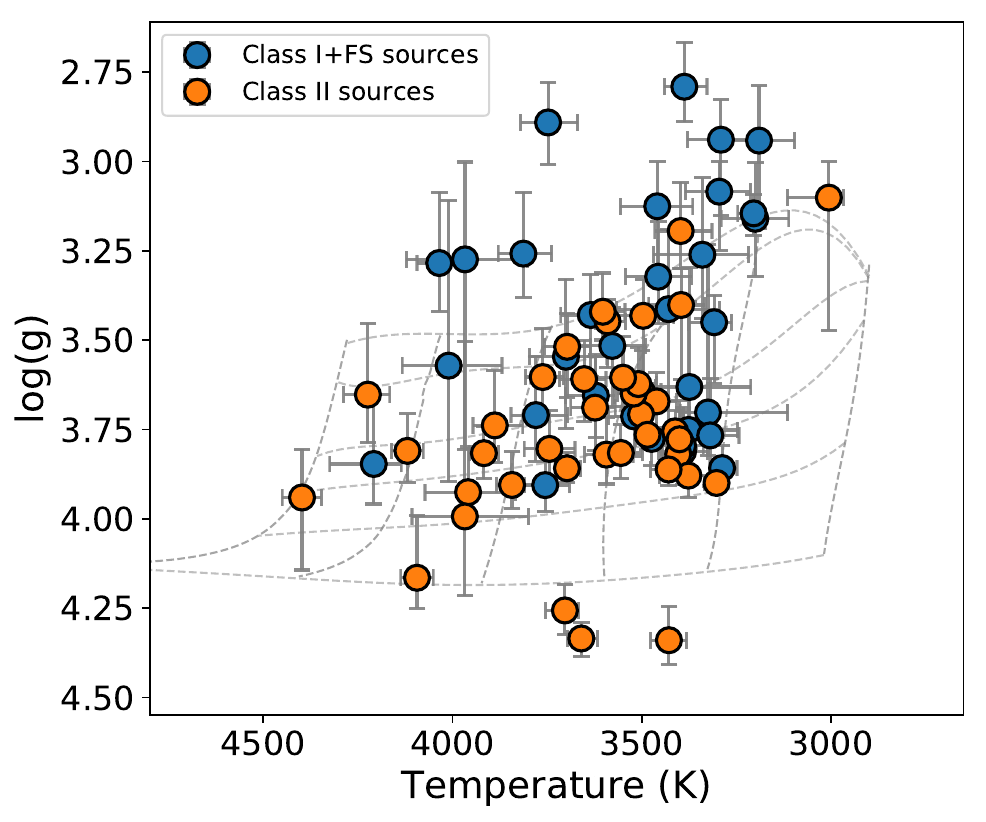}{0.5\textwidth}{(b)}
          }
\caption{In panel (a) we plot the temperatures and gravities for the $\Nstarsmodeled$ Class~I and FS sources in the HR diagram. Feiden's magnetic mass tracks (vertical) and isochrones (horizontal) are shown as gray dashed lines. Although most of the stars lie within the grid of model, 9 sources are above the 0.6 Myrs isochrone. Panel (b) shows a comparison between the position of the Class~I and FS sources from this work, and the 39 Class~II sources presented in \cite{Flores2022} marked in orange. Although there is an overlap for sources with ages between $\sim$0.6 to $\sim$5 Myrs, the Class~I and FS population extend to younger ages, and the Class~II population to older ages. Representative uncertainties for the temperatures and gravities are plotted for each data point. \label{fig:HR-diagram}}
\end{figure*}

To estimate the masses and ages of the young stars, we combine our gravity and temperature measurements with the magnetic stellar evolutionary models from \cite{Feiden2016} (Figure \ref{fig:HR-diagram}). In \cite{Flores2022}, we found that infrared temperatures are better indicators of the masses of young stars than optical temperatures for stars in the mass range of 0.3~$M_\odot$ to 1.3~$M_\odot$. This is equivalent to saying that infrared temperatures are  better tracers of the effective temperatures of young stars. Therefore, it is justified that we use our $K$-band derived temperatures as proxies for the effective temperatures of the young sources in the magnetic evolutionary models.



From panel (a) of Figure \ref{fig:HR-diagram}, we see that all the Class~I and FS sources in our sample are younger than 5 Myrs. In fact, 14 sources ($44\%$ of the sample) are 0.6 Myrs or younger as they sit on the youngest isochrone in the Feiden magnetic model or are above it. Four Class I and FS sources ($13\%$) have ages of 1 Myr, while  14 sources in our sample ($44\%$) have ages of 2-5 Myr. 
These results suggest that the observed Class~I and FS sources have a median age of about 0.6 to 1~Myr and that about half of the Class~I and FS sources in our sample are extremely young (proto)stars (age $\leq 0.6$~Myr). Our derived ages are similar to those derived by \cite{Heiderman2015} and \cite{Carney2016} ($\sim 0.3-0.6$~Myr) for embedded Class I source using statistical analysis of the number of stars in different evolutionary phases, a.k.a. `counting stars'.

From their position in the HR diagram and from visually extrapolating the mass tracks, we can infer that all the Class I and FS sources in our sample have masses between $0.3~M_\odot$ and $1.3~M_\odot$. As mentioned above, however, mass and age measurements can only be computed for the 23 Class~I and FS sources with ages equal or older than 0.6 Myrs, where the Feiden models are valid. In columns 8 and 9 of Table \ref{table:Stellar_params}, we list the age and stellar masses for the 23 Class I and FS sources. The mean mass of this subset is $0.61\pm 0.05$~$M_\odot$, which is $\sim 20\%$ lower than the median mass of 34 Class~II sources with a median mass of $0.75\pm 0.05$~$M_\odot$ \citep{Flores2022}. 

Although panel (b) of Figure \ref{fig:HR-diagram} shows an overlap in the theoretical-HR diagram position of Class~I and FS and Class~II sources, both distributions significantly differ as Class~I and FS objects extend towards younger ages and lower masses. The younger average age of Class~I and FS sources compared to the Class~II sources and their slightly lower masses indicate that Class~I and FS sources are the progenitors of Class~II objects, as widely assumed \citep[e.g.,][]{White2007}. The small mass differences between the two classes could, if not being purely a sample selection bias, suggest that the main mass accretion phase occurred before the stars became Class~I sources, i.e., in the Class 0 phase \citep[as suggested by ][]{White2004, Fiorellino2021}. By comparing the masses of both populations we see, that in a statistical sense, only $\sim20\%$ of the mass of the stars is acquired in the transition between the Class~I and the Class~II phases. There are three caveats to consider. First, the sample of Class~I and FS sources were selected to have visible photospheric lines in their spectra, and therefore we might be biased towards a more evolved phase of Class I and FS sources. Second, the Class~II sources were obtained only from Taurus-Auriga and Ophiuchus, while the Class I and FS sources were obtained from many different star forming regions. Last, the Class II sources were chosen to have ALMA resolved disks, introducing other unforeseen biases in our selection criteria. For these reasons, the mass difference between the Class I + FS source and the Class II sources, should be taken with caution.

\subsection{Accretion and circumstellar material indicators}
\label{sec:accretion_and_lines}
One of the defining characteristics of Class~I and FS sources is their circumstellar material. A protostar is conventionally pictured as a young star with a disk and a massive envelope, where material from the envelope is transported to the disk and then fed to the star. Spectroscopic evidence of the circumstellar material can be found in the form of absorption and emission lines. In our sample of $\Nstars$ Class~I sources, we find the following tracers of circumstellar material: the hydrogen $\rm H_2$ emission line at 2.1220~$\micron$ and 2.2235~$\micron$ typically associated with shock emission \citep{Beck2008}, the Br$\gamma$ emission line at 2.166~$\micron$ associated with magnetospheric accretion \citep{Muzerolle1998,Hartmann2016}, CO overtone emission starting at 2.2935~$\micron$ thought to arise from the thermal inversion of atmospheric layers in the inner disk ($R<10 \, R_\star$) \citep{Doppmann2008,Calvet1991}, and the CO overtone rovibrational lines from the envelope or diffuse material around Class~I and FS sources.

Not all these tracers are detected in each Class~I and FS source, and there are sources without any of these spectroscopic features. We detected Br$\gamma$ accretion tracers in 31 ($60\%$) of the Class~I and FS sources. CO overtone emission is clearly detected in 6 sources, and tentative CO emission is seen in 3 other stars. A combination of disk CO emission and stellar CO absorption likely prevents us from clearly characterizing these emission features.

The two $\rm H_2$ emission lines we detect are thought to originate in a distant outflow region, up to a few hundred au away from the stellar surface \citep{Beck2008,Laos2021}. The $\rm H_2$~$v=1-0$~$s(1)$ emission line at 2.122~$\micron$ is detected in red28 ($54\%$) of our sources, while the $\rm H_2$~$v=1-0$~$s(0)$ line at 2.223~$\micron$ only appears in the spectra of 17 stars $33\%$).

We interpret the absence of photospheric lines in the spectra of some Class~I and FS sources as an indicator of high IR veiling. The spectra of the sources with no photospheric detections are often redder than the other sources, and for 4 out of the 7 stars with no photospheric lines, we detect the CO overtone in emission. A high IR veiling can be associated with disk emission, which might be due to a significant and possibly recent burst of accretion. In a large survey of Class I sources, \cite{Connelley2010} found a correlation between the CO and Br$\gamma$ features; when the CO overtone is observed in emission, the Br$\gamma$ line is in emission, and the infrared veiling is high. They proposed that enhanced UV photons emitted by the accreted material can heat the inner disk producing both the CO in emission and the high IR veiling, analogous to what is observed in the eruptive EXor-type objects \citep[e.g.,][]{Aspin2010}.

In addition to the above-mentioned accretion and outflow-related features, we observe narrow CO rovibrational lines of non-stellar origin in the spectra of Class~I and FS stars. The rovibrational R- and P-branch lines are detected in 32 ($62\%$) of the observed spectra. As discussed in more detail in the next section, these narrow lines are associated with absorption from foreground line-of-sight cloud and/or with envelope and disk material in the circumstellar environments of the stars.

\subsection{Foreground and circumstellar CO rovibrational lines}
\label{sec:results_co_narrow_lines}

\begin{figure*}[ht!]
\epsscale{1.15}
\plotone{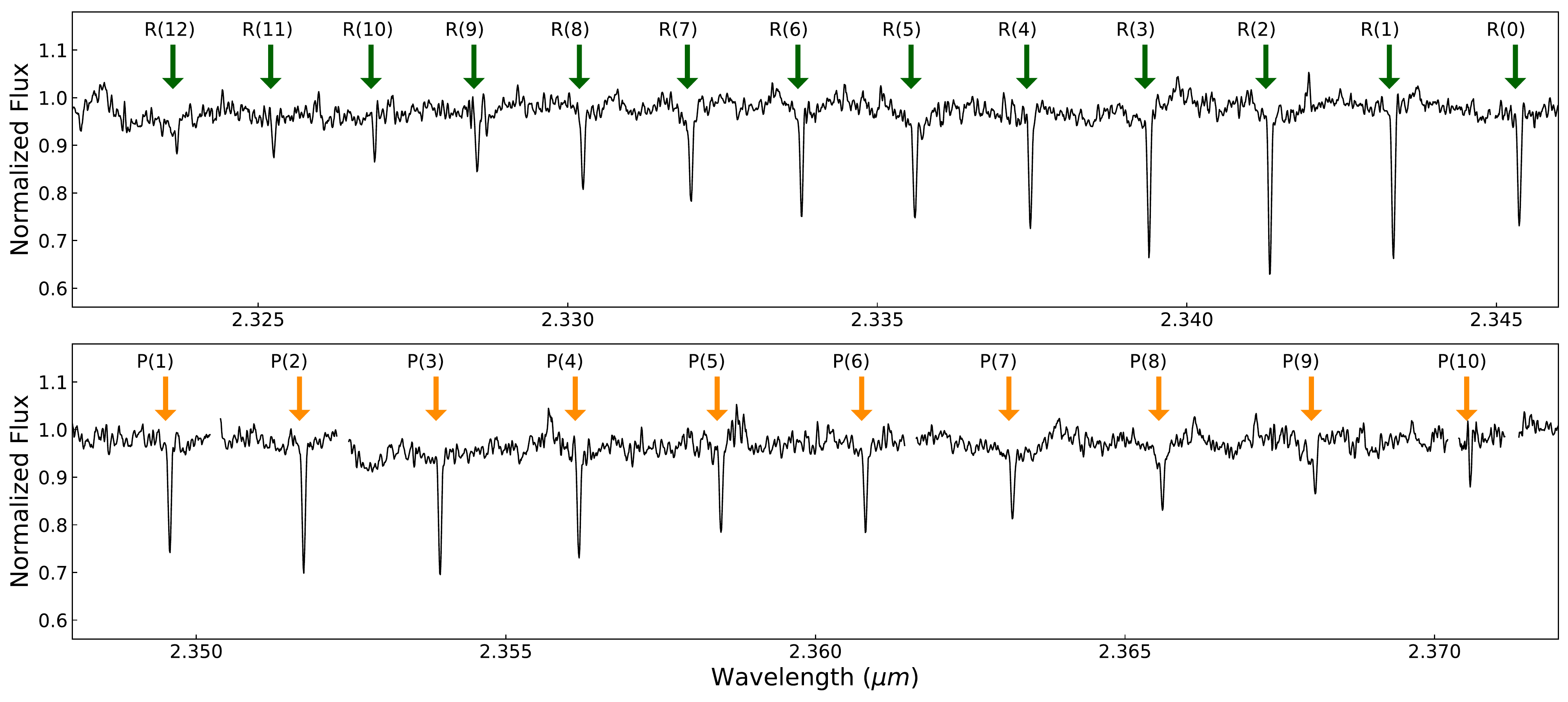}
\caption{Molecular rovibrational CO $v= 2-0$ overtone spectrum for IRAS16285-2355. Absorption lines from the R and P branches are marked with green and orange arrows and labeled accordingly. For IRAS16285-2355, photospheric CO absorption lines are detected at a $10\%$ depth level (at most) which are clearly separated from the narrow and deep R and P CO lines (depth up to $\sim40\%$ ). \label{fig:rotational_CO_lines-example}}
\end{figure*}

In the spectra of young stellar sources, we often find atmospheric CO overtone absorption lines which arise from the (proto)stellar photosphere. A good example of these CO photospheric lines is shown in the last panel of Figure \ref{fig:spectrum-example} starting at around 2.2935~$\mu$m. As discussed in the previous section, we have also detected a completely different type of CO absorption in the spectra of 32 out of our $\Nstars$ Class I and FS sources. These CO lines are much narrower than the CO photospheric lines (most unresolved at $R\sim$50,000) and they span a wavelength range from $\sim 2.3$~$\mu$m to $\sim 2.4$~$\mu$m (see Figure \ref{fig:rotational_CO_lines-example}). 

The narrow profile of these lines indicates that they are not affected by stellar rotational broadening. Additionally, their radial velocities (RV) often do not align with the measured stellar RV. Therefore, the origin of these narrow CO absorption lines is not from the photosphere, but rather from cool circumstellar and interstellar material directly in front of the observed young sources (resulting in absorption). In the literature, CO overtone and fundamental observations are often used to trace chemical enrichment in galaxies, to trace circumstellar material of high-mass embedded sources, and to measure individual carbon and oxygen isotope fractionation in the environment of young stellar objects \citep[e.g.,][]{Smith2009,Goto2003,Barr2020}. 

To the best of our knowledge, little work has been published on non-photospheric CO overtone lines in the spectra of low-mass young stars. The likely reason for this is that, at low-spectral resolution, the photospheric absorption lines of low-mass young stars can be confused with the non-photospheric CO overtone lines. Thus, high spectral resolution observations with broad wavelength range coverage are required, which are scarce.

In our case, due to the large wavelength coverage and high-respectral resolution that iSHELL provides, we can use the physical and kinematic properties of the non-stellar molecular CO overtone lines to study the presence of foreground contamination clouds, cold protostellar envelopes, and/or warm circumstellar disk material in the vicinity of the young stars. To distinguish whether CO lines arise from unrelated foreground clouds or circumstellar material, we calculate the absorbing material's excitation temperature, radial velocity, and line widths. 

Below, we analyze the non-stellar CO rovibrational lines detected in the spectra of the Class~I and FS sources to determine which sources display envelope or disk-like structures in their surroundings. In our analysis, we use the rotational diagram (or Boltzmann population) method to measure the excitation temperature and column density of the absorbing gas. This method has been extensively used in the literature \citep[e.g., ][]{Mitchell1990,Brittain2003,Goto2003} and we briefly describe it in Appendix \ref{appendix:rotational_diagrams}.

In Figure \ref{fig:rotational_CO_lines-example}, we display the spectrum of IRAS16285-2355, where the narrow CO $v=2-0$ rovibrational R- and P- branch lines are seen and labeled. As explained in Appendix \ref{appendix:rotational_diagrams}, the temperature and column densities are obtained by fitting linear functions to the rotational diagrams. This process is performed assuming local thermodynamic equilibrium (LTE), i.e., assuming that a single excitation temperature represents well the temperature of a parcel of gas. In the rotational diagram, the slope of the linear fit to the lower-level column densities corresponds to the inverse of the excitation temperature of the gas. In some instances, a single linear fit represents well the physical conditions of the observed features, while in other cases, the population diagram needs to be fit with two straight lines, which often intersect at $E_J/K_B\sim 50$~K. This behavior indicates that the observed CO arises from physically separated intervening gas, possibly gas clouds and/or disk material along the line-of-sight with different excitation temperatures \citep[e.g.,][]{Mitchell1990,Giannakopoulou1997,Goto2003}.

\begin{figure}
\gridline{\fig{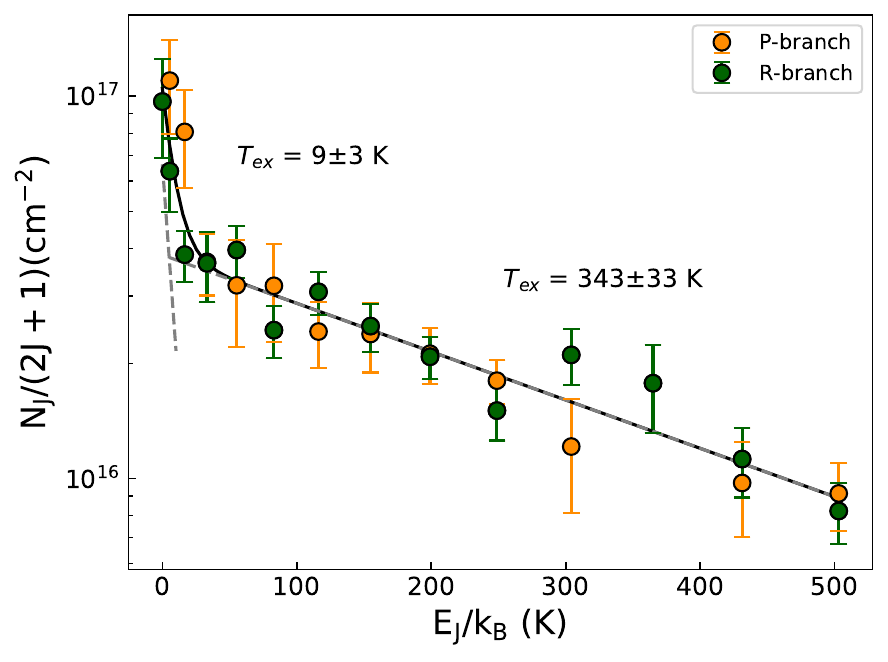}{0.45\textwidth}{(a)}
          }
\gridline{\fig{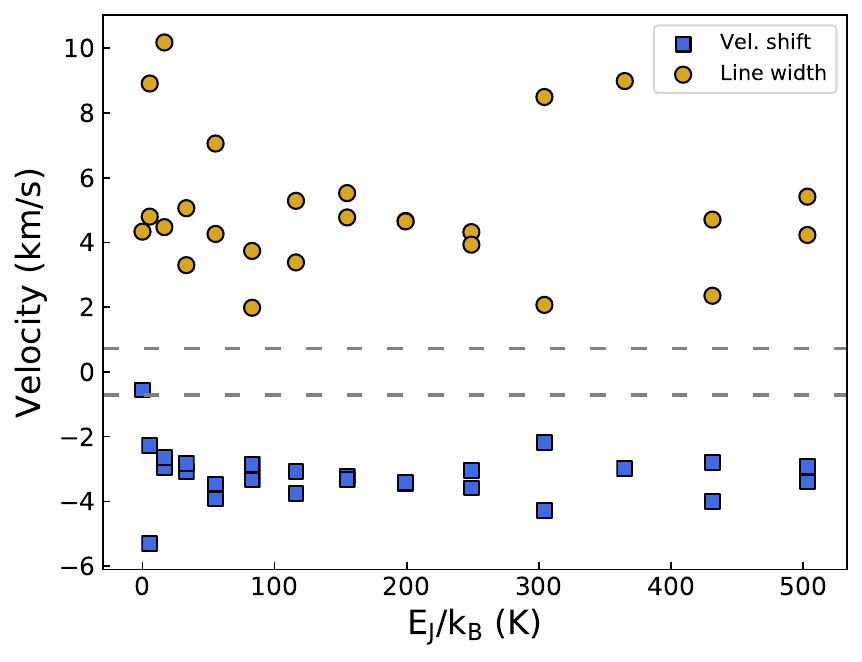}{0.45\textwidth}{(b)}
          }
\caption{ (a) Rotation diagram for $^{12}$CO $v=2-0$ for DG Tau. Green data points correspond to the R-branch measurements, orange are the P-branch. Two excitation temperatures are found with temperatures of 343~K and 9~K. (b) Kinematic structure of the CO lines detected in the spectrum of DG Tau. The velocity difference between the CO lines and the central source is plotted as blue squares. The horizontal dashed lines correspond to the RV uncertainty calculated for DG Tau from the fit to photospheric lines. Golden circles correspond to the width of the CO lines in $\rm km \, s^{-1}$. Typical uncertainties for velocity shift are $0.3 \, \rm km \, s^{-1}$ We interpret the temperature and kinematics of the lines as out-flowing CO gas with a projected velocity of about $-3 \, \rm km \, s^{-1}$  \label{fig:rotational_diagram}}
\end{figure}

In panel (a) of Figure \ref{fig:rotational_diagram}, we show the rotational diagram for the FS source DG~Tau. For this source, two straight lines were necessary to fit the lower-level column densities, which indicates the presence of warm and cold gas along the line-of-sight. In panel (b), we show the kinematic structure of the individual CO lines. We plot the relative RV of the CO line with respect to the star's RV, as well as the CO line widths (calculated from the line fits) as a function of the line energy. 


For this source, we interpret the significantly blue-shifted absorption of the cold and warm gas, as material escaping the central source at projected velocities of $ -3$~$\rm km\, s^{-1}$. This interpretation is consistent with direct ALMA observations of the gas around DG Tau, which shows winds and a jet-like structure driving a molecular outflow \citep[e.g.,][]{Gudel2018}.

\figsetstart
\figsetnum{2}
\figsettitle{Rotational diagrams (a)}

\figsetend

\figsetstart
\figsetnum{3}
\figsettitle{Kinematic plot (b)}

\figsetend

For 28 out of the 32 sources with detected non-photospheric CO lines in their spectra, we were able to perform a similar temperature and kinematic analysis, which is further discussed in Section \ref{sec:discussion-circumstellar_environent}. In Table \ref{table:narrow_CO_lines-summary}, we list the results for each source, display the temperature and column density of a single or double-component, and the average velocity shift with respect to the central source (when available). Also, when possible, we indicate whether the temperature and kinematics of the gas are consistent with disk, envelope, foreground cloud, or infall/outflow material (see Section \ref{sec:discussion-circumstellar_environent}).

\begin{deluxetable*}{lcccccccc}
\tabletypesize{\scriptsize}
\tablewidth{0pt}
\tablecolumns{9}
\tablecaption{Foreground absorption lines in the spectra of the Class I and FS sources. \label{table:narrow_CO_lines-summary}}
\tablehead{
\colhead{Source name} & \multicolumn{3}{c}{Cold component} & \colhead{ } & \multicolumn{3}{c}{Warm component} & \colhead{Interpretation}\\
\cline{2-4}
\cline{6-8}
\colhead{} & \colhead{Temperature} & \colhead{Col. density} & \colhead{Med. velocity} & \colhead{ } &  \colhead{Temperature} & \colhead{Col. density} & \colhead{Med. velocity} & \colhead{} \\
\colhead{} & \colhead{(K)} & \colhead{($\rm cm^{-2}$)} & \colhead{($\rm km\, s^{-1}$)} & \colhead{ } &  \colhead{(K)} & \colhead{($\rm cm^{-2}$)} & \colhead{($\rm km\, s^{-1}$)} & \colhead{}
} 

\startdata
DG Tau        &          $ 9 \pm 3 $ & $ 2.41 \pm 0.8 \, \times \, 10^{17} $ &   $-2.8 \pm 0.7$ &  &       $ 343 \pm 33 $ & $ 4.79 \pm 0.35 \, \times \, 10^{18} $ &  $-3.3 \pm 0.7$  & outflow \\ 
$\rm [EC92]$ 92       &         $ 20 \pm 1 $ & $ 1.50 \pm 0.1 \, \times \, 10^{18} $ &  $-0.8 \pm 1.1$ &  &                      &                   &  &   envelope    \\ 
$\rm [EC92]$ 95       &         $ 21 \pm 1 $ & $ 2.89 \pm 0.19 \, \times \, 10^{18} $ &  $1.7 \pm 1.4$ &  &                      &                   &  &  cloud     \\ 
$\rm [EC92]$ 129      &         $ 22 \pm 4 $ & $ 2.86 \pm 0.31 \, \times \, 10^{18} $ &   &  &      $ 154 \pm 111 $ & $ 2.84 \pm 1.9 \, \times \, 10^{18} $ &    &  \\ 
GV Tau S     &         $ 10 \pm 3 $ & $ 5.63 \pm 1.7 \, \times \, 10^{17} $ &   $0.8 \pm 0.2$ &  & $ 338 \pm 37 $ & $ 8.07 \pm 0.62 \, \times \, 10^{18} $ &  $1.5 \pm 0.2$  &  infall \\ 
$\rm [GY92]$ 21       &         $ 42 \pm 6 $ & $ 4.25 \pm 0.34 \, \times \, 10^{18} $ &   &  & $ 372 \pm 50 $ & $ 1.60 \pm 0.2 \, \times \, 10^{19} $ &    & \\ 
$\rm [GY92]$ 235      &         $ 26 \pm 7 $ & $ 1.99 \pm 0.5 \, \times \, 10^{18} $ &   $-0.2 \pm 0.5$ &  &                      &                   &  &   envelope    \\ 
$\rm [GY92]$ 33       &   &                   &  &  &      $ 200 \pm 62 $ & $ 1.56 \pm 0.33 \, \times \, 10^{19} $ &   $-1.3 \pm 0.5$ &    outflow     \\ 
IRAS03301+3111 &        $ 26 \pm 12 $ & $ 2.17 \pm 0.44 \, \times \, 10^{18} $ &   $-0.8 \pm 0.5$ &  &                      &                   &  &   cloud    \\ 
IRAS04108+2803(E) &          $ 9 \pm 3 $ & $ 8.82 \pm 2.5 \, \times \, 10^{17} $ &   $-0.4 \pm 0.7$ &  & $ 394 \pm 85 $ & $ 1.50 \pm 0.21 \, \times \, 10^{19} $ &  $2.1 \pm 0.7$  & infall \\ 
IRAS04158+2805 &  &  &     &       &     $ 323 \pm 31 $ & $ 1.70 \pm 0.12 \, \times \, 10^{19} $ &      &       \\ 
IRAS04181+2654(M) &         $ 19 \pm 4 $ & $ 1.54 \pm 0.31 \, \times \, 10^{18} $ &   $-0.4 \pm 0.6$ &  &                      &                   &  &   envelope    \\ 
IRAS04181+2654(S) & &  &      &    &  $ 276 \pm 29 $ & $ 1.12 \pm 0.08 \, \times \, 10^{19} $ &   $0.8 \pm 0.5$   &    infall    \\ 
IRAS04295+2251 &         $ 20 \pm 2 $ & $ 1.52 \pm 0.15 \, \times \, 10^{18} $ &   $0.1 \pm 1.7$ &  &                      &                   &  &    envelope   \\ 
IRAS04591-0856 &  &       &                   &  &       $ 182 \pm 20 $ & $ 1.23 \pm 0.08 \, \times \, 10^{19} $ & $0.6 \pm 0.5$                 &            infall       \\ 
IRAS05311-0631 &         $ 29 \pm 2 $ & $ 1.16 \pm 0.08 \, \times \, 10^{18} $ &   &  &                      &                   &  &       \\ 
IRAS05555-1405(4) &   &                   &  &                  &   $ 63 \pm 8 $ & $ 1.34 \pm 0.13 \, \times \, 10^{18} $ & $-0.9 \pm 1.2$   &         disk       \\ 
IRAS16285-2358  &  &       &                   &  &       $ 291 \pm 57 $ & $ 9.69 \pm 1.6 \, \times \, 10^{18} $ & $1.5 \pm 0.8$                  &         infall         \\ 
IRAS16285-2355  &         $ 20 \pm 2 $ & $ 2.01 \pm 0.15 \, \times \, 10^{18} $ &   &  & $ 174 \pm 19 $ & $ 4.93 \pm 0.64 \, \times \, 10^{18} $ &    & \\ 
IRAS16316-1540 &  &       &                   &  &        $ 69 \pm 11 $ & $ 3.00 \pm 0.47 \, \times \, 10^{18} $ &                  &                   \\ 
$\rm [TS84]$ IRS5 (NE)   &  &       &                   &  &        $ 69 \pm 22 $ & $ 5.27 \pm 1.0 \, \times \, 10^{18} $ &  $1.8 \pm 1.3$      &          infall         \\ 
EM* SR24A       &        $ 30 \pm 13 $ & $ 7.97 \pm 1.8 \, \times \, 10^{17} $ &  $1.2 \pm 1.0$ &  & $ 140 \pm 47 $ & $ 1.80 \pm 0.88 \, \times \, 10^{18} $ &  $2.4 \pm 1.0$  & infall \\ 
VSSG17     &         $ 22 \pm 2 $ & $ 2.01 \pm 0.18 \, \times \, 10^{18} $ &  $-0.7 \pm 1.0$ &  &                      &                   &  &   envelope    \\ 
WL20(E)      &         $ 21 \pm 6 $ & $ 2.11 \pm 0.53 \, \times \, 10^{18} $ &   $0.6 \pm 0.9$ &  &                      &                   &  &   envelope    \\ 
WL20(W)      &         $ 25 \pm 7 $ & $ 2.21 \pm 0.44 \, \times \, 10^{18} $ &   $-0.8 \pm 0.7$ &  &                      &                   &  &   cloud    \\ 
WL4        &         $ 28 \pm 6 $ & $ 2.98 \pm 0.49 \, \times \, 10^{18} $ &    &  &                      &                   &  &       \\ 
WLY2-54    &         $ 21 \pm 3 $ & $ 2.29 \pm 0.21 \, \times \, 10^{18} $ &   &  & $ 231 \pm 94 $ & $ 3.24 \pm 1.1 \, \times \, 10^{18} $ &   &  \\ 
YLW 45  &         $ 23 \pm 3 $ & $ 1.79 \pm 0.24 \, \times \, 10^{18} $ &   &  &                      &                   &  &      \\ 
\enddata
\tablecomments{We report temperatures in columns 2 and 5, column densities in columns 3 and 6, and median velocities in columns 4 and 7. There is tentative evidence of a cold component for sources [GY92]~33, IRAS04158+2805, IRAS4181+2654(S), IRAS4591-0856, and IRAS16285-2358 but the limited S/N of the narrow CO lines prevent a definite assessment of them.}
\end{deluxetable*}

\section{Discussion}
\label{sec:discussion}
\subsection{Comparing stellar parameters of Class I + FS vs. Class II sources}
\label{subsec:class_I_vs_class_II}

We compare the stellar parameters of the Class~I and FS sources derived in Section \ref{sec:stellar_params} (and summarized in Table \ref{table:Stellar_params}), against the results published by \cite{Flores2022}  for a sample of 39 Class~II sources from Taurus and Ophiuchus. 

We use the non-parametric Kolmogorov–Smirnov (KS) test to assess whether the distribution of stellar parameters for Class~I and FS sources are different from the stellar parameters of Class~II sources ($T_{K-band}$, $\log{g}$, $\langle B \rangle$, $v\sin{i}$, IR-veiling). The null hypothesis of this test is that the distributions come from the same parent population. We set the statistical significance threshold for this test to be $\alpha=0.05$, i.e., the null hypothesis is accepted if $p>0.05$. In Table \ref{table:Kolmogorov–Smirnov}, we show the results of the KS test for the five stellar parameters mentioned before. We find that the magnetic field strength and infrared veiling distributions between both populations are statistically indistinguishable, while the gravity, temperature, and projected rotational velocity are different. 

\begin{figure*}
\gridline{\fig{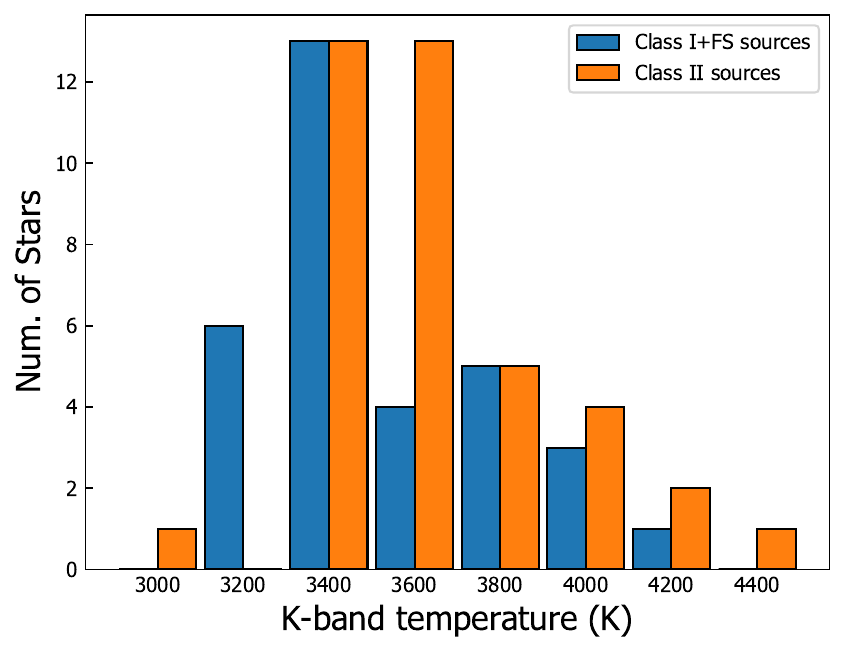}{0.45\textwidth}{(a)}
          \fig{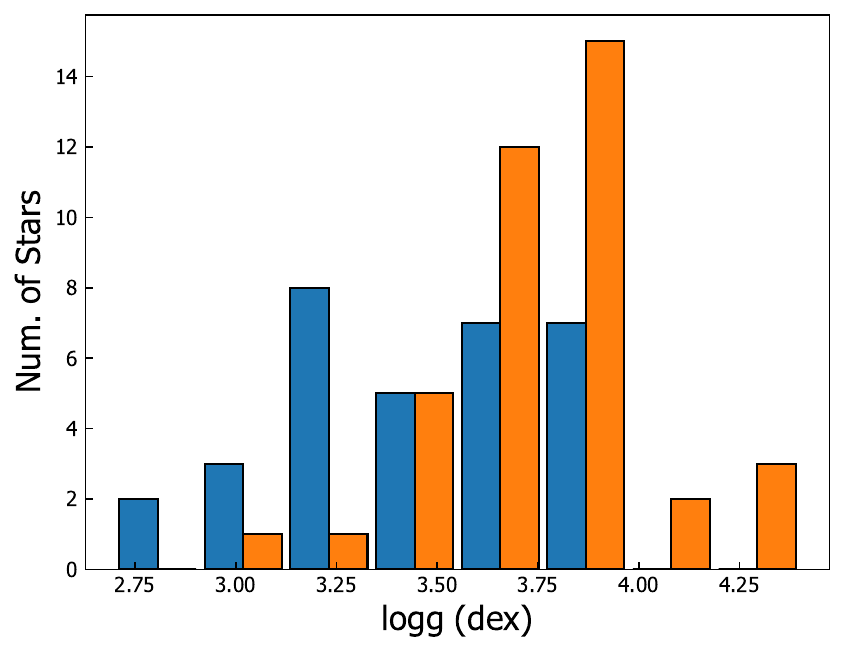}{0.45\textwidth}{(b)}
          }
\gridline{\fig{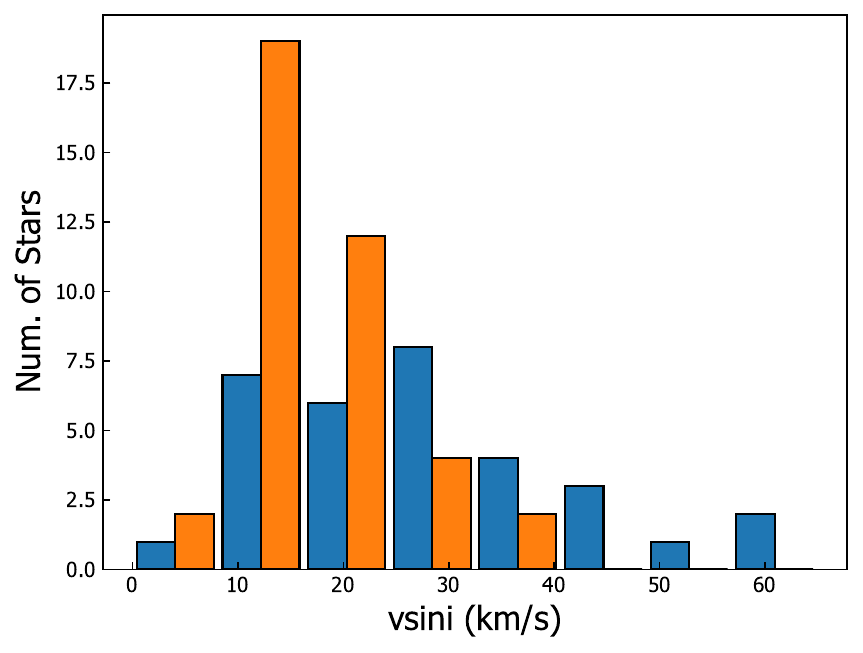}{0.45\textwidth}{(c)}
          \fig{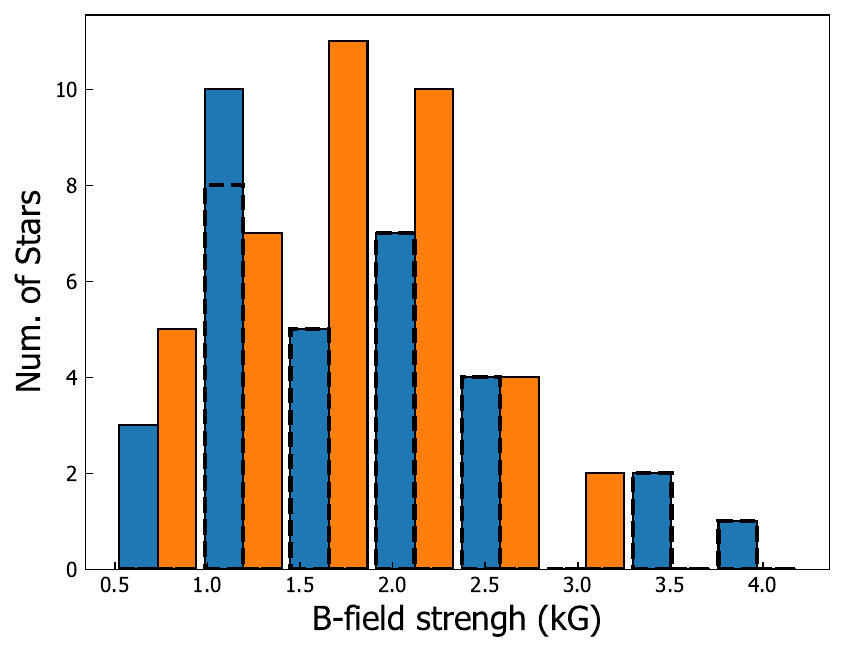}{0.45\textwidth}{(d)}
          }
\caption{We compare the distribution of our sample of Class~I sources against the results presented in \cite{Flores2022} of a sample of 39 Class~II sources
(a) $K$-band temperature, (b) surface gravity in log of $\rm cm \, s^{-2}$, (c) projected rotational velocity, and (d) the average magnetic field strength of the stars. Using black dashed lines, we also show the histogram excluding the five sources for which we only know their magnetic field upper limits. Eight bins are shown in each distribution. The distribution of 
the parameters for the Class~I and II sources in panels (a), (b), and (c) are significantly different from each other. We interpret the lower gravity and temperature of Class~I objects, as these sources being younger and less massive than the Class~II sources, respectively.
\label{fig:histogram_comparison}}
\end{figure*}

The median gravity and standard error of the mean for the Class~I and FS sources $\log{g}=3.48\pm 0.06$, is significantly lower than the median value and standard error of the mean for the Class~II objects, $\log{g}=3.76 \pm 0.04$, despite a significant overlap in their histograms (see Figure \ref{fig:histogram_comparison}). In particular, the largest peak of the gravity distribution of the Class~I and FS sources at $\log{g} \sim 3.8$ in log of $cm \, s^{-2}$ coincides with the only peak of the gravities of the Class~II sources. This overlap indicates that the observational class (measured from the infrared spectral index) of a source might not always correspond to its evolutionary stage. The overlap between the classes may be further accentuated by the flexible cuts in the spectral index that separate the different young stellar classes. For example, some authors define the Class I sources as young stellar objects with an infrared index ($\alpha$) greater than 0, while others define it as $\alpha$ greater than 0.3. Additionally, the arbitrary wavelengths at which the spectral indices are measured (sometimes at 22 $\mu m$ and other times at 24 $\mu m$) can contribute to the overlap as well. Nevertheless, our comparison of the gravities of Class~I and FS vs. Class~II sources still shows significant differences in their distribution, with the gravities of the Class~I and FS objects going as low as $\log{g} \sim 2.8$, while the lowest gravity Class II source being only of $\log{g}=3.1$. 

The median temperature value for the Class~I and FS sources and its standard error of the mean $\tkband = 3443 \pm 47$~K is significantly lower than the median value for the Class~II stars  $\tkband = 3590 \pm 33$~K, and as shown in Table \ref{table:Kolmogorov–Smirnov} these distributions are statistically different. The difference in the distribution of gravities and temperatures agrees with the results presented in Section \ref{sec:stellar-params-from-feiden-models}, where we showed that the ages of the Class~I and FS are lower than the ages of the Class~II sources, and the average mass of the Class~I objects is about 20$\%$ lower than the average mass of the Class~II stars. As previously discussed, these characteristics point to an evolutionary sequence from the Class~I and FS sources to the Class~II phase as originally postulated by \cite{Lada1987}, with the caveat that a fraction of Class~I sources overlaps on the HR-diagram with the Class~II objects.


The KS test reveals that the distribution of magnetic field strengths and infrared $K$-band veilings are statistically indistinguishable between the Class~I $+$ FS sources and Class~II sources. The distributions of projected rotational velocities $v\sin{i}$, however, are statistically different with $p\sim 8 \times 10^{-4}$.  As shown in panel (c) of  Figure \ref{fig:histogram_comparison}, the distribution of velocities for the Class~I and FS sources extends to higher rotational velocities of up to 66~$\rm km s^{-1}$, while the Class~II stars only go up to 33~$\rm km s^{-1}$. The median $v\sin{i}$ value for the Class~I and FS sources is $26 \pm 2.6 \, \rm km \,s^{-1}$, significantly higher than the Class~II sources of $16 \pm 1.1 \, \rm km \,s^{-1}$ \citep[also previously reported in, ][]{Covey2005}.

Overall, these results show that although there is an apparent evolutionary sequence from Class~I and FS sources to the Class~II sources and the former have larger radii and projected rotational velocities, the magnetic field strengths of the young stars remain about the same. These findings are in line with a scenario in which the magnetic field strength of low-mass (proto)stars is generated by a saturated dynamo-like process \citep[e.g.,][]{Reiners2009}, similar to what is observed in fast rotating M-dwarfs \citep{Kochukhov2021}. Furthermore, our results confirm
that the younger Class I and FS sources host strong magnetic fields on their photospheres, and therefore they likely accrete disk
material through a magnetosphere similarly to their more evolved Class II counterparts.

\begin{deluxetable}{lcc}
\tablewidth{10pt}
\tablecolumns{2}
\tablecaption{Kolmogorov–Smirnov test for distribution of stellar parameters between Class I and FS sources vs. Class II sources. \label{table:Kolmogorov–Smirnov}}
\tablehead{
\colhead{Stellar parameter} & \colhead{\textit{p}-value} & \colhead{distribution} \\
\colhead{} & \colhead{} & \colhead{interpretation}}
\startdata
$\tkband$& $6.2 \times 10^{-3}$ & different \\
$\log{g}$ & $3.1 \times 10^{-3}$ & different \\
$\langle B \rangle$ & 0.47 & same \\
\textit{v} $\sin{i}$& $7.8 \times 10^{-4}$ & different \\
IR veiling & 0.69 & same \\
\enddata
\tablecomments{We set the statistical significance threshold to be $\alpha$=0.05, i.e., the null hypothesis is accepted if \textit{p}-value $>0.05$}.\\
\end{deluxetable}

\subsection{Correlation of stellar parameters with magnetic field strength}

We have analyzed the statistical correlation between the measured magnetic field strengths and the other derived stellar parameters for our sample of Class I and FS sources, but also for the sample of Class II sources presented in \cite{Flores2022}. In all cases, we use the Spearman rank-order correlation to test whether or not two variables are correlated. 

For the sample Class I and FS sources, we find correlation $r$-values and significance \textit{p}-values of $r=$0.16, $p=$0.45 between the magnetic field strength and the temperature; $r=$0.08, $p=$0.70 between the magnetic field strength and the gravity; and $r=$-0.15, $p=$0.46 between the magnetic field strength and the infrared veiling. In all three cases, the large \textit{p}-values suggest that there is a high probability $\textit{p}>0.05$ (over $5\%$) that the correlations happen by random chance (i.e., the null hypothesis is true). In the sample of Class I and FS sources, however, there is a marginal correlation between the magnetic field strength and the projected rotational velocity of the Class~I sources ($r$=0.42 and $p$=0.03). This correlation must be viewed with caution, given our limited sample size.

For the sample of Class II sources, we find different results. We find a non-correlation between magnetic field strength and gravity with r=0.20, p=0.21, and also between magnetic field strength and projected rotational velocity r=-0.07, p=0.69. Significant correlations are found between magnetic field strength and temperature r=0.42, p=0.01, and between magnetic field strength and infrared veiling r=0.48, p=0.02.
 
We then combined both samples, i.e., Class I, FS, and II sources, and applied the same metric to understand which of the correlations remained valid for the larger sample of 65 sources (26 Class I and FS sources $+$ 39 Class II sources). By applying the Spearman rank-order test, we found that the only statistically significant correlation is between magnetic field strength and $K$-band temperature (r=0.27, p=0.03). There are several implications for the Class I $+$ FS group and the Class II group. First, the correlation between projected rotational velocity and magnetic field strength either disappears as stars age or the number of Class I and FS sources are too small to make it reliable. Similarly, the correlation between magnetic field strength and infrared veiling found for the Class II sources also disappears when we combine all the sources. Secondly, the positive correlation between temperature and magnetic field strength in Class II sources dominates over the non-correlation in Class I and FS sources and appears valid for the whole sample of young stellar objects.

\cite{Flores2022} showed a dependency between magnetic field strength and the difference between optical and infrared temperature in Class II sources. The explanation for this phenomenon was related to magnetically induced spots on the surface of the stars. Since few optical temperature measurements exist for most of the Class I and FS sources presented here, it is hard to directly test whether these sources are also affected by star spots. 
 
In Table \ref{table:Spearman-correlation}, we summarize the Spearman values and the false alarm probability \textit{p}-values for the $K$-band temperature, surface gravity, IR veiling, and the projected rotational velocity parameters.
\begin{deluxetable}{lcc}
\tablewidth{20pt}
\tablecolumns{3}
\tablecaption{Spearman rank-order correlation between the magnetic field strengths and other stellar properties. \label{table:Spearman-correlation}}
\tablehead{
\colhead{Quantities compared} & \colhead{r} & \colhead{\textit{p}-value}}
\startdata
Only Class I and FS sources & &\\
\hline
$\langle B \rangle$ vs. $\tkband$& 0.16 & 0.43 \\
$\langle B \rangle$ vs. $\log{g}$ & 0.07 & 0.72 \\
$\langle B \rangle$ vs. IR veiling & -0.15 & 0.48 \\
$\langle B \rangle$ vs. \textit{v} $\sin{i}$& 0.41 & 0.03 \\
\hline
Sample of Class I, FS, and II sources & &\\
\hline
$\langle B \rangle$ vs. $\tkband$& 0.26 & 0.03 \\
$\langle B \rangle$ vs. $\log{g}$ & 0.09 & 0.46 \\
$\langle B \rangle$ vs. IR veiling & 0.22 & 0.08 \\
$\langle B \rangle$ vs. \textit{v} $\sin{i}$& 0.17 & 0.18 \\
\enddata
\end{deluxetable}

\subsection{Photospheric and emission lines in our sample of Class I and FS sources}

We put in context the absorption and emission line statistics of our sources by comparing them to previous infrared studies of Class~0, Class I+FS, and Class~II sources. As described in Sections \ref{sec:stellar_params} and \ref{sec:accretion_and_lines}, we find that of our Class~I and FS sources $85 \pm 13 \%$ have photospheric lines, $60\pm 11\%$ show the Br$\gamma$ line in emission, 12 to 17$\%$ display CO in emission, and $54\pm 10\%$ show at least one $\rm H_2$ line in emission.

Six Class~0 sources from Perseus and Orion were reported by \cite{Laos2021} using MOSFIRE at Keck in the $K$-band at a spectral resolution of $R\sim2,400$. While they did not detect photospheric features in any of their Class 0 sources, they were able to detect CO in emission in 4 of their sources ($67\pm 33$~\%), Br$\gamma$ line in emission in four sources ($67\pm 33$~\%), and $\rm H_2$ emission lines in all six sources. \cite{Greene2018} reported the first Class 0 source with detected photospheric lines to date, S68N. In addition to the absorption lines, they detected the $\rm H_2$ $\nu=$ 1-0 $S(1) \, 2.122~\micron$ line in emission, but the Br$\gamma$ emission was not detected. Recently, \cite{LeGouellec2024} conducted a K-band survey using MOSFIRE at Keck and found photospheric lines in six Class 0 sources in a sample of 24 sources ($25\pm 10$~\%).

52 Class I and FS source K-band spectra were obtained by \cite{Doppmann2005} at $R\sim18,000$ from Taurus, Ophiuchus, Serpens, and Corona Australis. Of their sample, 41 sources exhibited photospheric lines ($79\pm12\%$), 23 showed $\rm H_2$ lines ($44\pm9\%$), 34 showed Br$\gamma$ emission ($65\pm 11 \%$), and eight sources had CO in emission ($15\pm 5\%$). In the largest NIR spectroscopic survey of Class~I and FS sources, \cite{Connelley2010} observed 110 sources between 0.8 and 2.4~$\micron$ at $R\sim 1,200$ with IRTF/SpeX. Although their sample included 10 FU~Ori-like objects and other types of young stellar sources (e.g., 3 Herbig Ae stars), their study is a complete sample to compare our data with. \cite{Connelley2010} detected strong atomic absorption lines in 50 of their objects ($45\pm 6\%$) and CO in absorption in 63 of their sources ($57\pm 7\%$). They detected $\rm H_2$ emission line in 47 sources ($43\pm 6\%$), Br$\gamma$ in 83 sources ($75\pm 8\%$), and CO in emission in 25 sources ($23\pm 5\%$). 

39 Class~II sources were analyzed and presented in \cite{Flores2022}, in which we detected the $\rm H_2$ $\nu=$ 1-0 $S(1) \, 2.122~\micron$ emission line in 9 sources ($23\pm 8\%$) and Br$\gamma$ in emission in 34 of the stars ($85\pm 15\%$). However, due to the design of their sample, \cite{Flores2022} did not include stars with CO in emission and only selected sources with deep photospheric lines.

These studies show that photospheric lines are highly uncommon in Class 0 sources, likely caused by their highly embedded nature. One of the impediments to placing better statistics on this class is that Class 0 sources are by definition extremely faint at near infrared wavelengths. Photospheric lines become more common for the Class I and FS sources, with an occurrence rate of $\sim 50-80 \%$. Sources classified as Class II sources will primarily exhibit deep photospheric lines. This is not surprising, as most of the envelope has already dissipated, and typically the protoplanetary disks do not block the stellar light (except for edge-on disks).

We find that the occurrence of Br$\gamma$ emission line in Class I and FS sources are consistent within uncertainties (assuming Poisson statistics) with the ones reported for other studies of Class 0 and II sources. It must be mentioned, however, that we are only comparing the occurrence of these lines and not the strength of the Br$\gamma$ emission line or the mass accretion rates. Thus, the conclusions we can draw are limited. 

The CO overtone emission in the $K$-band spectra of young stellar objects is more common ($\sim70\%$) in the Class 0 phase, it decreases to about  10-20\% for the Class I and FS phase, and becomes negligible towards the Class II phase. The $\rm H_2$ emission lines follow a similar pattern, in which about half of the spectrum of Class 0 and Class I sources display this molecular hydrogen line, but in the Class II phase, only about one-fourth of the sources do. Since the $\rm H_2$ is known to be a shock excited tracer, it reveals that the environment of Class 0 and I+FS sources are alike, but large differences can be found with the Class II sources, where the envelope and birth material have mostly dissipated.

\subsection{The circumstellar environment of Class I and FS sources}
\label{sec:discussion-circumstellar_environent}

In section \ref{sec:results_co_narrow_lines}, we computed the temperature and kinematics of gas along the line-of-sight of Class~I and FS sources from the narrow non-stellar rovibrational CO lines. In Table \ref{table:narrow_CO_lines-summary}, we  summarize the results, and in this section, we provide an interpretation of these observations. 

All the observations can be divided into having a single or a two-temperature component gas absorption. The two gas components can be associated with either cold gas (T$_{ex}<60$~K) or warm gas (T$_{ex}>60$~K), with a loose boundary separating both definitions. The cold component we detect has a typical temperature of $\sim$10-30~K, while the warm gas has a range of values from $\sim$70 to over 300~K. In general, the cold components have lower column densities ranging from $\sim10^{17}$ to $\sim10^{18}$ $\rm cm^{-2}$, while the warm components have about one order of magnitude higher column densities, i.e., tracing denser material, with values from $\sim10^{18}$ to $\sim10^{19}$ $\rm cm^{-2}$. For 19 of the 28 sources in Table \ref{table:narrow_CO_lines-summary}, we were able to measure the velocity of the central source from the fit to the photospheric lines (see Section \ref{sec:stellar_params}), and therefore we computed the relative velocity between the stellar sources and the CO foreground lines.

The CO gas with a single cold temperature measurement and an inconsistent radial velocity with respect to the central source was classified as foreground cloud absorption. If the radial velocity was consistent with the velocity of the central star, then we classified it as envelope absorption. We classified the CO gas absorption as coming from a circumstellar disk when we detect warm gas with relative velocity consistent with the central source (to ensure physical association).

When the warm component of the gas has a red-shifted velocity with respect to the central source, we interpret this as gas collapsing or infalling towards the central source. For the well-studied DG Tau star \citep[e.g.,][]{Mitchell1997, Zapata2015}, we detect signs of outflowing material, where the lines are blue-shifted with respect to the central star, as shown in Figure \ref{fig:rotational_diagram}. A sign of outflow was also detected for [GY92]~33.

We are not able to interpret the nature of the gas around stars without visible photospheric lines, given that we cannot obtain a relative RV shift between source and gas absorbing material. The temperature and kinematic plots for the sources can be found in the online version of this paper as figure sets.

In the sample of Class~II sources presented in \cite{Flores2022}, the presence of CO (non-photospheric) narrow lines only occurs in 7 out of their 39 Class~II stars ($18\pm 7\%$). In our sample of Class~I and FS objects, the occurrence is significantly higher, with 32/$\Nstars$ ($60\pm11\%$) showing narrow CO rovibrational lines. Since both samples of stars come mostly from Taurus and Ophiuchus, we can exclude the possibility that most of the CO absorption lines result from chance-alignment with foreground clouds. In general, the analysis of narrow CO overtone lines provides an alternative method to trace the near circumstellar environment in embedded sources using high-resolution $K$-band spectroscopic observations.

\section{Conclusion}
\label{sec:conclusion}

We have obtained iSHELL $K$-band observations of $\Nstars$ Class~I and FS sources from the Taurus-Auriga, Ophiuchus, Orion, Serpens, and Corona Australis star-forming regions. We used a magnetic radiative transfer code to measure the magnetic field strength, temperature, gravity, rotational velocity, and infrared veiling for $\Nstarsmodeled$  Class~I and FS sources. We have combined our stellar parameters with a magnetic evolutionary model to derive ages and masses for these sources. In addition, we have detected a myriad of circumstellar, accretion, and outflow indicators in the spectra of the stars from absorption and emission lines. We have employed a rotational diagram method to detect the presence of envelope and disk material around some Class~I and FS sources. A summary of our findings is as follows:

\begin{enumerate}
\item We present the largest sample of magnetic fields measured for Class~I and FS sources to date, which range from  $<1.0$~kG to 4.1~kG. The magnetic field strengths we measure are overall consistent with the magnetic field values of Class~II sources in the literature. By comparing our sources to the sample of 39 Class~II sources from \cite{Flores2022}, we found that the distributions of their magnetic field strengths are statistically indistinguishable. These results confirm that Class~I and FS sources host strong magnetic fields on their photospheres with values similarly to those found on Class~II sources.

\item We find that the distribution of gravities and temperatures between Class~I + FS and Class~II sources are statistically different, although a significant overlap exist. The median gravities of Class I and FS sources ($\log{g}=3.48$) are 0.3 dex lower than those of Class II sources ($\log{g}=3.8$). Moreover, some Class I and FS sources can have gravity values as low as $\log{g}=2.8$.  When combined with stellar evolutionary models, our findings suggest that Class~I and FS objects are younger and less massive than the Class~II sources. Using \cite{Feiden2016} magnetic evolutionary models, we find that about half of the Class I and FS sources have ages younger or equal to 0.6 Myrs old. 

\item The statistics of the $\rm H_2$, and CO emission lines in our sample are similar to other Class~I and FS source surveys. These lines are less frequent or equal than in studies of Class 0, and significantly higher than for the sample of Class~II stars. We interpret the variation in the occurrence of the $\rm H_2$ emission line as clear evidence of a change in the circumstellar environment of the different classes. The occurrence of Br$\gamma$, on the other hand, is consistent with uncertainties in all the classes.

\item We detected narrow CO non-photospheric lines in the spectra of our young sources. A detailed analysis reveals that most of these lines originate from a disk, an envelope, or from infalling material, and just a few correspond to foreground intervening clouds. This analysis reaffirms that Class~I and FS sources have a rich circumstellar environment, a known characteristic of young stellar objects.

\end{enumerate}

In conclusion, we present the first measurements of magnetic field strengths for a large sample of Class I and FS sources. The gravities measured for these sources indicate that many of these sources are younger than Class II sources and could correspond to protostars with ages of a few hundred thousand years. The detection of kG magnetic fields at these early ages, along with accretion signatures from emission lines, suggests that the magnetospheric accretion process starts early in the life of a protostar and continues throughout the T Tauri phase until the circumstellar material has completely dissipated.


\newpage

\begin{acknowledgments}
\section*{ACKNOWLEDGEMENTS}
Christian Flores, Michael S. Connelley, and Adwin Boogert acknowledge support from the NASA Infrared Telescope Facility, which is operated by the University of Hawaii under contract 80HQTR19D0030 with the National Aeronautics and Space Administration.The technical support and advanced computing resources from University of Hawaii Information Technology Services – Cyberinfrastructure, funded in part by the National Science Foundation MRI award $\# 1920304$, are gratefully acknowledged. We are particularly grateful to the iSHELL team for all the help provided during the data reduction process. This research has made use of NASA's Astrophysics Data System Bibliographic Services. This research has made use of the SIMBAD database, operated at CDS, Strasbourg, France. This work has made use of the VALD database, operated at Uppsala University, the Institute of Astronomy RAS in Moscow, and the University of Vienna. This publication makes use of data products from the Two Micron All Sky Survey, which is a joint project of the University of Massachusetts and the Infrared Processing and Analysis Center/California Institute of Technology, funded by the National Aeronautics and Space Administration and the National Science Foundation. This publication makes use of data products from the Wide-field Infrared Survey Explorer, which is a joint project of the University of California, Los Angeles, and the Jet Propulsion Laboratory/California Institute of Technology, funded by the National Aeronautics and Space Administration. This research has made use of the NASA/IPAC Infrared Science Archive, which is funded by the National Aeronautics and Space Administration and operated by the California Institute of Technology.
\end{acknowledgments}

%

\newpage

\vspace{10mm}
\facility{IRTF, IRSA}


\vspace{10mm}

\software{astropy \citep{2013A&A...558A..33A}
          MoogStokes \citep{Deen2013}, 
          emcee \citep{Foreman-Mackey2013}, 
          Spextools \citep{Cushing2004}, 
          xtellcor \citep{vacca2003}
          }



\appendix

\section{Rotational Diagrams}
\label{appendix:rotational_diagrams}

To analyze the detected CO rovibrational absorption lines in our Class~I and FS sources, we follow the analysis of the CO fundamental $v=1-0$ lines detailed in \cite{Barr2020} and \cite{Smith2009} but applied to the first CO overtone $v=2-0$ observations. 

The observed spectra were continuum normalized as described in Section \ref{sec:observations}. We used the HITRAN \citep{Gordon2021} CO database to find the position of the CO $v=2-0$ overtone R- and P- branch lines and then fitted a Gaussian profile convolved with the iSHELL $K$-band 0\farcs75 instrumental response to accurately characterize each line. To exclude the broader CO absorption lines from the stellar photosphere (when present), we simultaneously fit a polynomial baseline of order 5 along with the Gaussian profile. The fitting process was performed in an MCMC fashion running 200,000 models and discarding half of them as a burn-in phase. 

During the fitting process, we allowed the wavelength position, the Gaussian FWHM, a normalization factor, and the six parameters related to the 5th order polynomial to freely vary within the given priors. We assumed flat priors for each parameter and allowed the wavelength shift to vary within $3~\rm \AA$ ($\sim~40 \, \rm km \, s^{-1}$). We visually checked the quality of the fits for each individual line and also a successful convergence of the MCMC runs. 

\begin{figure*}
\gridline{\fig{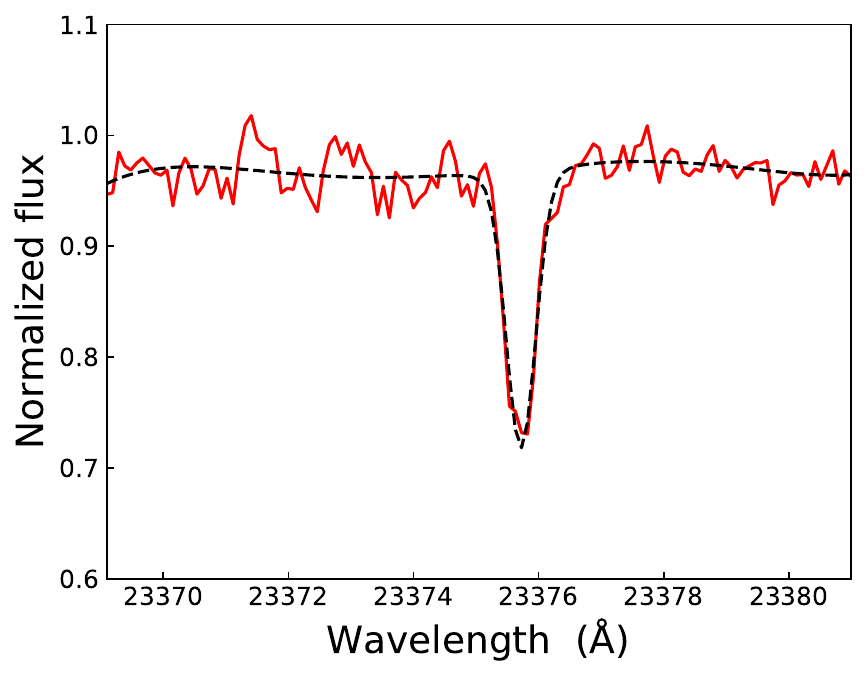}{0.3\textwidth}{R(4)}
          \fig{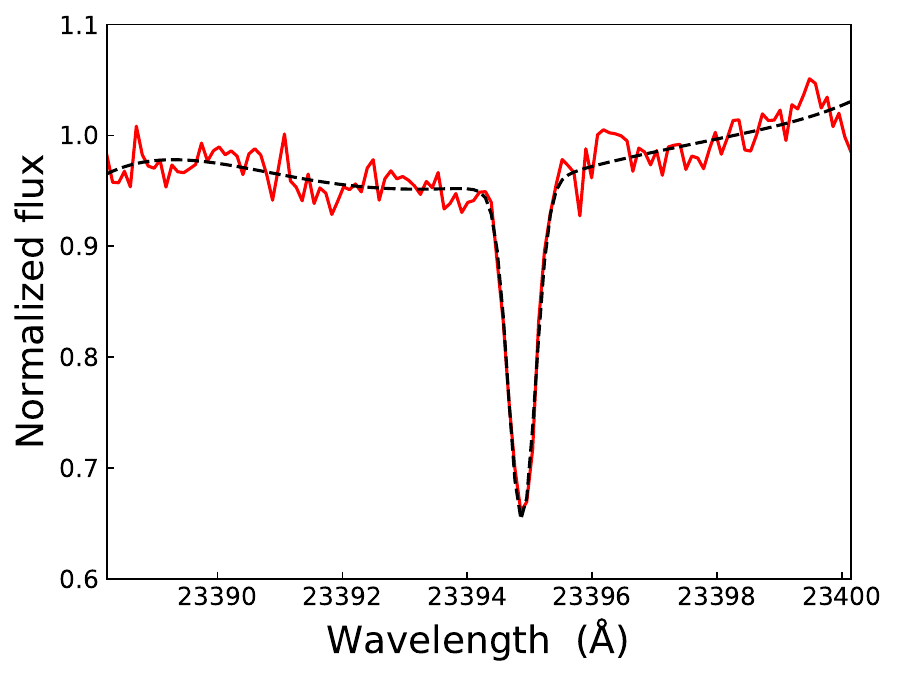}{0.3\textwidth}{R(3)}
          \fig{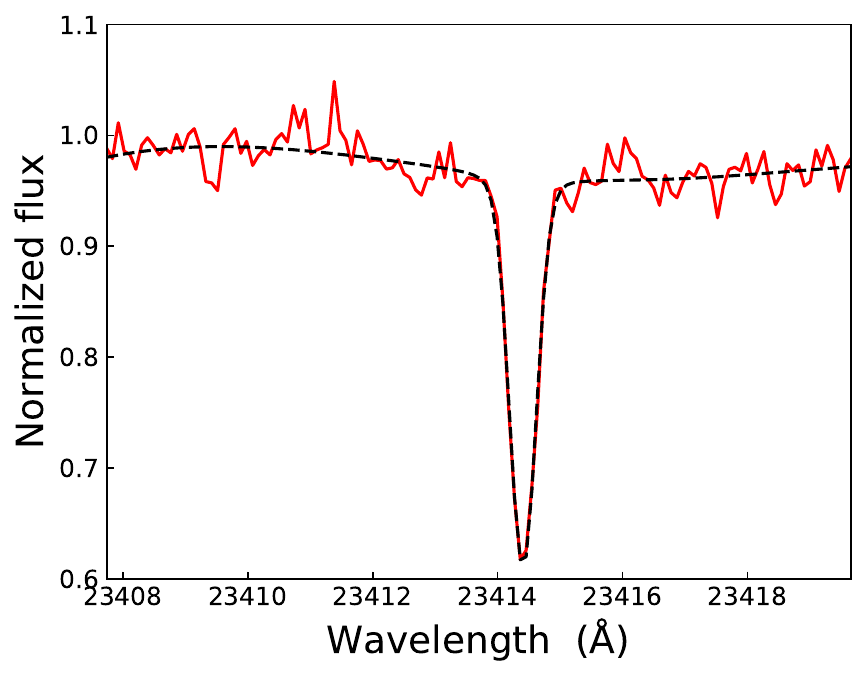}{0.3\textwidth}{R(2)}
          }
\gridline{\fig{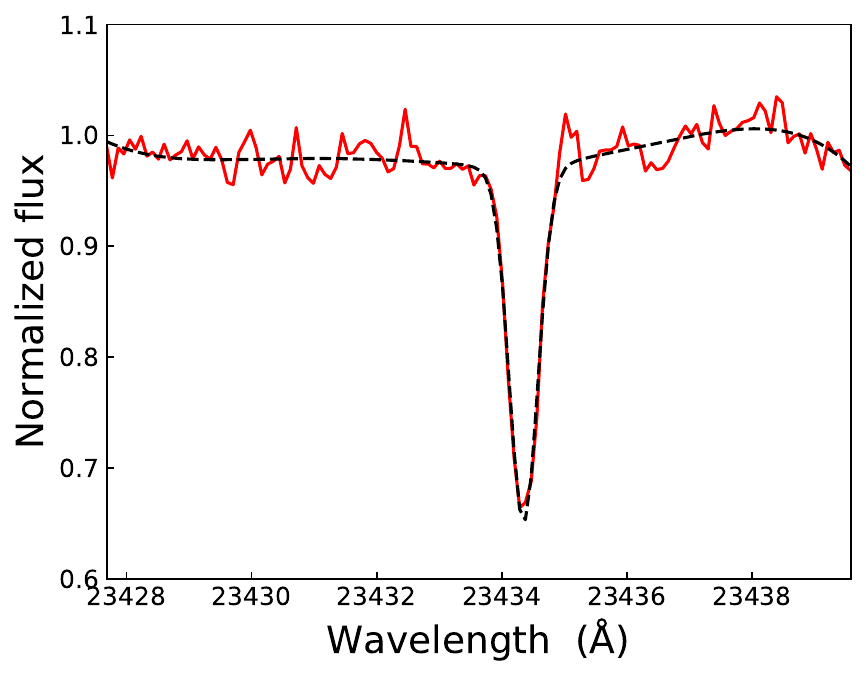}{0.3\textwidth}{R(1)}
          \fig{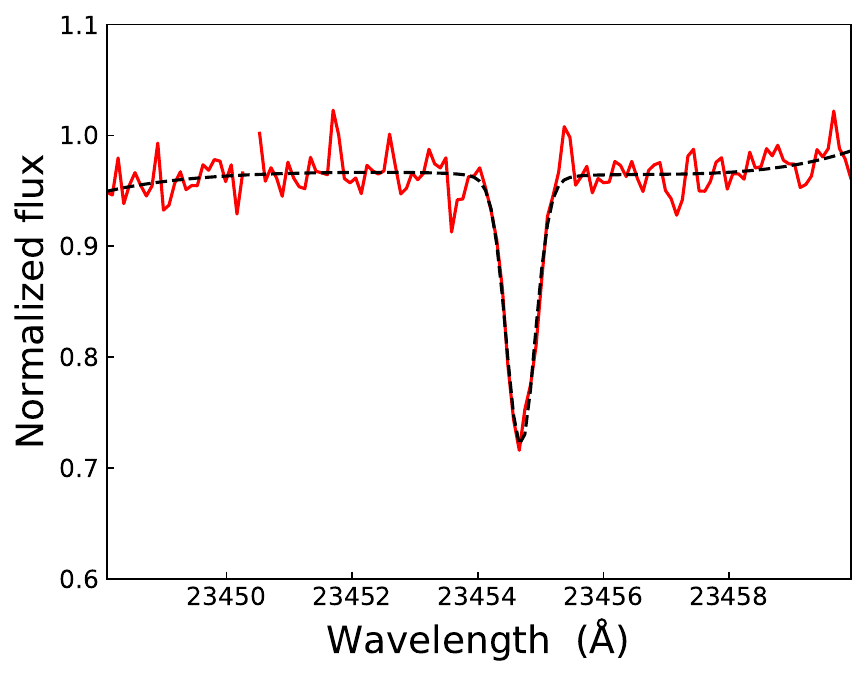}{0.3\textwidth}{R(0)}
          \fig{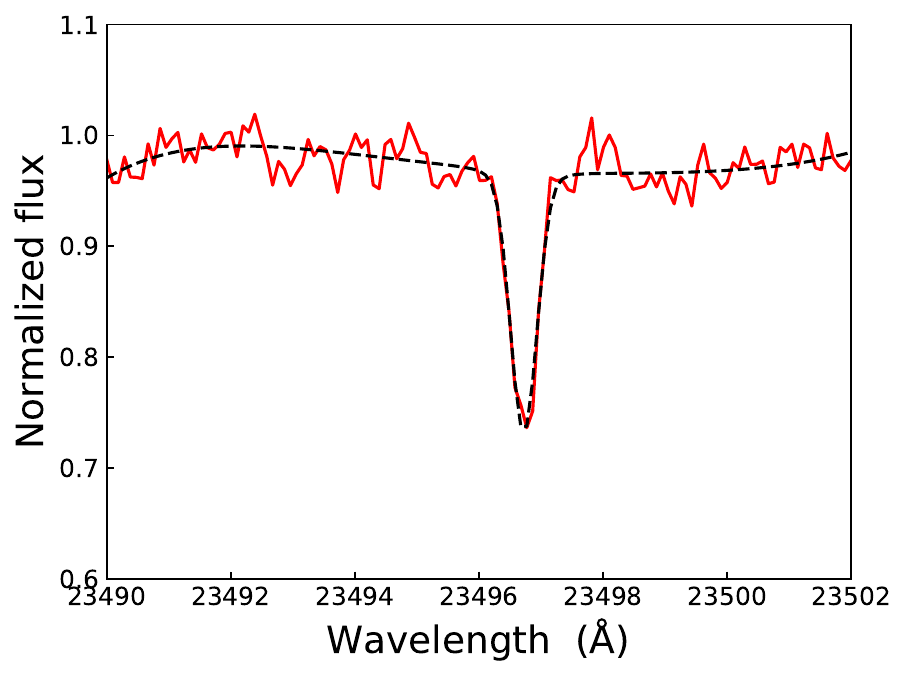}{0.3\textwidth}{P(1)}
          }
\gridline{\fig{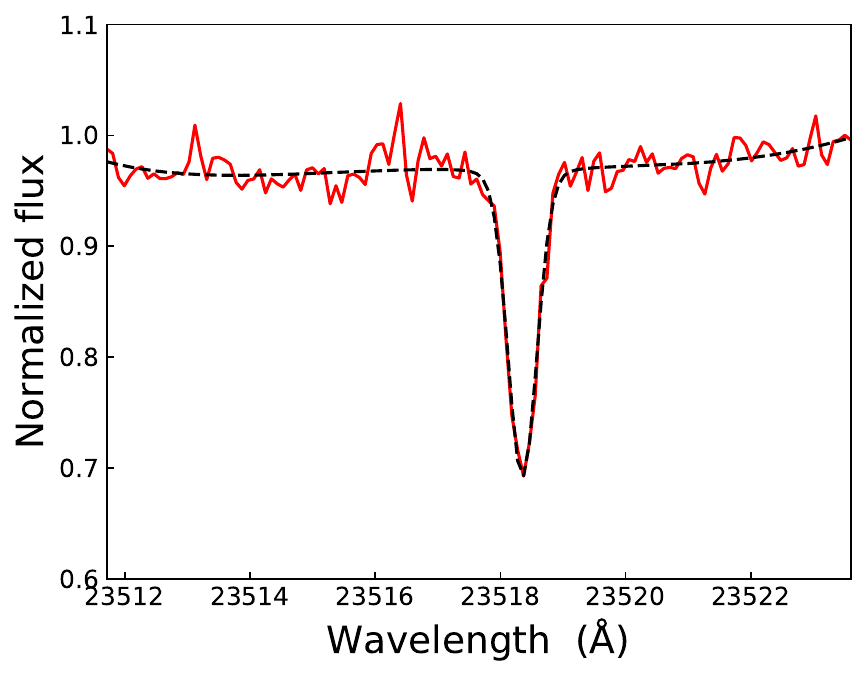}{0.3\textwidth}{P(2)}
          \fig{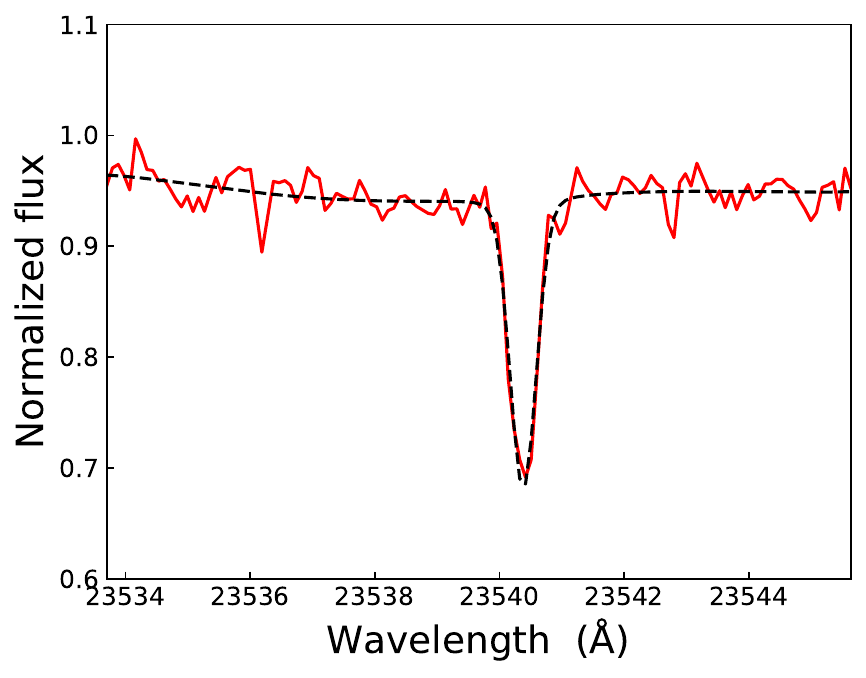}{0.3\textwidth}{P(3)}
          \fig{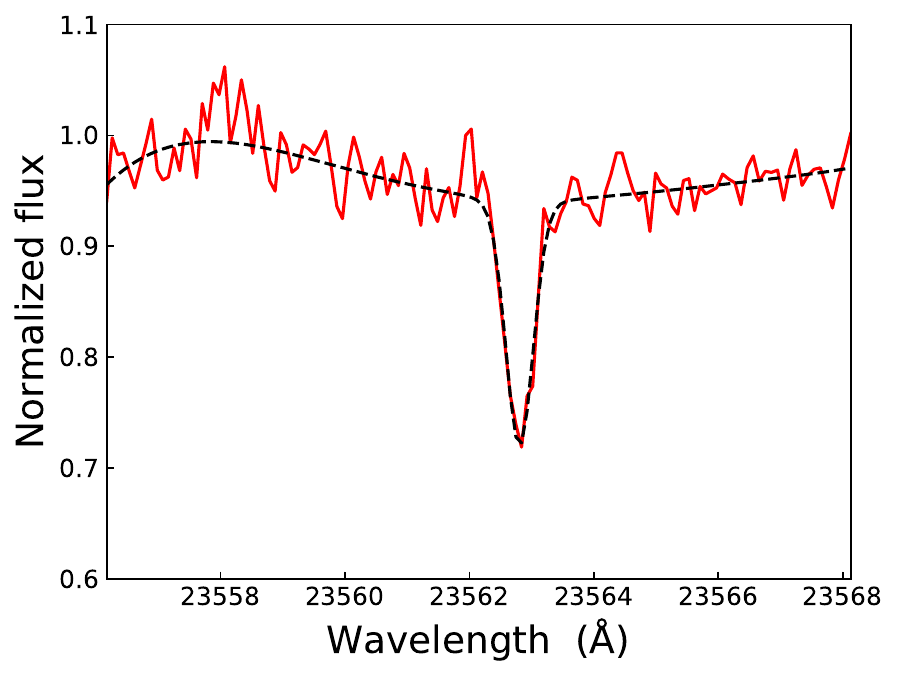}{0.3\textwidth}{P(4)}
          }
\caption{ Gaussian convolved with iSHELL profile plus baseline fitting for the R(4) to P(4) ro-vibrational CO lines in the spectra of IRAS16285-2355.
\label{fig:individual-line-fit-CO}}
\end{figure*}

In Figure \ref{fig:individual-line-fit-CO}, we show the fitting to the non-stellar R(4) to P(4) CO overtone lines in the spectrum of IRAS16285-2355 (shown in Figure \ref{fig:rotational_CO_lines-example}). The individual fits successfully reproduced the wavelength peak, FHWM, and normalization of the CO lines. Also, the 5th order polynomial function properly reproduces the underlying CO stellar absorption lines, ensuring that the depth of the narrow lines are correctly calculated.

After we fitted the individual lines, we transformed the CO line flux to an optical depth profile in velocity space to obtain the column density (in $cm^{-2}$) of the transitions using equation (\ref{eq:flux_to_optical_depth}).

\begin{eqnarray}
    \label{eq:flux_to_optical_depth}
    \tau(v) &= -\ln \left( \frac{flux}{continuum} \right)\\ \label{eq:lower-level-coldens}
    \frac{dN_l}{dv} &= \tau( v) \frac{g_l}{g_u} \frac{8\pi}{A \lambda^3}
\end{eqnarray}

In equation (\ref{eq:flux_to_optical_depth}), $\tau(v)$ is the optical depth profile in velocity space, $g_l$ and $g_u$ are the statistical weights (degeneracy) of the lower and upper states respectively, A is the Einstein coefficient for spontaneous emission of the transition, and $\lambda$ is the wavelength.

We used the Simpson integration algorithm in scipy to integrate equation (\ref{eq:lower-level-coldens}) and obtain the lower energy level column density $N_l$. To derive an excitation temperature $T_{ex}$ and total column density $N$ for a parcel of gas in local thermodynamic equilibrium, we use the Boltzmann equation (\ref{eq:Boltzman}). 

\begin{equation}
\label{eq:Boltzman}
    \ln\frac{N_l}{g_l} = \ln (N) - \ln (Z(T_{ex})) - \frac{E_l}{kT}
\end{equation}

Here, $E_l$ is the energy of the lower state, and $Z(T)$ is the partition function at the excitation temperature. The partition function for CO were obtained from the HITRAN database.

The rotation diagram was populated using our derived lower-level column densities and quantum mechanical parameters for each CO transition. Then a single or double straight line were fitted to the data (as shown in Figure \ref{fig:rotational_diagram}) to derive the excitation temperature and column density. In the rotation diagram, the inverse of the slope of the lines relates to the temperature of the absorbing gas, thus the steeper slopes correspond to cooler temperatures. The results are presented in Figure \ref{fig:rotational_diagram} as figure set.

\begin{deluxetable*}{lccccccc}
\tabletypesize{\scriptsize}
\tablewidth{0pt}
\tablecolumns{7}
\tablecaption{Emission and absorption lines in the spectra of the Class I and FS sources. \label{table:emission_and_absorption_lines}}
\tablehead{
\colhead{Source name} & \colhead{Thin CO } & \colhead{$H_2$ at 2.1220 } & \colhead{$H_2$ at 2.2235} & \colhead{CO emission} & \colhead{Br $\gamma$ emission} & {Photospheric} & {Stellar} \\
\colhead{} & \colhead{rovibrational bands} & \colhead{$\micron$} & \colhead{$\micron$} & \colhead{} & \colhead{} & \colhead{lines} & {parameters}}
\startdata
2MASSJ16263682 &  & & & & \checkmark  & \checkmark & Fringy spectrum \\ \hline
DG Tau & \checkmark  & \checkmark & \checkmark & maybe & \checkmark & \checkmark & \checkmark \\ \hline
EC92 & \checkmark  & \checkmark & & & & \checkmark & \checkmark \\ \hline
EC95 & \checkmark  & & & &  & \checkmark & \checkmark \\ \hline
EC129 & \checkmark & \checkmark & & \checkmark & \checkmark  & \checkmark & Shallow lines \\ \hline
Elia 2-32 & \checkmark & & & & & \checkmark & \checkmark \\ \hline
GV Tau S  & \checkmark  & \checkmark & \checkmark & \checkmark & \checkmark & \checkmark & \checkmark\\ \hline
GY21 & \checkmark  & \checkmark & \checkmark & & & & No photospheric lines \\ \hline
GY224 & \checkmark & \checkmark & & & \checkmark & & No photospheric lines  \\ \hline
GY235 & \checkmark  & \checkmark & & & \checkmark & \checkmark & \checkmark \\ \hline
GY284 &   & & & & & \checkmark & \checkmark \\ \hline
GY33 & \checkmark & & & &  & \checkmark & \checkmark \\ \hline
Haro 6-13 & \checkmark  & & & & \checkmark & \checkmark & \checkmark \\ \hline
Haro 6-28 A &  & & & & \checkmark & \checkmark & \checkmark \\ \hline
IRAS03220+3035N &  & & & \checkmark & \checkmark & & No photospheric lines \\ \hline
IRAS03220+3035S &  & & & & & \checkmark & Shallow lines \\ \hline
IRAS03260+3111B &  & & & & & \checkmark & \checkmark \\ \hline
IRAS03301+3111 & \checkmark  & \checkmark & \checkmark & maybe  & \checkmark & \checkmark & \checkmark  \\ \hline
IRAS04108+2803E & \checkmark  & \checkmark & \checkmark & & \checkmark & \checkmark & \checkmark  \\ \hline
IRAS04108+2803W &  & & & & & \checkmark & Cold source  \\ \hline
IRAS04113+2758S &  & & & & & \checkmark & \checkmark \\ \hline
IRAS04158+2805 & \checkmark & \checkmark & \checkmark & & & \checkmark & Shallow lines \\ \hline
IRAS04181+2654M & \checkmark & \checkmark & \checkmark & & & \checkmark & \checkmark \\ \hline
IRAS04181+2654S & \checkmark & \checkmark & \checkmark & & \checkmark & \checkmark & \checkmark  \\ \hline
IRAS04295+2251 & \checkmark & \checkmark & \checkmark & & \checkmark & \checkmark & \checkmark \\ \hline
IRAS04369+2539 & \checkmark  & & & \checkmark & \checkmark & & No photospheric lines  \\ \hline
IRAS04489+3042 & \checkmark  & \checkmark & & & \checkmark & \checkmark & \checkmark \\ \hline
IRAS04591-0856 & \checkmark & \checkmark & \checkmark & & & \checkmark & \checkmark \\ \hline
IRAS05256+3049 &  & \checkmark & \checkmark & \checkmark & \checkmark & & No photospheric lines  \\ \hline
IRAS05311-0631 & \checkmark  & & & & \checkmark  & \checkmark & Shallow lines  \\ \hline
IRAS05379-0758(2) &  & \checkmark  & & & \checkmark & \checkmark & \checkmark  \\ \hline
IRAS05513-1024 &  & & & \checkmark  & \checkmark  & & No photospheric lines  \\ \hline
IRAS05555-1405(4) & \checkmark  & \checkmark  & & & \checkmark  & \checkmark & \checkmark  \\ \hline
IRAS16235-2416 &  & & & & & \checkmark & Hot source? \\ \hline
IRAS16285-2355 & \checkmark & & & & \checkmark & \checkmark & Shallow lines \\ \hline
IRAS16285-2358 &  & \checkmark & \checkmark & & \checkmark & \checkmark & \checkmark \\ \hline
IRAS16288-2450W2 & & & & & & \checkmark & \checkmark \\ \hline
IRAS16316-1540 & \checkmark & \checkmark & \checkmark & & & \checkmark & Shallow lines \\ \hline
IRAS18270-0153 &  & \checkmark & \checkmark & & & \checkmark & Low S/N \\ \hline
IRAS19247+2238-1 &  & \checkmark & & & \checkmark & \checkmark & \checkmark  \\ \hline
IRAS19247+2238-2 &  & \checkmark & & & \checkmark & \checkmark & \checkmark \\ \hline
RCraIRS5 NNE & \checkmark  & \checkmark & \checkmark & & \checkmark & \checkmark & \checkmark \\ \hline
SR24B &  & \checkmark & \checkmark & & \checkmark & \checkmark & Unresolved binary \\ \hline
SR24A  & \checkmark & \checkmark & & maybe & \checkmark & \checkmark & \checkmark \\ \hline
V347 Aur &  & & & & \checkmark & \checkmark & \checkmark \\ \hline
VSSG 17 & \checkmark  & & & & \checkmark & \checkmark & \checkmark \\ \hline
WL20(E) & \checkmark & & & & & \checkmark & \checkmark \\ \hline
WL20(W) & \checkmark & & & & & \checkmark & \checkmark \\ \hline
WL4  & \checkmark & & & & & \checkmark & Unresolved binary \\ \hline
WLY2-42  &  & & & & & \checkmark & \checkmark \\ \hline
WLY2-54  & \checkmark  & \checkmark & \checkmark &  & \checkmark & & No photospheric lines \\ \hline
YLW 45 & \checkmark & \checkmark & & & \checkmark & & No photospheric lines \\ \hline
\enddata
\end{deluxetable*}

\section{Exclusion of aluminum lines in our models}
\label{appendix:aluminum-no-aluminum}

In Section \ref{sec:results}, we mentioned that a subset of our stars display strong water absorption lines in their spectra, which affect the continuum normalization and the choice of a pseudo-continuum during the synthetic fitting process. These stars are typically cooler than $\tkband \sim 3500$~K, have low rotational velocities, and low gravities. Since the water absorption lines are worse in the region redward of 2.2~\micron, we tried models that include and exclude the aluminum regions at 2.110~\micron \, and 2.117~\micron, for such stars.

To understand the effects of excluding the aluminum lines, we modeled 14 main- and post-main-sequence stars using the synthetic spectrum code described in Section \ref{sec:stellar_params}. Three of the standard stars in our sample are red giants with accurate gravities calculated from asteroseismology. Five sources are sub-giant stars selected from the APOKASC catalog \citep{Serenelli2017}, which have accurate gravities derived from asteroseismology and temperatures from APOGEE H-band spectroscopy. Five of the stars are M dwarfs with temperatures calculated from interferometric radii measurements; the remaining star is a K dwarf with stellar parameters measured from high-resolution optical spectroscopy. 
We selected our sample of standard stars to span a temperature range from $\sim$3\,400 K to $\sim$5\,400 K and gravities from 2.6 to 4.9, which encompass the temperature and gravity ranges of low-mass young stars \citep{Doppmann2005,Baraffe2015,Feiden2016}. All but the APOKASC sources presented in this section were already published in \cite{Flores2019}

Unlike young stars, most late-type main-sequence and post-main-sequence stars have a very small or non-detectable amount of circumstellar material around them. Therefore, during this analysis, we assumed that main-sequence and post-main-sequence stars have a negligible excess of infrared radiation, and we do not include the IR $K$-band veiling parameter in the calculations. Instead, we allowed the stellar metallicity [M/H] to vary. The stellar parameters we simultaneously fit for the sample of main and post-main sequence stars are T$_{\rm eff}$, log(g),  $\langle \rm B \rangle$ , $v\sin(i)$, $v_{\rm micro}$, and metallicity [M/H]. 

\begin{figure*}
\gridline{\fig{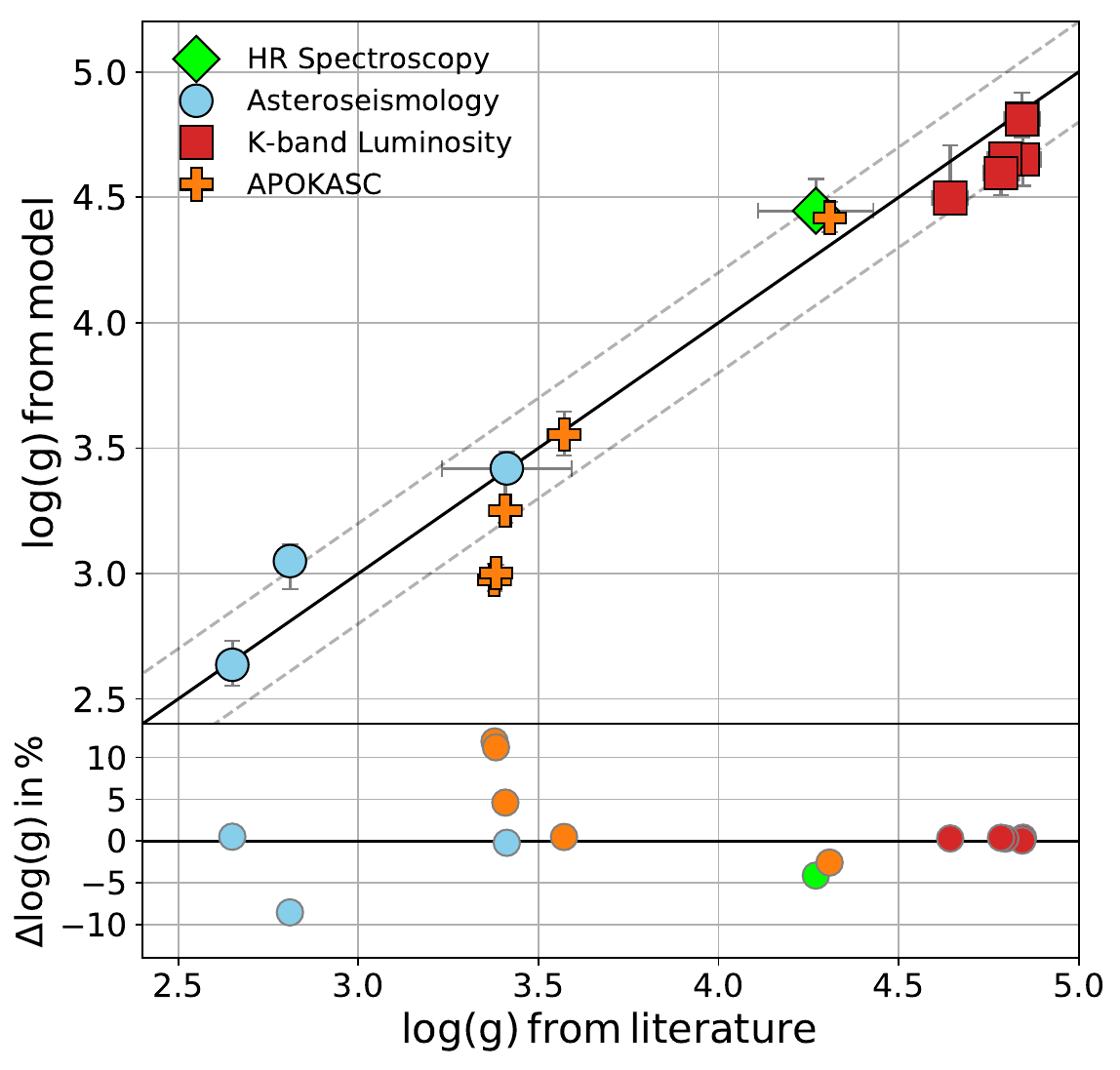}{0.45\textwidth}{(a) Comparison between literature gravities and our gravities derived from models where the aluminum regions at 2.110~$\micron$ and 2.117~$\micron$ are included.}
          \fig{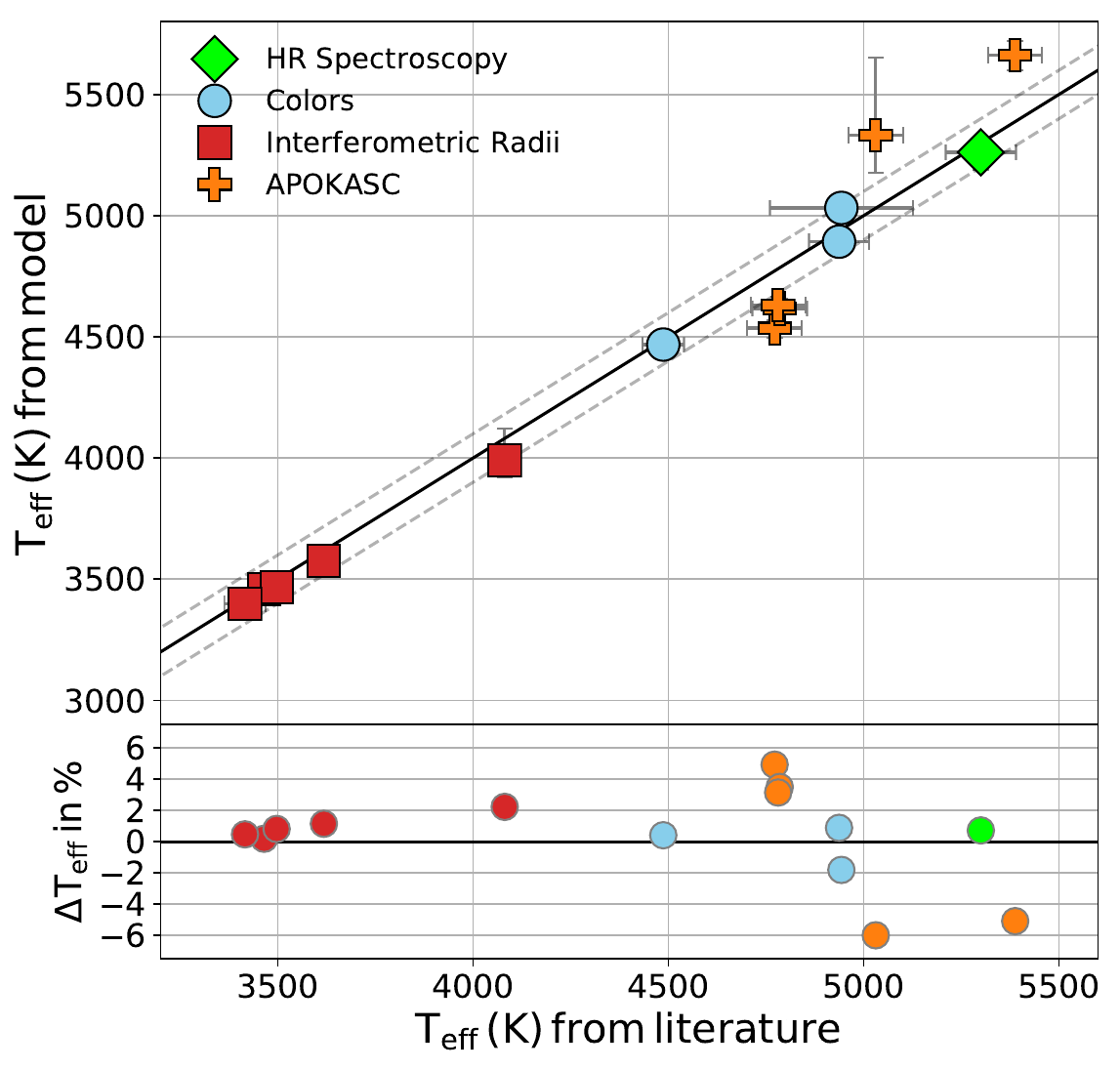}{0.45\textwidth}{(b) Comparison between literature temperatures and our temperature derived from models where the aluminum regions at 2.110~$\micron$ and 2.117~$\micron$ are included.}
          }
\gridline{\fig{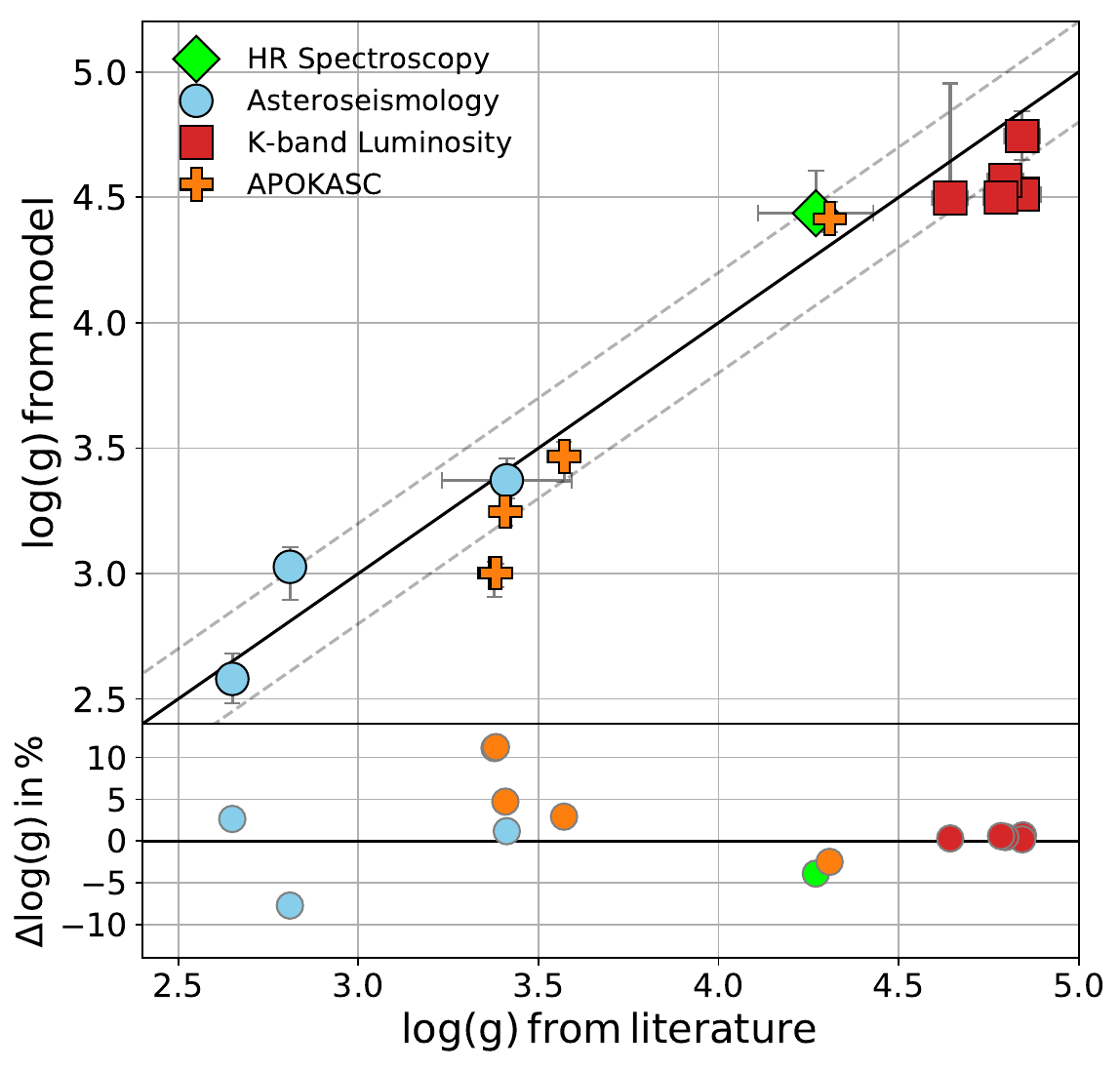}{0.45\textwidth}{(c) Same as panel a but when the aluminum regions are excluded.}
          \fig{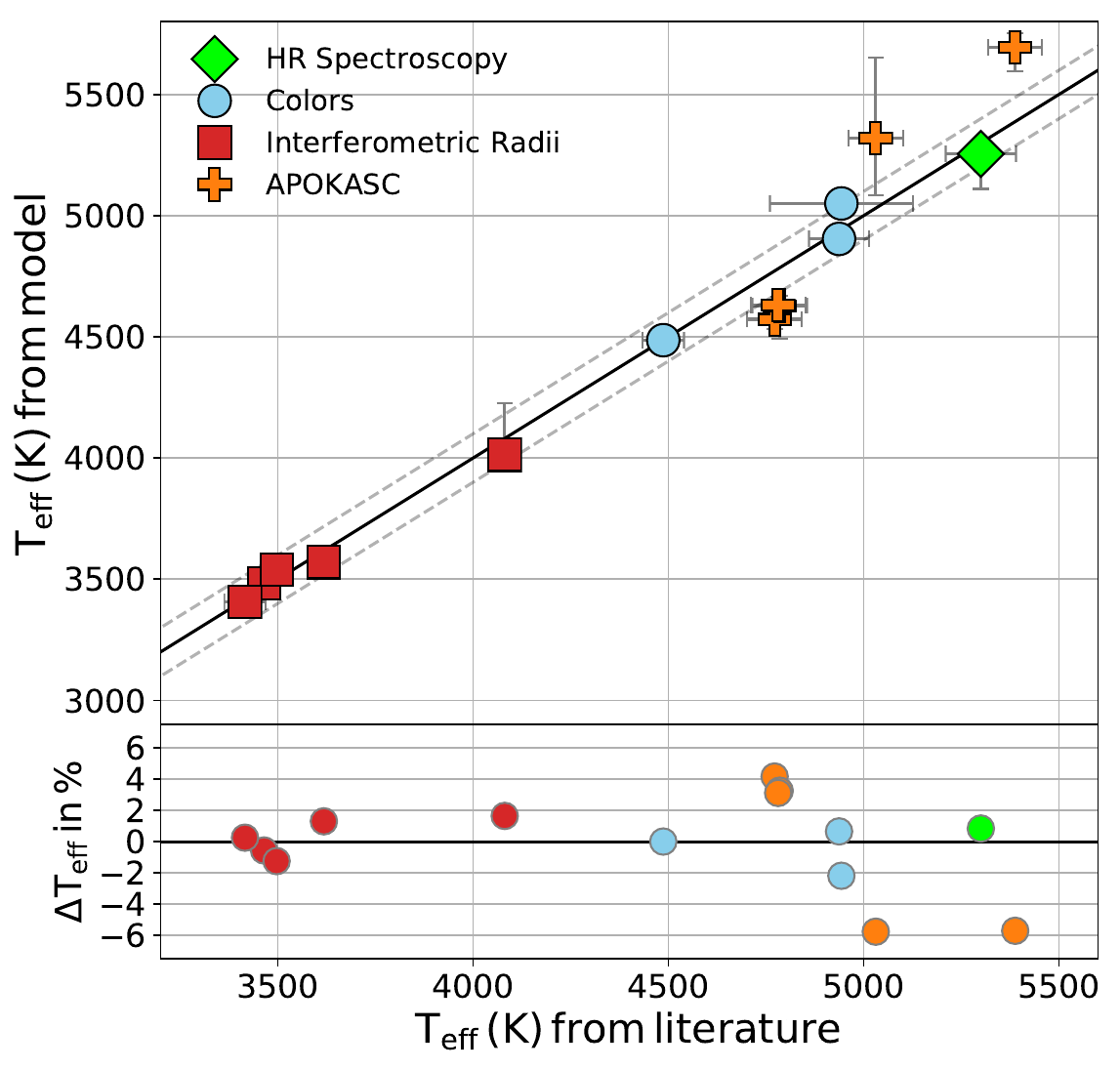}{0.45\textwidth}{(d) Same as panel b but when the aluminum regions are excluded.}
          }
\caption{In the first column (panels a and c), we show a comparison between the literature surface gravity vs. our derived surface gravity for the sample of main- and post-main-sequence stars. The solid black line corresponds to a one-to-one correlation between the model and the literature parameters. The dashed gray lines correspond to differences of $\pm$0.2 dex between both measurements. The second column (panels b and d) shows the comparison between the literature effective temperatures vs. our derived effective temperatures for the sample of main- and post-main-sequence stars. The solid black line corresponds to a one-to-one correlation between the model and the literature parameters. The dashed gray lines correspond to differences of $\pm$100~K between both measurements. In panels a and b, the aluminum regions are included in the fit, in panels c and d, we excluded them.
\label{fig:stellar-params-confirmation}}
\end{figure*}

In the first column of Figure \ref{fig:stellar-params-confirmation}, we show a comparison between log(g) derived from the literature and the log(g) values derived from our models.  In this figure, dwarf stars with log(g) values calculated from absolute $K$-band mass-luminosity relations \citep{Boyajian2012} are represented as squares, the dwarf star with log(g) value obtained from high-resolution optical spectroscopy \citep{Gonzalez1999} is shown as a diamond, and the giant stars with accurate gravities calculated from asteroseismology \citep{Huber2016} are shown as circles and crosses for the stars of the APOKASC sample. 

Panel (a) shows the gravity results of the models where we included the aluminum regions. A median gravity difference between literature and our values of $\Delta log{g} = 0.09$ in log of $cm\, s^{-2}$ or $\Delta log{g} = 2\%$ is measured in this case. In panel (c), we show a similar comparison but for models where the aluminum regions were excluded. In this case, the median gravity difference is $\Delta log{g} = 0.13$ in log of $cm\, s^{-2}$ or $\Delta log{g} = 3\%$. The difference between including and excluding the aluminum regions is at most $1\%$, and in both cases our gravities are accurate to high precision (in both cases the standard deviation of the difference is $5\%$).

The second column of Figure \ref{fig:stellar-params-confirmation}, shows how literature temperature compares to our derived temperature for the main and post-main sequence stars. The temperature values for the sources are calculated from interferometric radii measurements (squares), high-resolution spectroscopy (cross and diamond), or from colors (circles). Panel (b) shows the results in which we used the aluminum regions, while panel (d) show the results in which these were excluded. When the aluminum regions are included the median temperature difference is $\Delta T = 21$~K, but when excluded it is of  $\Delta T = 33$~K. In both cases, the median temperature difference is at most 1\%, and the standard deviation of the temperature difference is  3\%. More generally, our derived temperature values agree within 6\% with the literature temperature values regardless of whether the aluminum regions are included or not. 

\section{Residuals and Chi-squared tests}
\label{appendix:chi-squared}
To test the goodness of fit in our models, we report the reduced $\chi^2$ metric associated with each source in the last column of Table \ref{table:Stellar_params}. To compute this reduced $\chi^2$, we counted the number of degrees of freedoms as $N-M$, where N is the number of data points, typically $N \sim 3500$ with a slightly larger number of data points used for rapidly rotating sources ($N \sim 3650$) to accommodate for the broaden lines. M refers to the number of fitted parameters during the second model iteration, i.e., $M=5$. As shown in Table \ref{table:Stellar_params}, the reduced $\chi^2$ values range from 1.01  to 2.81, with a median value of 1.93. We have additionally computed the average residual when data and models are subtracted. We find that these residuals range between 0.99 to 1.69 times the measured noise in their stellar spectra,  with a median value of 1.34.

\begin{figure*}
\gridline{\fig{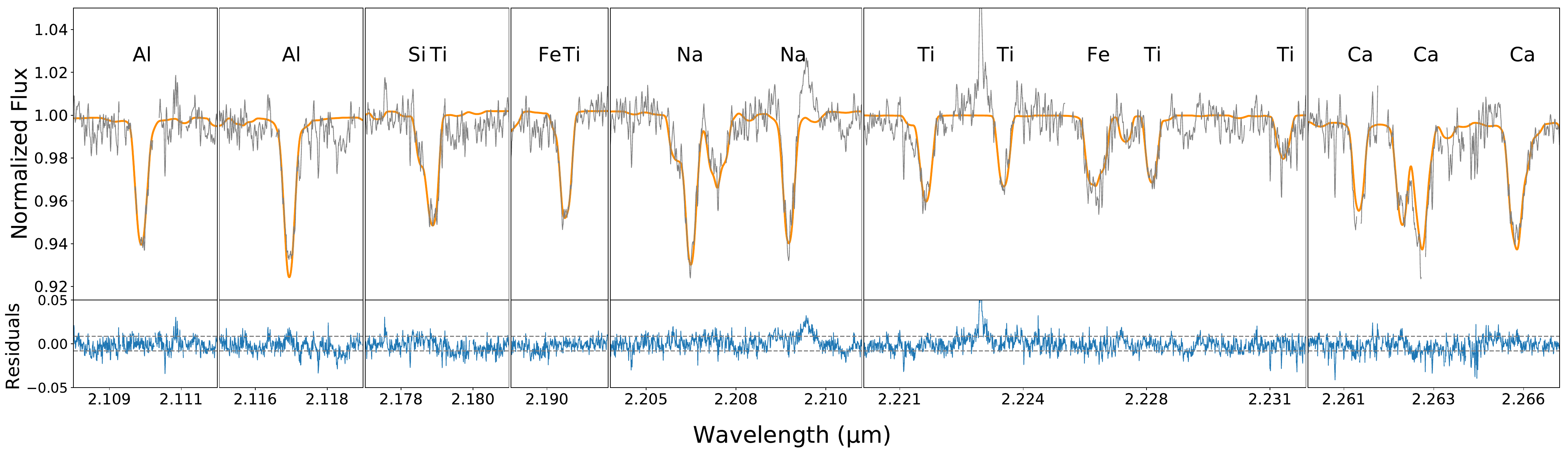}
{1.0\textwidth}{(a) DG Tau - median residual = 1.02$\sigma$ - reduced $\chi^2$ = 1.01}
   }
\gridline{\fig{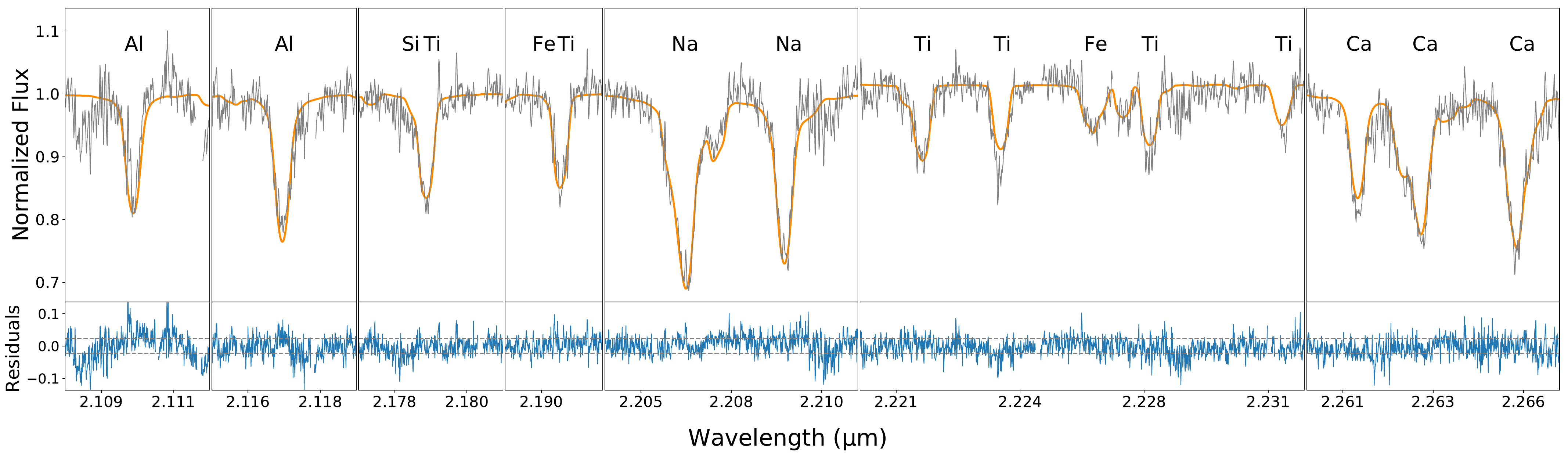}{1.0\textwidth}{(b) WL20(W) -  median residual = 1.46$\sigma$ - reduced $\chi^2$ = 2.06}
          }
\gridline{\fig{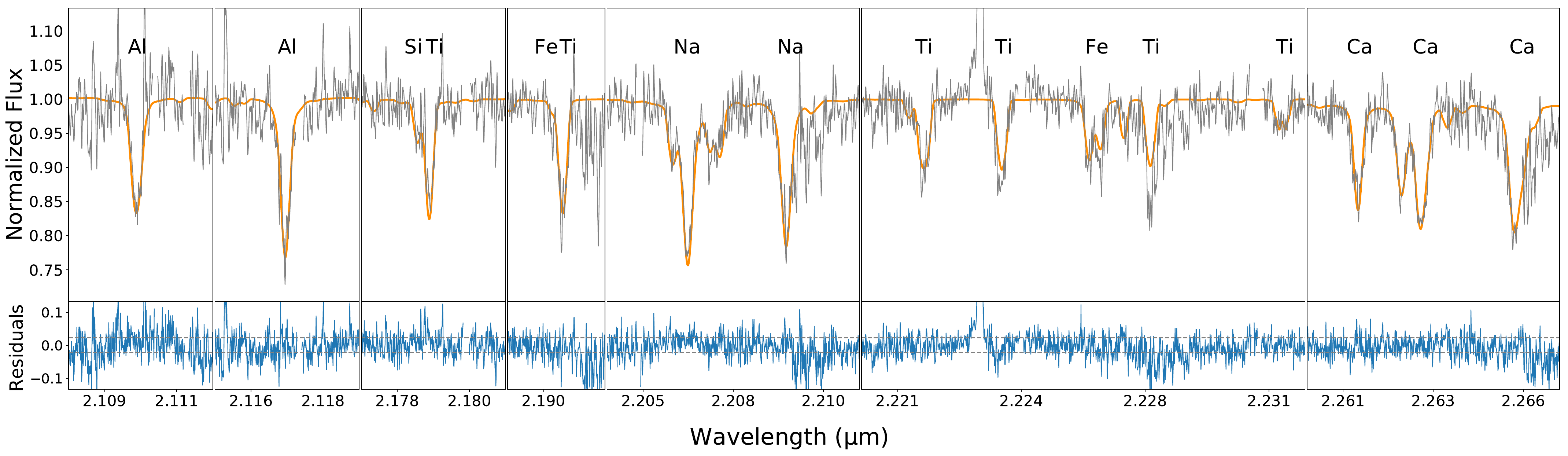}{1.0\textwidth}{(c) IRAS04108+2803(E) -  median residual = 1.69$\sigma$ - reduced $\chi^2$ = 2.61}}

\caption{MoogStokes models vs. spectra for different sources emphasising the residuals (data - model). Three cases were selected to illustrate the level of residuals in our work (a) DG Tau, (b) WL20(W), and (c) IRAS04108+2803(E). The sources are sorted from lower to higher residuals from top to bottom. In the residual box, we displayed a $\pm 1\sigma$ noise for each source as dashed gray lines. \label{fig:residual_plot}}
\end{figure*}

In Figure \ref{fig:residual_plot}, we show three representative examples of the residuals obtained by subtracting the data from the model. For each source, we have indicated the reduced $\chi^2$ value as well as the median residual. DG Tau has one of the lowest median residuals and reduced $\chi^2$ values (top panel). WL20(W) has a reduced $\chi^2$ value close to the median of the sample (middle panel), and IRAS04108+2803(E) is among the highest reduced $\chi^2$ values (bottom panel). Often, the residuals are produced because we have neglected water lines in the spectrum of the stars, which affect more strongly the cooler sources at wavelengths redward of $\sim 2.2~\mu m$. For example, this can be seen in the left-most panels for WL20(W) and IRAS04108+2803(E). Other model simplifications, such as considering a single temperature model for such thermally complex sources or using a single magnetic field strength component to represent complex magnetic structures, likely also contribute to the observed residual. Further work could address these problems with more detailed models. In the residual and $\chi^2$ calculations, we have omitted the $H_2$ emission line at 2.2235~$\mu m$ as our model does not attempt to reproduce any emission line.


\section{Magnetic field and rotational velocities}
\label{appendix:magnetic_fields_rotation}
In our sample of young stars, we find some fast-rotating sources, with projected rotational values up to $v\sin{i}=63$ $\rm km \, s^{-1}$. Since large rotational velocities broaden photospheric lines, we empirically investigate whether our measured magnetic field strengths and gravities are reliable under such high rotation conditions. To carry out this test, we obtained spectra of young sources (Class III sources in the \citealt{Lada1987} classification scheme) with nice deep lines, which rotate slowly and have a variety of B-field strengths. We selected stars with relatively low $v\sin{i}$ (so we can broaden them) and with magnetic field strengths ranging from 0.7 to 3 kG to cover the range of observed magnetic field strength in our sample of Class I sources. We obtained spectra of TWA4, TWA13S, TWA9A, and HIP16563SE. We analyzed the Class III sources in a similar fashion as the Class I sources in this study, although only $\sim50000$ models were run for each configuration (about half of the number run in the Class I sources). The measured magnetic field strength, $\log{g}$, temperature, and micro-turbulence are presented in Table \ref{table:classIII-test}.

\begin{figure*}
\gridline{\fig{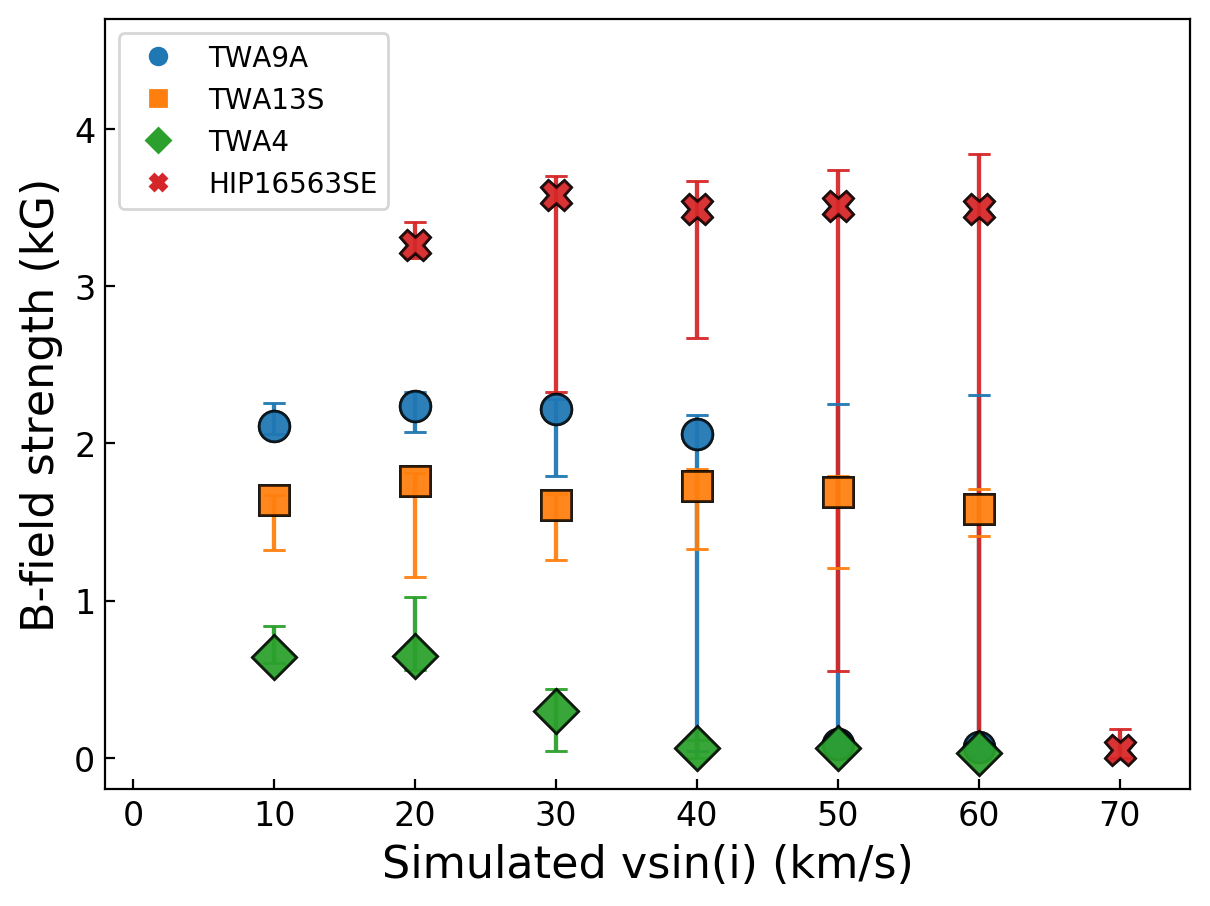}{0.45\textwidth}{(a)}
          \fig{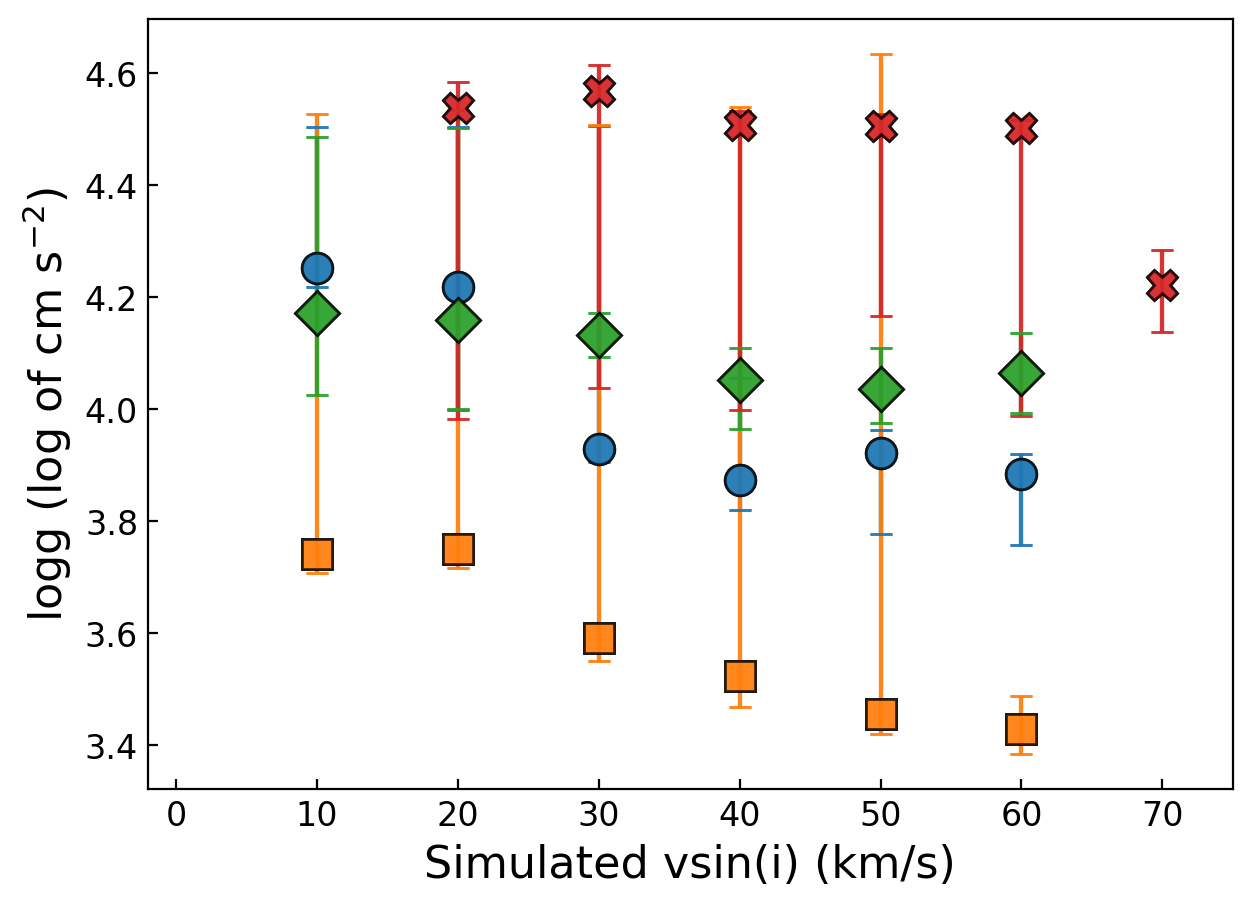}{0.45\textwidth}{(b)}
          }
\gridline{\fig{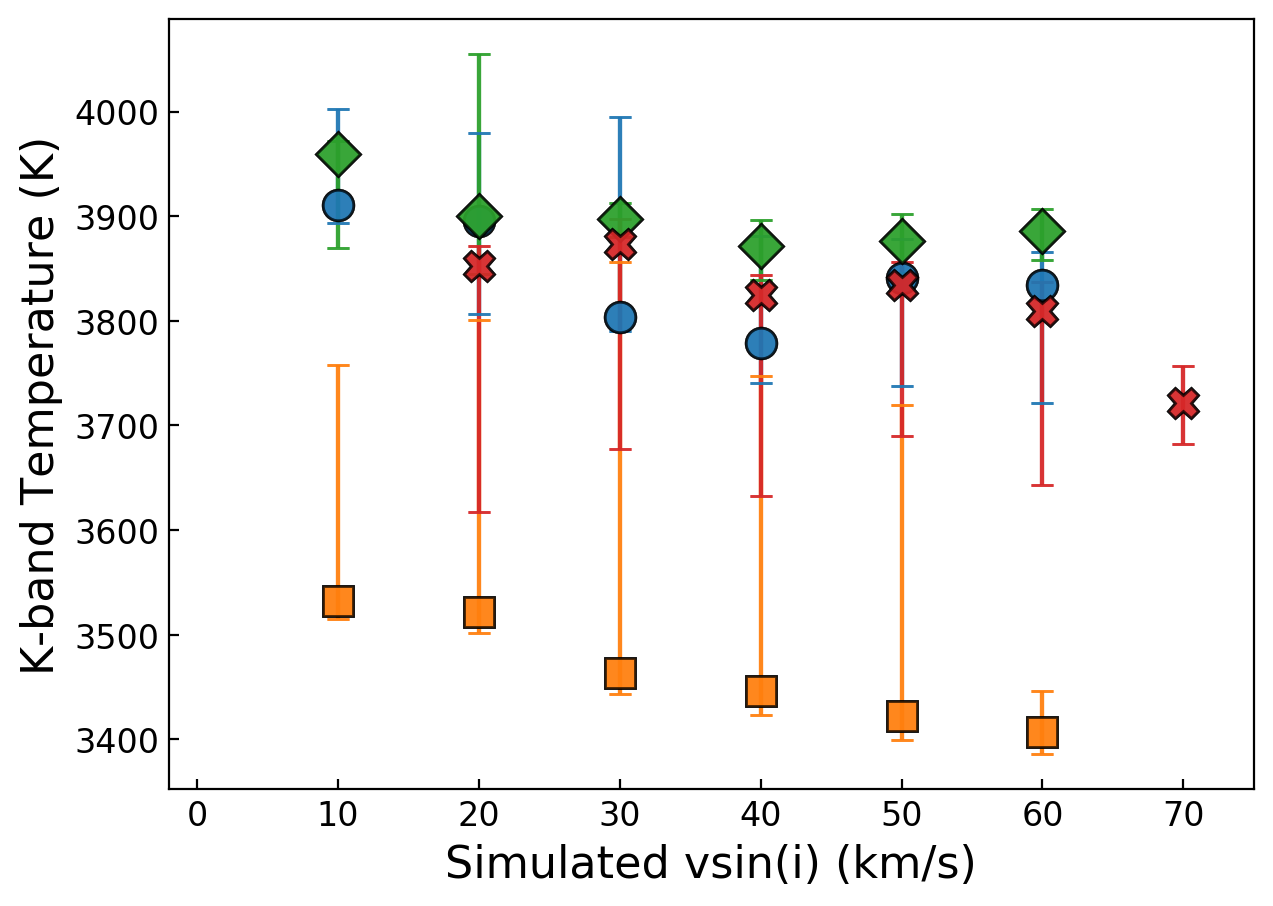}{0.45\textwidth}{(c)}
          \fig{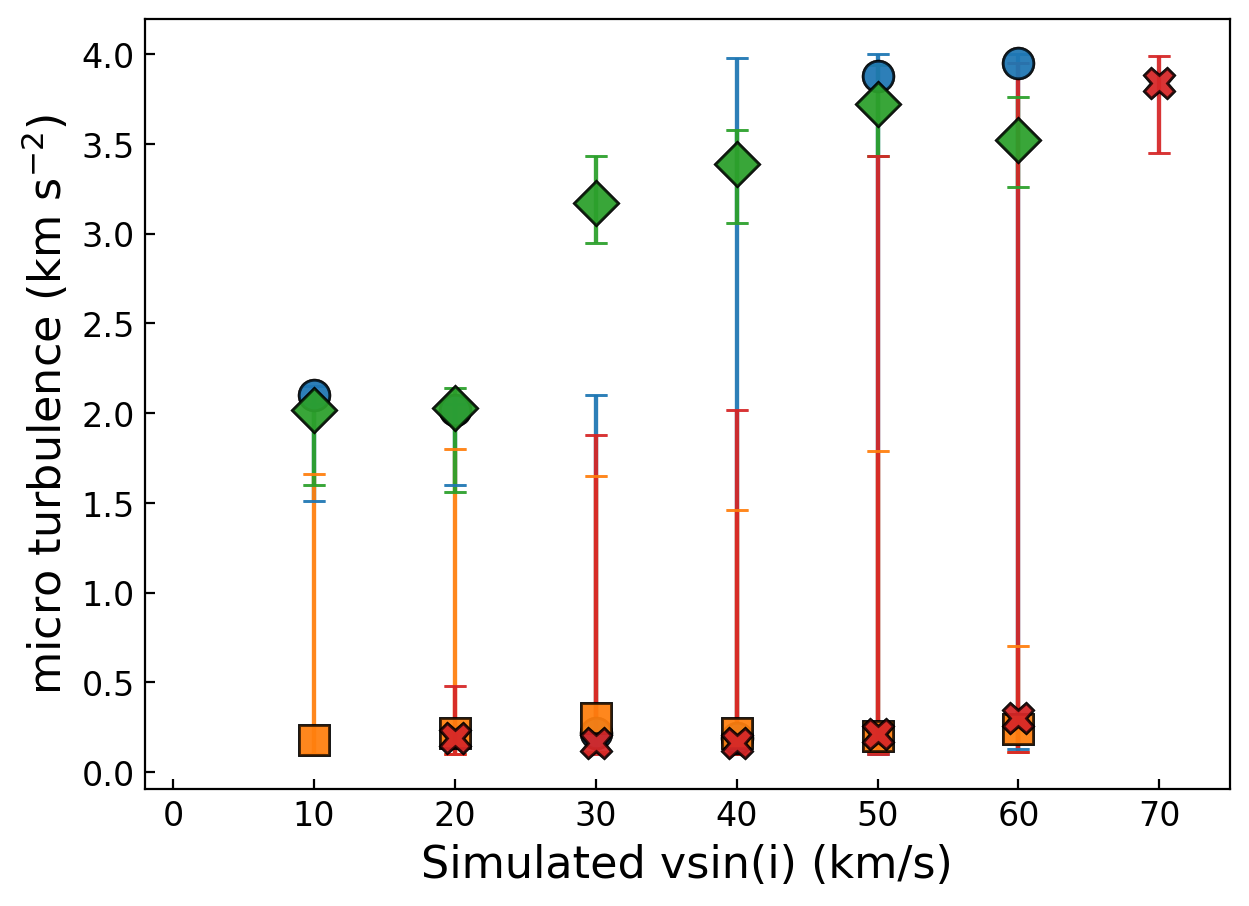}{0.45\textwidth}{(d)}
          }
\caption{Testing the accuracy of stellar parameter determination as a function of stellar rotation. We artificially broadened the spectrum of four Class III sources and tried to re-infer their stellar properties 
(a) the average magnetic field strength of the stars, (b) surface gravity in log of $\rm cm \, s^{-2}$, (c) $K$-band temperature, and (d) the micro turbulence of the stars. The intrinsic projected rotational velocity of TW9A, TWA13S and TWA4 is close to $\sim10$~$\rm km \, s^{-1}$, thus the broadening happened between 20 to 60 $\rm km \, s^{-1}$. For HIP16563SE the intrinsic projected rotational velocity is close to $\sim20$~$\rm km \, s^{-1}$, thus the broadening happened between 30 to 70 $\rm km \, s^{-1}$.
\label{fig:ClassIII-tests}}
\end{figure*}

We artificially broadened the spectra of the Class III sources with a rotation kernel following \cite{Gray1992} and tried to re-infer their magnetic field strengths, gravities, temperatures, and micro-turbulence values. In this exercise, we particularly aimed to see at what rotational velocity our code can no longer retrieve the input magnetic field strength of the star, but also to see if other stellar parameters get affected when stars rotate too fast. For sources with B-field below $\sim2$~kG, we broadened their spectrum up to 60~$\rm km \, s^{-1}$. For the source with a magnetic field strength of $\sim3$~kG, we broadened its spectrum up to 70~$\rm km \, s^{-1}$, hoping that stronger magnetic fields can be detected even at faster-rotating velocities. The results of the test are plotted in Figure \ref{fig:ClassIII-tests}.

Several conclusions can be obtained from Figure \ref{fig:ClassIII-tests}, some of these are: 1) Magnetic field strength of $\langle \rm B \rangle \lesssim 1$~kG cannot be recovered in a star that rotates faster than $\sim20$~$\rm km \, s^{-1}$, green data points in panel (a). 2) We can measure magnetic field strengths in stars rotating up to $\sim60$~$\rm km \, s^{-1}$ if their B-fields are sufficiently strong, e.g., red and also orange data points in panel (a). 3) The limited sample of stars selected shows no perfect 1:1 correlation between the maximum measurable B-field and its rotational velocity. In this test, TWA9A has a larger magnetic field strength than TWA13S. However, at 50~$\rm km \, s^{-1}$, the magnetic field of TWA9A can no longer be detected while the one for TWA13S can be measured up to $60$~$\rm km \, s^{-1}$.

As previously shown by \cite{Basri1992}, the effective Land\'e-g factor of spectral lines alone is an inadequate way to predict the magnetic sensitivity. The number of lines that a line splits into, i.e., the Zeeman structure of the line and the strength of the individual components once separated, play a large role in the shape and EW of the magnetic line profiles. Although it is difficult to quantify these effects separately, they allow magnetic field measurements even in stars with large rotational velocities. Because of this complexity in the magnetic sensitivity of spectral lines, it is possible that specific $\log{g}$ and temperature (or perhaps the precise line formation region in the stellar atmosphere) could influence the highest retrievable B-field strength for a given rotational velocity.

Furthermore, the uncertainties shown in Figure \ref{fig:ClassIII-tests} are not straightforward to understand as the solution space of the B-field becomes bimodal at velocities of 40-50~$\rm km \, s^{-1}$, which accounts for the extensive range of possible magnetic field strengths seen in panel (a). For example, some of the walkers converge to the initial (true) magnetic field value, while others go to zero (thus the significant uncertainties at higher $v\sin{i}$). Further exploration of this will be needed, but it is well beyond the scope of this paper.

As for the effect of the artificial spectral broadening on other stellar parameters, we can check panels (b), (c), and (d) in Figure \ref{fig:ClassIII-tests}. First, we see that not all the derived parameters get affected similarly when the spectra get broadened. Both gravity (b) and temperature (c) seem to be affected in similar ways. Both tend to decrease in values as the spectra are artificially broadened, which could be understood if temperature and gravity correlate for these sources, as in the case of some Class I sources too (see corner plot Figure \ref{fig:cornerplot}). The gravities are under-predicted by $\sim$0.2-0.4 dex when stars rotate between 3-8 times faster than their original value. Temperatures are under-predicted by 100~K to 200~K when the spectra are artificially broadened. The micro-turbulence value is the most affected parameter in this test and seems to be the least reliable for fast-rotating stars. In fact, by looking at panels (a) and (d), it seems like when the magnetic field strengths of the stars are no longer measurable, the value of the $v_{micro}$ reaches the peak value of $v_{micro}=4.0$~$\rm km \, s^{-1}$. This can be noticed in HIP16563SE at $v\sin{i} = 70$~$\rm km \, s^{-1}$, TWA9A at $v\sin{i} = 50$~$\rm km \, s^{-1}$, and TWA4 at $v\sin{i} = 40$~$\rm km \, s^{-1}$.

We conclude this experiment by noting that using iSHELL spectra in the K2 mode at $R\sim50,000$, along with the modeling procedure discussed in Section \ref{sec:stellar_params}, it is possible to measure magnetic field strengths in stars that rotate moderately fast. In particular, we confirm that a magnetic field strength of $\sim$3~kG can be measured in sources rotating up to $v\sin{i} = 60$~$\rm km \, s^{-1}$; strengths of $\sim$1.5-2~kG in stars rotating at least at $v\sin{i} = 40$~$\rm km \, s^{-1}$, and $\sim$1.0~kG for rotations of at least $v\sin{i} = 20$~$\rm km \, s^{-1}$. Although this is not an extensive analysis of the maximum possible measurable magnetic field strength as a function of rotational velocity, it is a guideline in our study. More importantly, only three sources in our sample break these rules: IRAS04181+2654(M), IRAS05555-1405(4), and SR24A. Thus their magnetic field strength should be taken very cautiously.


\begin{deluxetable*}{lccccccc}
\tablewidth{10pt}
\tablecolumns{8}
\tablecaption{Variation of stellar parameters as a function of artificially increased rotational velocity. \label{table:classIII-test}}
\tablehead{
\colhead{Stellar parameters} & \colhead{$\sim$10~$\rm km\, s^{-1}$} & \colhead{20~$\rm km\, s^{-1}$} & \colhead{30~$\rm km\, s^{-1}$} & \colhead{40~$\rm km\, s^{-1}$} & \colhead{50~$\rm km\, s^{-1}$} & \colhead{60~$\rm km\, s^{-1}$} & \colhead{70~$\rm km\, s^{-1}$}}
\startdata
\cutinhead{TWA4}
$\langle B \rangle$ & $0.64^{+0.20}_{-0.04}$ & $0.65^{+0.37}_{-0.09}$ & $0.30^{+0.14}_{-0.26}$ & $0.06^{+0.05}_{-0.06}$ & $0.06^{+0.06}_{-0.06}$ & $0.03^{+0.06}_{-0.03}$ & - \\
$\log{g}$ & $4.17^{+0.31}_{-0.15}$ & $4.16^{+0.34}_{-0.16}$ & $4.13^{+0.04}_{-0.04}$ & $4.05^{+0.06}_{-0.09}$ & $4.04^{+0.07}_{-0.06}$ & $4.06^{+0.07}_{-0.07}$ & -\\
$\tkband$& $3959^{+12}_{-90}$ & $3900^{+154}_{-55}$ & $3897^{+15}_{-17}$ & $3872^{+24}_{-33}$ & $3876^{+26}_{-31}$ & $3886^{+21}_{-29}$ & - \\
$v_{micro}$ & $2.02^{+0.03}_{-0.42}$ & $2.03^{+0.11}_{-0.47}$ & $3.17^{+0.26}_{-0.22}$ & $3.39^{+0.19}_{-0.33}$ &  $3.72^{+0.22}_{-0.29}$ &  $3.52^{+0.24}_{-0.26}$ & - \\
\cutinhead{TWA13S}
$\langle B \rangle$ & $1.64^{+0.03}_{-0.32}$  & $1.76^{+0.05}_{-0.61}$ & $1.61^{+0.07}_{-0.35}$ & $1.73^{+0.11}_{-0.40}$ & $1.69^{+0.10}_{-0.48}$ & $1.58^{+0.13}_{-0.17}$ & -\\
$\log{g}$ & $3.74^{+0.79}_{-0.03}$ & $3.75^{+0.78}_{-0.03}$ & $3.59^{+0.92}_{-0.04}$ & $3.52^{+1.01}_{-0.05}$ &  $3.45^{+1.18}_{-0.05}$ & $3.43^{+0.06}_{-0.04}$ & -\\
$\tkband$ & $3533^{+225}_{-18}$ & $3522^{+279}_{-20}$ & $3463^{+393}_{-20}$ & $3446^{+302}_{-22}$ & $3423^{+297}_{-23}$ & $3406^{+39}_{-21}$ & -\\
$v_{micro}$ & $0.18^{+1.48}_{-0.08}$ & $0.22^{+1.58}_{-0.12}$ & $0.30^{+1.35}_{-0.19}$ & $0.22^{+1.24}_{-0.12}$ & $0.20^{+1.59}_{-0.09}$ & $0.24^{+0.46}_{-0.13}$ & -\\
\cutinhead{TWA9A}
$\langle B \rangle$ & $2.11^{+0.15}_{-0.05}$ & $2.24^{+0.09}_{-0.17}$ & $2.22^{+0.06}_{-0.43}$ & $2.06^{+0.12}_{-2.02}$ & $0.09^{+2.16}_{-0.09}$ & $0.07^{+2.24}_{-0.07}$ & - \\
$\log{g}$ & $4.25^{+0.25}_{-0.04}$ & $4.22^{+0.29}_{-0.22}$ & $3.93^{+0.58}_{-0.02}$ & $3.87^{+0.18}_{-0.05}$ & $3.92^{+0.04}_{-0.14}$ & $3.88^{+0.07}_{-0.13}$ & - \\
$\tkband$ & $3910^{+91}_{-18}$ & $3895^{+84}_{-88}$ & $3804^{+191}_{-13}$ & $3779^{+102}_{-38}$ & $3841^{+37}_{-104}$ & $3834^{+31}_{-113}$ & -\\
$v_{micro}$ & $2.10^{+0.05}_{-0.59}$ & $2.02^{+0.03}_{-0.42}$ & $0.22^{+1.88}_{-0.11}$ & $0.19^{+3.79}_{-0.08}$ & $3.88^{+0.12}_{-3.78}$ & $3.95^{+0.05}_{-3.82}$ & -\\
\cutinhead{HIP16563SE}
$\langle B \rangle$ & - & $3.26^{+0.15}_{-0.08}$ & $3.58^{+0.12}_{-1.25}$ & $3.49^{+0.18}_{-0.82}$ & $3.51^{+0.23}_{-2.96}$ & $3.49^{+0.35}_{-3.48}$ & $0.05^{+0.13}_{-0.05}$ \\
$\log{g}$ & - & $4.54^{+0.05}_{-0.56}$ & $4.57^{+0.05}_{-0.53}$ & $4.51^{+0.03}_{-0.51}$ & $4.50^{+0.02}_{-0.34}$ & $4.50^{+0.02}_{-0.51}$ & $4.22^{+0.06}_{-0.08}$ \\
$\tkband$ & - & $3852^{+19}_{-235}$ & $3873^{+24}_{-196}$ & $3824^{+19}_{-191}$ & $3834^{+22}_{-145}$ & $3810^{+84}_{-167}$ & $3721^{+35}_{-40}$\\
$v_{micro}$ & - & $0.19^{+0.29}_{-0.09}$ & $0.16^{+1.72}_{-0.06}$ & $0.16^{+1.86}_{-0.06}$ & $0.21^{+3.22}_{-0.11}$ & $0.30^{+3.65}_{-0.19}$ & $3.84^{+0.15}_{-0.39}$\\
\enddata
\tablecomments{The first column marks the natural velocity of the stars. The measured value of the $v\sin{i}$ for the TWA4, TWA13S, TWA9A, and HIP16563SE are: $9.3^{+0.70}_{-0.58}$, $13.12^{+1.35}_{-0.26}$, $11.53^{+1.50}_{-1.38}$, and $22.47^{+0.85}_{-3.24}$~$\rm km \, s^{-1}$, respectively}.\\
\end{deluxetable*}

\section{corner plots}
\label{appendix:cornerplot}
\begin{figure*}[ht!]
\epsscale{1.}
\plotone{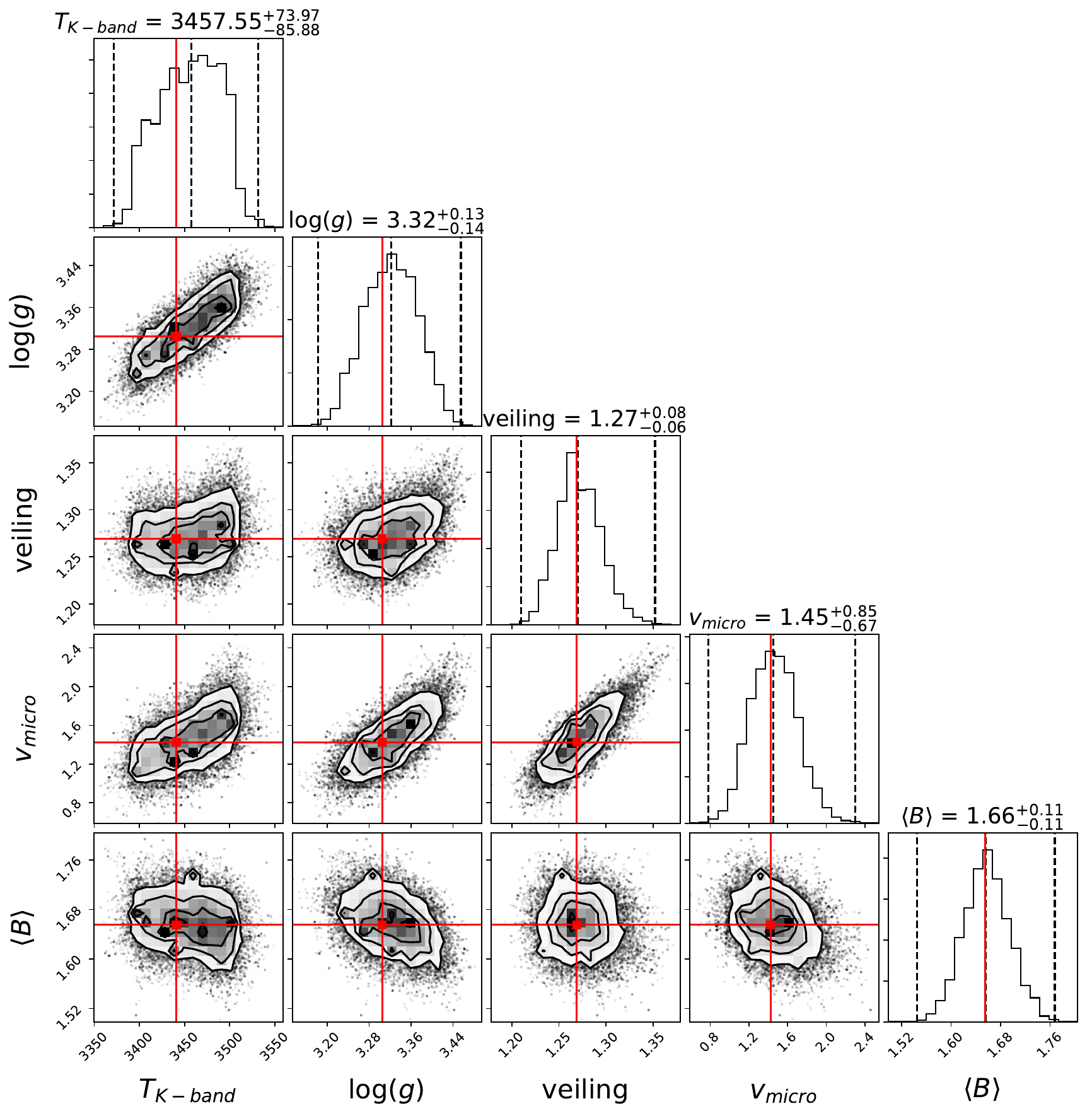}
\caption{We present the results for the second MCMC run for IRAS03301+3111 (rotational velocity is not included). The corner plot shows only 50\% of the models, i.e., about 50000 models. The central black lines correspond to the median of the distributions, while the black lines on the side are the 3$\sigma$ uncertainties. The specific best fit model is traced by a red line. This corner plot also corresponds to the solution used to compute Figure \ref{fig:spectrum-example}.}  \label{fig:cornerplot}
\end{figure*}

In Figure \ref{fig:cornerplot}, we present the corner plot for IRAS03301+3111 made from its second MCMC run. The corner plot for the rest of the sources can be accessed in the online journal as figure sets. As explained in Section \ref{sec:stellar_params}, the second MCMC step does no include the CO region (starting at $\sim2.293\, \micron$) neither it fits the $v\sin{i}$ parameter. For every source, the corner plots only include the last 50\% of the $\sim$100,000 models. We visually assessed the convergence of the MCMC runs, which typically happens within the first 10$\%$ of each run. There are small correlation between some stellar parameters, for example, for this source the $v_{micro}$ weakly correlates with $T_{K-band}$, $\log{g}$, and veiling. The largest degeneracy for this source is between the $T_{K-band}$ and $\log{g}$. Interestingly, no much correlation can be seen between the $\log{g}$ and the magnetic field strength.

For most of the stars some correlations are present and the reader can see each particular case in the Figure set available online. In a few cases, the corner plots are not single-peaked and display more complex structures. For a limited number of Class I and Flat Spectrum sources the models reached a lower limit or upper limit value for the $v_{micro}$ (see also Table \ref{table:Stellar_params}). The values quoted on top of the 1-D histograms correspond to the median of the distribution (central black dashed line), the side black dashed lines represent the 3$\sigma$ uncertainties. Red lines point to the specific best fit model of the MCMC chain. This best-fit model is well within the 3$\sigma$ uncertainty for each stellar parameter in every case.

\figsetstart
\figsetnum{4}
\figsettitle{Cornerplots of MCMC runs}

\figsetend

\bibliography{sample631}{}

\begin{thebibliography}{}
\expandafter\ifx\csname natexlab\endcsname\relax\def\natexlab#1{#1}\fi
\providecommand{\url}[1]{\href{#1}{#1}}
\providecommand{\dodoi}[1]{doi:~\href{http://doi.org/#1}{\nolinkurl{#1}}}
\providecommand{\doeprint}[1]{\href{http://ascl.net/#1}{\nolinkurl{http://ascl.net/#1}}}
\providecommand{\doarXiv}[1]{\href{https://arxiv.org/abs/#1}{\nolinkurl{https://arxiv.org/abs/#1}}}

\bibitem[{{Andr\'e} {et~al.}(1993){Andr\'e}, {Ward-Thompson}, \&
  {Barsony}}]{Andre1993}
{Andr\'e}, P., {Ward-Thompson}, D., \& {Barsony}, M. 1993, \apj, 406, 122,
  \dodoi{10.1086/172425}

\bibitem[{{Aspin} {et~al.}(2010){Aspin}, {Reipurth}, {Herczeg}, \&
  {Capak}}]{Aspin2010}
{Aspin}, C., {Reipurth}, B., {Herczeg}, G.~J., \& {Capak}, P. 2010, \apjl, 719,
  L50, \dodoi{10.1088/2041-8205/719/1/L50}

\bibitem[{{Astropy Collaboration} {et~al.}(2013){Astropy Collaboration},
  {Robitaille}, {Tollerud}, {Greenfield}, {Droettboom}, {Bray}, {Aldcroft},
  {Davis}, {Ginsburg}, {Price-Whelan}, {Kerzendorf}, {Conley}, {Crighton},
  {Barbary}, {Muna}, {Ferguson}, {Grollier}, {Parikh}, {Nair}, {Unther},
  {Deil}, {Woillez}, {Conseil}, {Kramer}, {Turner}, {Singer}, {Fox}, {Weaver},
  {Zabalza}, {Edwards}, {Azalee Bostroem}, {Burke}, {Casey}, {Crawford},
  {Dencheva}, {Ely}, {Jenness}, {Labrie}, {Lim}, {Pierfederici}, {Pontzen},
  {Ptak}, {Refsdal}, {Servillat}, \& {Streicher}}]{2013A&A...558A..33A}
{Astropy Collaboration}, {Robitaille}, T.~P., {Tollerud}, E.~J., {et~al.} 2013,
  \aap, 558, A33, \dodoi{10.1051/0004-6361/201322068}

\bibitem[{{Baraffe} {et~al.}(2015){Baraffe}, {Homeier}, {Allard}, \&
  {Chabrier}}]{Baraffe2015}
{Baraffe}, I., {Homeier}, D., {Allard}, F., \& {Chabrier}, G. 2015, \aap, 577,
  A42, \dodoi{10.1051/0004-6361/201425481}

\bibitem[{{Barr} {et~al.}(2020){Barr}, {Boogert}, {DeWitt}, {Montiel},
  {Richter}, {Lacy}, {Neufeld}, {Indriolo}, {Pendleton}, {Chiar}, \&
  {Tielens}}]{Barr2020}
{Barr}, A.~G., {Boogert}, A., {DeWitt}, C.~N., {et~al.} 2020, \apj, 900, 104,
  \dodoi{10.3847/1538-4357/abab05}

\bibitem[{{Basri} {et~al.}(1992){Basri}, {Marcy}, \& {Valenti}}]{Basri1992}
{Basri}, G., {Marcy}, G.~W., \& {Valenti}, J.~A. 1992, \apj, 390, 622,
  \dodoi{10.1086/171312}

\bibitem[{{Beck} {et~al.}(2008){Beck}, {McGregor}, {Takami}, \&
  {Pyo}}]{Beck2008}
{Beck}, T.~L., {McGregor}, P.~J., {Takami}, M., \& {Pyo}, T.-S. 2008, \apj,
  676, 472, \dodoi{10.1086/527528}

\bibitem[{{Boyajian} {et~al.}(2012){Boyajian}, {von Braun}, {van Belle},
  {McAlister}, {ten Brummelaar}, {Kane}, {Muirhead}, {Jones}, {White},
  {Schaefer}, {Ciardi}, {Henry}, {L{\'o}pez-Morales}, {Ridgway}, {Gies}, {Jao},
  {Rojas-Ayala}, {Parks}, {Sturmann}, {Sturmann}, {Turner}, {Farrington},
  {Goldfinger}, \& {Berger}}]{Boyajian2012}
{Boyajian}, T.~S., {von Braun}, K., {van Belle}, G., {et~al.} 2012, \apj, 757,
  112, \dodoi{10.1088/0004-637X/757/2/112}

\bibitem[{{Brittain} {et~al.}(2003){Brittain}, {Rettig}, {Simon}, {Kulesa},
  {DiSanti}, \& {Dello Russo}}]{Brittain2003}
{Brittain}, S.~D., {Rettig}, T.~W., {Simon}, T., {et~al.} 2003, \apj, 588, 535,
  \dodoi{10.1086/373987}

\bibitem[{{Calvet} {et~al.}(1991){Calvet}, {Patino}, {Magris}, \&
  {D'Alessio}}]{Calvet1991}
{Calvet}, N., {Patino}, A., {Magris}, G.~C., \& {D'Alessio}, P. 1991, \apj,
  380, 617, \dodoi{10.1086/170618}

\bibitem[{{Carney} {et~al.}(2016){Carney}, {Y{\i}ld{\i}z}, {Mottram}, {van
  Dishoeck}, {Ramchandani}, \& {J{\o}rgensen}}]{Carney2016}
{Carney}, M.~T., {Y{\i}ld{\i}z}, U.~A., {Mottram}, J.~C., {et~al.} 2016, \aap,
  586, A44, \dodoi{10.1051/0004-6361/201526308}

\bibitem[{{Cieza} {et~al.}(2019){Cieza}, {Ru{\'\i}z-Rodr{\'\i}guez}, {Hales},
  {Casassus}, {P{\'e}rez}, {Gonzalez-Ruilova}, {C{\'a}novas}, {Williams},
  {Zurlo}, {Ansdell}, {Avenhaus}, {Bayo}, {Bertrang}, {Christiaens}, {Dent},
  {Ferrero}, {Gamen}, {Olofsson}, {Orcajo}, {Pe{\~n}a Ram{\'\i}rez},
  {Principe}, {Schreiber}, \& {van der Plas}}]{Cieza2019}
{Cieza}, L.~A., {Ru{\'\i}z-Rodr{\'\i}guez}, D., {Hales}, A., {et~al.} 2019,
  \mnras, 482, 698, \dodoi{10.1093/mnras/sty2653}

\bibitem[{{Connelley} \& {Greene}(2010)}]{Connelley2010}
{Connelley}, M.~S., \& {Greene}, T.~P. 2010, \aj, 140, 1214,
  \dodoi{10.1088/0004-6256/140/5/1214}

\bibitem[{{Connelley} \& {Greene}(2014)}]{Connelley2014}
---. 2014, \aj, 147, 125, \dodoi{10.1088/0004-6256/147/6/125}

\bibitem[{{Covey} {et~al.}(2005){Covey}, {Greene}, {Doppmann}, \&
  {Lada}}]{Covey2005}
{Covey}, K.~R., {Greene}, T.~P., {Doppmann}, G.~W., \& {Lada}, C.~J. 2005, \aj,
  129, 2765, \dodoi{10.1086/429736}

\bibitem[{{Cushing} {et~al.}(2004){Cushing}, {Vacca}, \&
  {Rayner}}]{Cushing2004}
{Cushing}, M.~C., {Vacca}, W.~D., \& {Rayner}, J.~T. 2004, \pasp, 116, 362,
  \dodoi{10.1086/382907}

\bibitem[{{Deen}(2013)}]{Deen2013}
{Deen}, C.~P. 2013, \aj, 146, 51, \dodoi{10.1088/0004-6256/146/3/51}

\bibitem[{{Donati} {et~al.}(2019){Donati}, {Bouvier}, {Alencar}, {Hill},
  {Carmona}, {Folsom}, {M{\'e}nard}, {Gregory}, {Hussain}, {Grankin}, {Moutou},
  {Malo}, {Takami}, {Herczeg}, \& {MaTYSSE Collaboration}}]{Donati2019}
{Donati}, J.~F., {Bouvier}, J., {Alencar}, S.~H., {et~al.} 2019, \mnras, 483,
  L1, \dodoi{10.1093/mnrasl/sly207}

\bibitem[{{Doppmann} {et~al.}(2005){Doppmann}, {Greene}, {Covey}, \&
  {Lada}}]{Doppmann2005}
{Doppmann}, G.~W., {Greene}, T.~P., {Covey}, K.~R., \& {Lada}, C.~J. 2005, \aj,
  130, 1145, \dodoi{10.1086/431954}

\bibitem[{{Doppmann} {et~al.}(2008){Doppmann}, {Najita}, \&
  {Carr}}]{Doppmann2008}
{Doppmann}, G.~W., {Najita}, J.~R., \& {Carr}, J.~S. 2008, \apj, 685, 298,
  \dodoi{10.1086/590328}

\bibitem[{{Evans} {et~al.}(2009){Evans}, {Dunham}, {J{\o}rgensen}, {Enoch},
  {Mer{\'\i}n}, {van Dishoeck}, {Alcal{\'a}}, {Myers}, {Stapelfeldt}, {Huard},
  {Allen}, {Harvey}, {van Kempen}, {Blake}, {Koerner}, {Mundy}, {Padgett}, \&
  {Sargent}}]{Evans2009}
{Evans}, Neal~J., I., {Dunham}, M.~M., {J{\o}rgensen}, J.~K., {et~al.} 2009,
  \apjs, 181, 321, \dodoi{10.1088/0067-0049/181/2/321}

\bibitem[{{Feiden}(2016)}]{Feiden2016}
{Feiden}, G.~A. 2016, \aap, 593, A99, \dodoi{10.1051/0004-6361/201527613}

\bibitem[{{Fiorellino} {et~al.}(2021){Fiorellino}, {Manara}, {Nisini},
  {Ramsay}, {Antoniucci}, {Giannini}, {Biazzo}, {Alcal{\`a}}, \&
  {Fedele}}]{Fiorellino2021}
{Fiorellino}, E., {Manara}, C.~F., {Nisini}, B., {et~al.} 2021, \aap, 650, A43,
  \dodoi{10.1051/0004-6361/202039264}

\bibitem[{{Flores} {et~al.}(2019){Flores}, {Connelley}, {Reipurth}, \&
  {Boogert}}]{Flores2019}
{Flores}, C., {Connelley}, M.~S., {Reipurth}, B., \& {Boogert}, A. 2019, \apj,
  882, A75, \dodoi{10.3847/1538-4357/ab35d4}

\bibitem[{{Flores} {et~al.}(2022){Flores}, {Connelley}, {Reipurth}, \&
  {Duch{\^e}ne}}]{Flores2022}
{Flores}, C., {Connelley}, M.~S., {Reipurth}, B., \& {Duch{\^e}ne}, G. 2022,
  \apj, 925, 21, \dodoi{10.3847/1538-4357/ac37bd}

\bibitem[{{Flores} {et~al.}(2020){Flores}, {Reipurth}, \&
  {Connelley}}]{Flores2020}
{Flores}, C., {Reipurth}, B., \& {Connelley}, M.~S. 2020, \apj, 898, 109,
  \dodoi{10.3847/1538-4357/ab9e67}

\bibitem[{{Foreman-Mackey} {et~al.}(2013){Foreman-Mackey}, {Hogg}, {Lang}, \&
  {Goodman}}]{Foreman-Mackey2013}
{Foreman-Mackey}, D., {Hogg}, D.~W., {Lang}, D., \& {Goodman}, J. 2013, \pasp,
  125, 306, \dodoi{10.1086/670067}

\bibitem[{{Giannakopoulou} {et~al.}(1997){Giannakopoulou}, {Mitchell},
  {Hasegawa}, {Matthews}, \& {Maillard}}]{Giannakopoulou1997}
{Giannakopoulou}, J., {Mitchell}, G.~F., {Hasegawa}, T.~I., {Matthews}, H.~E.,
  \& {Maillard}, J.-P. 1997, \apj, 487, 346, \dodoi{10.1086/304574}

\bibitem[{{Gonzalez} {et~al.}(1999){Gonzalez}, {Wallerstein}, \&
  {Saar}}]{Gonzalez1999}
{Gonzalez}, G., {Wallerstein}, G., \& {Saar}, S.~H. 1999, \apjl, 511, L111,
  \dodoi{10.1086/311847}

\bibitem[{Gordon {et~al.}(2022)Gordon, Rothman, Hargreaves, Hashemi, Karlovets,
  Skinner, Conway, Hill, Kochanov, Tan, Wcisło, Finenko, Nelson, Bernath,
  Birk, Boudon, Campargue, Chance, Coustenis, Drouin, Flaud, Gamache, Hodges,
  Jacquemart, Mlawer, Nikitin, Perevalov, Rotger, Tennyson, Toon, Tran,
  Tyuterev, Adkins, Baker, Barbe, Canè, Császár, Dudaryonok, Egorov,
  Fleisher, Fleurbaey, Foltynowicz, Furtenbacher, Harrison, Hartmann, Horneman,
  Huang, Karman, Karns, Kassi, Kleiner, Kofman, Kwabia–Tchana, Lavrentieva,
  Lee, Long, Lukashevskaya, Lyulin, Makhnev, Matt, Massie, Melosso,
  Mikhailenko, Mondelain, Müller, Naumenko, Perrin, Polyansky, Raddaoui,
  Raston, Reed, Rey, Richard, Tóbiás, Sadiek, Schwenke, Starikova, Sung,
  Tamassia, Tashkun, {Vander Auwera}, Vasilenko, Vigasin, Villanueva, Vispoel,
  Wagner, Yachmenev, \& Yurchenko}]{Gordon2021}
Gordon, I., Rothman, L., Hargreaves, R., {et~al.} 2022, Journal of Quantitative
  Spectroscopy and Radiative Transfer, 277, 107949,
  \dodoi{https://doi.org/10.1016/j.jqsrt.2021.107949}

\bibitem[{{Goto} {et~al.}(2003){Goto}, {Usuda}, {Takato}, {Hayashi},
  {Sakamoto}, {Gaessler}, {Hayano}, {Iye}, {Kamata}, {Kanzawa}, {Kobayashi},
  {Minowa}, {Nedachi}, {Oya}, {Pyo}, {Saint-Jacques}, {Suto}, {Takami},
  {Terada}, \& {Mitchell}}]{Goto2003}
{Goto}, M., {Usuda}, T., {Takato}, N., {et~al.} 2003, \apj, 598, 1038,
  \dodoi{10.1086/378978}

\bibitem[{{Gray}(1992)}]{Gray1992}
{Gray}, D.~F. 1992, {The observation and analysis of stellar photospheres.},
  Vol.~20

\bibitem[{{Greene} {et~al.}(2018){Greene}, {Gully-Santiago}, \&
  {Barsony}}]{Greene2018}
{Greene}, T.~P., {Gully-Santiago}, M.~A., \& {Barsony}, M. 2018, \apj, 862, 85,
  \dodoi{10.3847/1538-4357/aacc6c}

\bibitem[{{Greene} {et~al.}(1994){Greene}, {Wilking}, {Andre}, {Young}, \&
  {Lada}}]{Greene1994}
{Greene}, T.~P., {Wilking}, B.~A., {Andre}, P., {Young}, E.~T., \& {Lada},
  C.~J. 1994, \apj, 434, 614, \dodoi{10.1086/174763}

\bibitem[{{G{\"u}del} {et~al.}(2018){G{\"u}del}, {Eibensteiner}, {Dionatos},
  {Audard}, {Forbrich}, {Kraus}, {Rab}, {Schneider}, {Skinner}, \&
  {Vorobyov}}]{Gudel2018}
{G{\"u}del}, M., {Eibensteiner}, C., {Dionatos}, O., {et~al.} 2018, \aap, 620,
  L1, \dodoi{10.1051/0004-6361/201834271}

\bibitem[{{Gustafsson} {et~al.}(2008){Gustafsson}, {Edvardsson}, {Eriksson},
  {J{\o}rgensen}, {Nordlund}, \& {Plez}}]{Gustafsson2008}
{Gustafsson}, B., {Edvardsson}, B., {Eriksson}, K., {et~al.} 2008, \aap, 486,
  951, \dodoi{10.1051/0004-6361:200809724}

\bibitem[{{Hartmann} {et~al.}(2016){Hartmann}, {Herczeg}, \&
  {Calvet}}]{Hartmann2016}
{Hartmann}, L., {Herczeg}, G., \& {Calvet}, N. 2016, \araa, 54, 135,
  \dodoi{10.1146/annurev-astro-081915-023347}

\bibitem[{{Heiderman} \& {Evans}(2015)}]{Heiderman2015}
{Heiderman}, A., \& {Evans}, Neal~J., I. 2015, \apj, 806, 231,
  \dodoi{10.1088/0004-637X/806/2/231}

\bibitem[{{Huber} {et~al.}(2016){Huber}, {Bryson}, {Haas}, {Barclay},
  {Barentsen}, {Howell}, {Sharma}, {Stello}, \& {Thompson}}]{Huber2016}
{Huber}, D., {Bryson}, S.~T., {Haas}, M.~R., {et~al.} 2016, \apjs, 224, 2,
  \dodoi{10.3847/0067-0049/224/1/2}

\bibitem[{{Johns-Krull}(2007)}]{Johns-Krull2007}
{Johns-Krull}, C.~M. 2007, \apj, 664, 975, \dodoi{10.1086/519017}

\bibitem[{{Johns-Krull} {et~al.}(2009){Johns-Krull}, {Greene}, {Doppmann}, \&
  {Covey}}]{Johns-Krull2009}
{Johns-Krull}, C.~M., {Greene}, T.~P., {Doppmann}, G.~W., \& {Covey}, K.~R.
  2009, \apj, 700, 1440, \dodoi{10.1088/0004-637X/700/2/1440}

\bibitem[{{Kenyon} {et~al.}(1990){Kenyon}, {Hartmann}, {Strom}, \&
  {Strom}}]{Kenyon1990}
{Kenyon}, S.~J., {Hartmann}, L.~W., {Strom}, K.~M., \& {Strom}, S.~E. 1990,
  \aj, 99, 869, \dodoi{10.1086/115380}

\bibitem[{{Kochukhov}(2021)}]{Kochukhov2021}
{Kochukhov}, O. 2021, \aapr, 29, 1, \dodoi{10.1007/s00159-020-00130-3}

\bibitem[{{Lada}(1987)}]{Lada1987}
{Lada}, C.~J. 1987, in Star Forming Regions, ed. M.~{Peimbert} \& J.~{Jugaku},
  Vol. 115, 1

\bibitem[{{Lada} \& {Wilking}(1984)}]{Lada1984}
{Lada}, C.~J., \& {Wilking}, B.~A. 1984, \apj, 287, 610, \dodoi{10.1086/162719}

\bibitem[{{Laos} {et~al.}(2021){Laos}, {Greene}, {Najita}, \&
  {Stassun}}]{Laos2021}
{Laos}, S., {Greene}, T.~P., {Najita}, J.~R., \& {Stassun}, K.~G. 2021, \apj,
  921, 110, \dodoi{10.3847/1538-4357/ac1f1b}

\bibitem[{{Lavail} {et~al.}(2017){Lavail}, {Kochukhov}, {Hussain}, {Alecian},
  {Herczeg}, \& {Johns-Krull}}]{Lavail2017}
{Lavail}, A., {Kochukhov}, O., {Hussain}, G.~A.~J., {et~al.} 2017, \aap, 608,
  A77, \dodoi{10.1051/0004-6361/201731889}

\bibitem[{{Le Gouellec} {et~al.}(2024){Le Gouellec}, {Greene}, {Hillenbrand},
  \& {Yates}}]{LeGouellec2024}
{Le Gouellec}, V. J.~M., {Greene}, T.~P., {Hillenbrand}, L.~A., \& {Yates}, Z.
  2024, arXiv e-prints, arXiv:2401.16532, \dodoi{10.48550/arXiv.2401.16532}

\bibitem[{{Leinert} {et~al.}(1993){Leinert}, {Zinnecker}, {Weitzel},
  {Christou}, {Ridgway}, {Jameson}, {Haas}, \& {Lenzen}}]{Leinert1993}
{Leinert}, C., {Zinnecker}, H., {Weitzel}, N., {et~al.} 1993, \aap, 278, 129

\bibitem[{{L{\'o}pez-Valdivia} {et~al.}(2021){L{\'o}pez-Valdivia}, {Sokal},
  {Mace}, {Kidder}, {Hussaini}, {Nofi}, {Prato}, {Johns-Krull}, {Oh}, {Lee},
  {Park}, {Oh}, {Kraus}, {Kaplan}, {Llama}, {Mann}, {Kim}, {Gully-Santiago},
  {Lee}, {Pak}, {Hwang}, \& {Jaffe}}]{Lopez-Valdivia2021}
{L{\'o}pez-Valdivia}, R., {Sokal}, K.~R., {Mace}, G.~N., {et~al.} 2021, \apj,
  921, 53, \dodoi{10.3847/1538-4357/ac1a7b}

\bibitem[{{McLean} {et~al.}(1998){McLean}, {Becklin}, {Bendiksen}, {Brims},
  {Canfield}, {Figer}, {Graham}, {Hare}, {Lacayanga}, {Larkin}, {Larson},
  {Levenson}, {Magnone}, {Teplitz}, \& {Wong}}]{McLean1998}
{McLean}, I.~S., {Becklin}, E.~E., {Bendiksen}, O., {et~al.} 1998, in Society
  of Photo-Optical Instrumentation Engineers (SPIE) Conference Series, Vol.
  3354, Infrared Astronomical Instrumentation, ed. A.~M. {Fowler}, 566--578,
  \dodoi{10.1117/12.317283}

\bibitem[{{Mitchell} {et~al.}(1990){Mitchell}, {Maillard}, {Allen}, {Beer}, \&
  {Belcourt}}]{Mitchell1990}
{Mitchell}, G.~F., {Maillard}, J.-P., {Allen}, M., {Beer}, R., \& {Belcourt},
  K. 1990, \apj, 363, 554, \dodoi{10.1086/169365}

\bibitem[{{Mitchell} {et~al.}(1997){Mitchell}, {Sargent}, \&
  {Mannings}}]{Mitchell1997}
{Mitchell}, G.~F., {Sargent}, A.~I., \& {Mannings}, V. 1997, \apjl, 483, L127,
  \dodoi{10.1086/310750}

\bibitem[{{Muzerolle} {et~al.}(1998){Muzerolle}, {Hartmann}, \&
  {Calvet}}]{Muzerolle1998}
{Muzerolle}, J., {Hartmann}, L., \& {Calvet}, N. 1998, \aj, 116, 2965,
  \dodoi{10.1086/300636}

\bibitem[{{Pakhomov} {et~al.}(2019){Pakhomov}, {Ryabchikova}, \&
  {Piskunov}}]{Pakhomov2019}
{Pakhomov}, Y.~V., {Ryabchikova}, T.~A., \& {Piskunov}, N.~E. 2019, Astronomy
  Reports, 63, 1010, \dodoi{10.1134/S1063772919120047}

\bibitem[{{Rayner} {et~al.}(2016){Rayner}, {Tokunaga}, {Jaffe}, {Bonnet},
  {Ching}, {Connelley}, {Kokubun}, {Lockhart}, \& {Warmbier}}]{Rayner2016}
{Rayner}, J., {Tokunaga}, A., {Jaffe}, D., {et~al.} 2016, Society of
  Photo-Optical Instrumentation Engineers (SPIE) Conference Series, Vol. 9908,
  {iSHELL: a construction, assembly and testing}, 990884,
  \dodoi{10.1117/12.2232064}

\bibitem[{{Rayner} {et~al.}(2022){Rayner}, {Tokunaga}, {Jaffe}, {Bond},
  {Bonnet}, {Ching}, {Connelley}, {Cushing}, {Kokubun}, {Lockhart}, {Vacca}, \&
  {Warmbier}}]{Rayner2022}
---. 2022, \pasp, 134, 015002, \dodoi{10.1088/1538-3873/ac3cb4}

\bibitem[{{Reiners} {et~al.}(2009){Reiners}, {Basri}, \&
  {Browning}}]{Reiners2009}
{Reiners}, A., {Basri}, G., \& {Browning}, M. 2009, \apj, 692, 538,
  \dodoi{10.1088/0004-637X/692/1/538}

\bibitem[{{Rieke} {et~al.}(2004){Rieke}, {Young}, {Engelbracht}, {Kelly},
  {Low}, {Haller}, {Beeman}, {Gordon}, {Stansberry}, {Misselt}, {Cadien},
  {Morrison}, {Rivlis}, {Latter}, {Noriega-Crespo}, {Padgett}, {Stapelfeldt},
  {Hines}, {Egami}, {Muzerolle}, {Alonso-Herrero}, {Blaylock}, {Dole}, {Hinz},
  {Le Floc'h}, {Papovich}, {P{\'e}rez-Gonz{\'a}lez}, {Smith}, {Su}, {Bennett},
  {Frayer}, {Henderson}, {Lu}, {Masci}, {Pesenson}, {Rebull}, {Rho}, {Keene},
  {Stolovy}, {Wachter}, {Wheaton}, {Werner}, \& {Richards}}]{Rieke2004}
{Rieke}, G.~H., {Young}, E.~T., {Engelbracht}, C.~W., {et~al.} 2004, \apjs,
  154, 25, \dodoi{10.1086/422717}

\bibitem[{{Rothman} {et~al.}(2010){Rothman}, {Gordon}, {Barber}, {Dothe},
  {Gamache}, {Goldman}, {Perevalov}, {Tashkun}, \& {Tennyson}}]{Rothman2010}
{Rothman}, L.~S., {Gordon}, I.~E., {Barber}, R.~J., {et~al.} 2010, \jqsrt, 111,
  2139, \dodoi{10.1016/j.jqsrt.2010.05.001}

\bibitem[{{Ryabchikova} {et~al.}(2015){Ryabchikova}, {Piskunov}, {Kurucz},
  {Stempels}, {Heiter}, {Pakhomov}, \& {Barklem}}]{Ryabchikova2015}
{Ryabchikova}, T., {Piskunov}, N., {Kurucz}, R.~L., {et~al.} 2015, \physscr,
  90, 054005, \dodoi{10.1088/0031-8949/90/5/054005}

\bibitem[{{Serenelli} {et~al.}(2017){Serenelli}, {Johnson}, {Huber},
  {Pinsonneault}, {Ball}, {Tayar}, {Silva Aguirre}, {Basu}, {Troup}, {Hekker},
  {Kallinger}, {Stello}, {Davies}, {Lund}, {Mathur}, {Mosser}, {Stassun},
  {Chaplin}, {Elsworth}, {Garc{\'\i}a}, {Handberg}, {Holtzman}, {Hearty},
  {Garc{\'\i}a-Hern{\'a}ndez}, {Gaulme}, \& {Zamora}}]{Serenelli2017}
{Serenelli}, A., {Johnson}, J., {Huber}, D., {et~al.} 2017, \apjs, 233, 23,
  \dodoi{10.3847/1538-4365/aa97df}

\bibitem[{{Skrutskie} {et~al.}(2006){Skrutskie}, {Cutri}, {Stiening},
  {Weinberg}, {Schneider}, {Carpenter}, {Beichman}, {Capps}, {Chester},
  {Elias}, {Huchra}, {Liebert}, {Lonsdale}, {Monet}, {Price}, {Seitzer},
  {Jarrett}, {Kirkpatrick}, {Gizis}, {Howard}, {Evans}, {Fowler}, {Fullmer},
  {Hurt}, {Light}, {Kopan}, {Marsh}, {McCallon}, {Tam}, {Van Dyk}, \&
  {Wheelock}}]{Skrutskie2006}
{Skrutskie}, M.~F., {Cutri}, R.~M., {Stiening}, R., {et~al.} 2006, \aj, 131,
  1163, \dodoi{10.1086/498708}

\bibitem[{{Smith} {et~al.}(2009){Smith}, {Pontoppidan}, {Young}, {Morris}, \&
  {van Dishoeck}}]{Smith2009}
{Smith}, R.~L., {Pontoppidan}, K.~M., {Young}, E.~D., {Morris}, M.~R., \& {van
  Dishoeck}, E.~F. 2009, \apj, 701, 163, \dodoi{10.1088/0004-637X/701/1/163}

\bibitem[{{Sokal} {et~al.}(2020){Sokal}, {Johns-Krull}, {Mace}, {Nofi},
  {Prato}, {Lee}, \& {Jaffe}}]{Sokal2020}
{Sokal}, K.~R., {Johns-Krull}, C.~M., {Mace}, G.~N., {et~al.} 2020, \apj, 888,
  116, \dodoi{10.3847/1538-4357/ab59d8}

\bibitem[{{Sullivan} {et~al.}(2019){Sullivan}, {Wilking}, {Greene}, {Lisalda},
  {Gibb}, \& {Ejeta}}]{Sullivan2019}
{Sullivan}, T., {Wilking}, B.~A., {Greene}, T.~P., {et~al.} 2019, \aj, 158, 41,
  \dodoi{10.3847/1538-3881/ab24c0}

\bibitem[{{Tobin} {et~al.}(2015){Tobin}, {Dunham}, {Looney}, {Li}, {Chandler},
  {Segura-Cox}, {Sadavoy}, {Melis}, {Harris}, {Perez}, {Kratter},
  {J{\o}rgensen}, {Plunkett}, \& {Hull}}]{Tobin2015}
{Tobin}, J.~J., {Dunham}, M.~M., {Looney}, L.~W., {et~al.} 2015, \apj, 798, 61,
  \dodoi{10.1088/0004-637X/798/1/61}

\bibitem[{{Vacca} {et~al.}(2003){Vacca}, {Cushing}, \& {Rayner}}]{vacca2003}
{Vacca}, W.~D., {Cushing}, M.~C., \& {Rayner}, J.~T. 2003, \pasp, 115, 389,
  \dodoi{10.1086/346193}

\bibitem[{{Vogt} {et~al.}(1994){Vogt}, {Allen}, {Bigelow}, {Bresee}, {Brown},
  {Cantrall}, {Conrad}, {Couture}, {Delaney}, {Epps}, {Hilyard}, {Hilyard},
  {Horn}, {Jern}, {Kanto}, {Keane}, {Kibrick}, {Lewis}, {Osborne},
  {Pardeilhan}, {Pfister}, {Ricketts}, {Robinson}, {Stover}, {Tucker}, {Ward},
  \& {Wei}}]{Vogt1994}
{Vogt}, S.~S., {Allen}, S.~L., {Bigelow}, B.~C., {et~al.} 1994, in Society of
  Photo-Optical Instrumentation Engineers (SPIE) Conference Series, Vol. 2198,
  Instrumentation in Astronomy VIII, ed. D.~L. {Crawford} \& E.~R. {Craine},
  362, \dodoi{10.1117/12.176725}

\bibitem[{{White} {et~al.}(2007){White}, {Greene}, {Doppmann}, {Covey}, \&
  {Hillenbrand}}]{White2007}
{White}, R.~J., {Greene}, T.~P., {Doppmann}, G.~W., {Covey}, K.~R., \&
  {Hillenbrand}, L.~A. 2007, in Protostars and Planets V, ed. B.~{Reipurth},
  D.~{Jewitt}, \& K.~{Keil}, 117.
\newblock \doarXiv{astro-ph/0604081}

\bibitem[{{White} \& {Hillenbrand}(2004)}]{White2004}
{White}, R.~J., \& {Hillenbrand}, L.~A. 2004, \apj, 616, 998,
  \dodoi{10.1086/425115}

\bibitem[{{Yang} {et~al.}(2005){Yang}, {Johns-Krull}, \& {Valenti}}]{Yang2005}
{Yang}, H., {Johns-Krull}, C.~M., \& {Valenti}, J.~A. 2005, \apj, 635, 466,
  \dodoi{10.1086/497070}

\bibitem[{{Zapata} {et~al.}(2015){Zapata}, {Lizano}, {Rodr{\'\i}guez}, {Ho},
  {Loinard}, {Fern{\'a}ndez-L{\'o}pez}, \& {Tafoya}}]{Zapata2015}
{Zapata}, L.~A., {Lizano}, S., {Rodr{\'\i}guez}, L.~F., {et~al.} 2015, \apj,
  798, 131, \dodoi{10.1088/0004-637X/798/2/131}

\end{thebibliography}
\bibliographystyle{aasjournal}



\end{document}